\documentclass[prd,aps,amsfonts,showpacs,longbibliography,notitlepage,nofootinbib,superscriptaddress]{revtex4-1}

\usepackage{graphicx}
\usepackage[usenames,dvipsnames]{xcolor}
\usepackage{amsmath,amssymb,graphics,amsthm,isomath}
\usepackage{bbm}
\usepackage{bm}

\usepackage{comment}

\usepackage[colorlinks=true, urlcolor=violet, linkcolor=blue, citecolor=red, hyperindex=true, linktocpage=true]{hyperref}

\allowdisplaybreaks

\numberwithin{thm}{section}

\makeatletter
\renewcommand{\p@subsection}{}
\renewcommand{\p@subsubsection}{}
\makeatother

\usepackage{xcolor}
\usepackage{mathtools}
\usepackage[compat=1.0.0]{tikz-feynman}
\usepackage{subcaption}

\def\i{\text{i}}

\newcommand{\tr}{\mathrm{tr}}

\newcommand{\abs}[1]{\lvert{#1}\rvert}

\usepackage{dsfont}

\usepackage[dvipsnames]{xcolor}
\usepackage{bm}
\usepackage{bbm}
\usepackage{physics}
\usepackage{graphicx}

\begin{document}

\title{Floquet circuits inspired by holographic matrix models}

\author{Yun Ma}
\email{yun.ma@colorado.edu}
\affiliation{Department of Physics and Center for Theory of Quantum Matter, University of Colorado, Boulder, CO 80309, USA}

\author{Andrew Lucas}
\email{andrew.j.lucas@colorado.edu}
\affiliation{Department of Physics and Center for Theory of Quantum Matter, University of Colorado, Boulder, CO 80309, USA}

\date{\today}

\begin{abstract}
We argue that near-term experiments with neutral atoms in movable optical tweezers can simulate circuits that mimic the Trotterized time-evolution of simple matrix models in quantum mechanics.  As a cartoon of this proposal, we study Floquet Clifford circuits which exhibit a number of signatures of fast scrambling.  One such illustration is a simplified Hayden-Preskill recovery protocol, in which stabilizer quantum error correction replaces postselection.
\end{abstract}

\maketitle
\tableofcontents

\section{Introduction}
Some many-body quantum systems are believed to be ``holographically" dual to quantum gravity in one higher dimension.   This is most concretely illustrated in the context of the AdS/CFT correspondence \cite{Maldacena1998LargeN}, which relates certain conformal field theories with ``large-$N$ matrix" degrees of freedom to a theory of quantum gravity in asymptotically anti de Sitter space.   Although this conjecture originates in string theory, holographic models are also interesting because they are believed to saturate a number of practical bounds on quantum information processing, most famously a ``fast scrambling" bound \cite{SekinoSusskind2008FastScramblers} \begin{equation}
    t_{\mathrm{s}} \gtrsim \log N.  \label{eq:introscrambling}
\end{equation}  on the time needed to spread information amongst $N$ qubits using only few-qubit interactions.   Although the appropriate meaning of ``fast scrambling" has been debated \cite{Bentsen2019FastScramblingSparseGraphs,Lucas2019QuantumManyBodyDynamicsStarGraph,Harrow2021SeparationOTOCEntanglement}, it is tempting to speculate that a quantum mechanical system exhibiting fast scrambling might, in some circumstances, itself be a model of quantum gravity.

Unfortunately, many models of quantum gravity are quite difficult to study in a quantum simulator.  It is impossible to saturate \eqref{eq:introscrambling} in any $d$-dimensional lattice model, due to the Lieb-Robinson Theorem \cite{LiebRobinson1972FiniteGroupVelocity, ourreview}.  The most popular model of quantum holography, the Sachdev-Ye-Kitaev (SYK) model \cite{SachdevYe1993GaplessSpinFluid,KitaevSuh2018SoftModeSYK,Sachdev2015BekensteinHawkingStrangeMetals,MaldacenaStanford2016RemarksSYK}, naively requires $\sim N^4$ all-to-all random interactions amongst $N$ degrees of freedom.  This is difficult to realize in an experimental platform, although proposals do exist \cite{BaumgartnerSonner2024QuantumSimulationSYKOpticalCavities, UhrichBandyopadhyaySauerweinSonnerBrantutHauke2023CavityQEDSYK, PikulinFranz2017BlackHoleChip, ChewEssinAlicea2017ApproximatingSYKMajoranaWires, ChenIlanDeJuanPikulinFranz2018QuantumHolographyGraphene, Danshita2017SYKColdAtoms, Garc_a_lvarez_2017, Babbush2019SYKQubitization, Wei2021OpticalLatticeSYK}.

In this paper, we will instead explore a simple model of quantum dynamics \emph{inspired} by the original models of holography, namely matrix models \cite{BanksFischlerShenkerSusskind1997MatrixModel}.  See
\cite{BerensteinHanadaHartnoll2009MultiMatrixEmergentGeometry,Anninos2015LargeNMatrices,Hartnoll2018TopologicalOrderMatrixIsing,Hartnoll2017MatrixQMFromQubits,Frenkel:2023aft}
for even simpler quantum mechanical matrix models that exhibit some features of an emergent spacetime.   In a matrix model, degrees of freedom such as $\Phi_{ij}$ carry ``matrix indices":  $1\le i,j\le N$; the Hamiltonian of a matrix model is of the \emph{schematic} form 
\begin{equation}
    H = \cdots +\frac{g}{N} \mathrm{tr}\left(\Phi^4\right) + \cdots . \label{eq:intromatrix}
    \end{equation}
If the degrees of freedom $\Phi_{ij}$ were arranged in a two-dimensional array, as may be appropriate in any experimental quantum simulation of the matrix model, such an interaction term would look completely non-local in space.  On the other hand, we have $N^2$ degrees of freedom with $N^4$ non-random interactions (at least in the $g$ term above), so the situation could be more promising than SYK, which would require $N^8$ interactions amongst $N^2$ degrees of freedom.

Still, because the $N^2$ degrees of freedom interact in a complicated way, it may feel hopeless to realize a matrix model natively in a quantum simulator.  A reasonable alternative is to simulate quantum dynamics in a matrix model by ``digitizing" it into a quantum circuit, and hope that the layout of the circuit could be efficiently realized in some platform.   

We will indeed argue that there is some room for cautious optimism -- by using Rydberg atom qubits stored in movable optical tweezer arrays \cite{Saffman2010QuantumInformationRydberg,Kaufman2021QuantumScienceTweezers,Bluvstein2021ControllingRydbergDynamics,Cong2022FaultTolerantRydbergQC,Bluvstein2022CoherentTransportProcessor,Jenkins2022YtterbiumTweezerQubits}
, the interaction connectivity of the matrix model Hamiltonian can be realized in a parallelizable fashion, using present-day capabilities.  To see why this is possible, let us imagine a simplified set-up where we wish to evaluate \begin{equation}
    \mathrm{tr}\left( \Phi \Phi^{\mathsf{T}} \Phi \Phi^{\mathsf{T}}\right) = \sum_{i,j,k,l=1}^N \Phi_{ij}\Phi_{kj}\Phi_{kl}\Phi_{il}. \label{eq:toyPhi}
\end{equation}
Let us store the matrix qubits $\Phi_{ij}$ in a two-dimensional array.  Implementing a gate where $|i-k|$ and $|j-l|$ are small is then ``natural": the Rydberg blockade strongly enhances interactions between nearby atoms in space.    To implement the full matrix model, it would suffice to permute the indices by rearranging rows and columns so that, eventually, all of the pairs $i,j,k,l$ in \eqref{eq:toyPhi} are brought close together.   We will explicitly describe one way to implement these permutations in a ``parallel way" in neutral atom arrays.

The above argument suggests that matrix models have a better interaction pattern than SYK.  Still, this does not mean that it is easy to simulate them.  Indeed, in the near term, it seems challenging to realize a large enough circuit depth to perform any faithful quantum simulation of ``digitized" matrix model dynamics, which would likely require 100s of sequential gates.  (Quantum computations with Rydberg atoms often hope for an order of magnitude fewer gates \cite{Saffman2016RydbergReview} before error correction is important). Therefore, before embarking on an exhaustive analysis of the resource requirements needed to simulate a matrix model using Rydberg atoms, in this paper we will sketch how such a simulation might be possible using a cartoon model that is easy to efficiently study using classical computers.  Our goal will be to propose a specific method for realizing the connectivity of a matrix model using experiments with Rydberg atom qubits that physically move atomic qubits in space, without requiring sophisticated atom-by-atom spatial rearrangements.  To do so, we will show that Floquet Clifford circuits can exhibit simplified versions of holographic dynamics like fast scrambling, while it is also easy to recover the scrambled information using a simplified version \cite{Vikram2026BidirectionalTeleportation} of the Hayden-Preskill protocol \cite{HaydenPreskill2007}.  Such simplified experiments may be important test cases for probing the quantum coherence of a simulator before attempting a more honest microscopic simulation.  

The outline of the paper is as follows:  in Section \ref{cartoon} we introduce a cartoon version of the story, explaining how to store degrees of freedom in a matrix model and how to realize fast scrambling with simple circuits.  Section \ref{clifford dynamics} proposes and analyzes a concrete Floquet circuit that uses the kinds of gates available in present-day neutral atom quantum processors.   Section \ref{double_layer} then describes some modifications to the model that make it particularly amenable to a near-term experiment.

\section{Cartoon model}\label{cartoon}
In this section, we begin by introducing a simple set of cartoons for discrete time ``digitized" matrix model dynamics.
\subsection{Set up of the problem}\label{setup}
We study a cartoon model of a quantum circuit with non-random interactions between qubits that have the interaction graph of a matrix model.  Let us first explain why we can (in principle) study circuits, rather than continuous time evolution, even if we wish to analyze a matrix model such as \eqref{eq:intromatrix}.   Let $N$ be some positive integer (we will usually assume it is divisible by 4, for reasons we will explain shortly). We can expand $\mathrm{tr}(\Phi^4)$ as
\begin{align}
   \tr(\Phi^4) &= \sum_{i,j,k,l=1}^N \Phi_{ij}\Phi_{jk}\Phi_{kl}\Phi_{li}.  \label{eq:trace_expansion}
\end{align}
  $N^4 = (N^2)^2$, so the number of interactions formally grows superextensively with the number of qubits.  We conjecture that one can, with reasonable accuracy,  simulate the matrix model as follows: naively approximate the time-evolution operator at each time step $\Delta t$ as
\begin{align}
    \exp(\tr(\Phi^4)\Delta t) &= \exp(\sum_{ijkl}\Phi_{ij}\Phi_{jk}\Phi_{kl}\Phi_{li}\Delta t) \approx \prod_{ijkl}\exp(\Phi_{ij}\Phi_{jk}\Phi_{kl}\Phi_{li}\Delta t),
    \label{eq:trotterization}
\end{align}
using some discretized circuit; the approximation in the last step corresponds to Trotterization.  The intuition is that the Trotterized matrix model dynamics is conceptually similar to 
\begin{align}
   \left(\prod_{\text{distinct fraction $p$ of $ijkl$}}\exp(\Phi_{ij}\Phi_{jk}\Phi_{kl}\Phi_{li}\frac{\Delta t}{p}) 
    \right)^{t/\Delta t}&\sim  \left(U\right)^{ct}.\label{eq:trott_approximation}
\end{align}
for unitary matrix $U$, some O(1) constant $c$, and sufficiently late time $t$.   Ideally, the fraction $p$ of gates implemented in each time step might be very small, even $\sim 1/N^2$ (each qubit interacts with a finite number of others at a time), but it can be changed during each time window so as to more faithfully ``sample" the interactions of the original matrix model. Crude justification for this idea comes from the study of ``sparse SYK models" \cite{XuSusskindSuSwingle2020SparseQuantumHolography} which exhibit similar physics to the all-to-all connected SYK, with fewer interactions.  Ultimately, we will simply take a sparse discretization as our starting point, leaving a detailed analysis of the legitimacy of this approximation to later work.

We now explain a concrete way to realize something in the spirit of \eqref{eq:trott_approximation}, with as little ``randomness" as possible.  (This will be highly desirable for any experimental implementation.)  We introduce a structured permutation of the row and column indices of the qubits in an $N\times N$ qubit matrix $Q$: for any row or column index $i\in\{1,2,...,N\}$, let the permuted index $\sigma(i)$ be
\begin{align}
    \sigma(i) = \begin{cases}
    \frac{i+1}{2}, & \text{if $i$ is odd}. \\
    \frac{i+N}{2}, & \text{if $i$ is even}. 
    \label{eq:sigma}
    \end{cases}
\end{align}
where $\sigma$ is the corresponding shuffling permutation, which groups all odd indices together and all even indices together, and is particularly simple to realize experimentally (see Section \ref{double_layer}).  Thus, the qubit at the $(i,j)$-entry in $Q$ is swapped to $(\sigma(i),\sigma(j))$-entry according to (\ref{eq:sigma}) after permutation. We can equivalently think of this operation as implementing the unitary gate $U_{\mathrm{perm}}$, defined via its action
\begin{equation}
U_{\mathrm{perm}}^\dagger \mathcal{O}_{ij} U_{\mathrm{perm}} = \mathcal{O}_{\sigma(i)\sigma(j)}
\end{equation}\label{eq:permutation_matrix}
on all single-qubit operators $\mathcal{O}$.

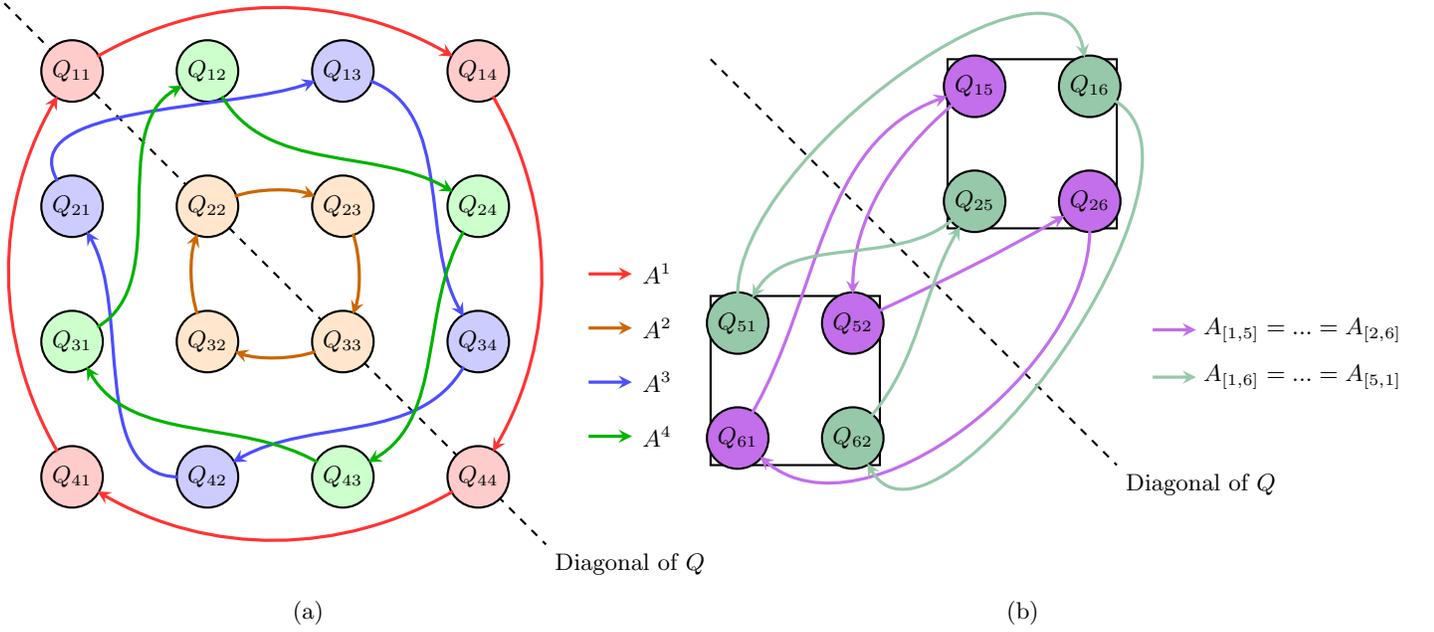
\begin{figure}[t]
    \vspace{-0.4in}
    \centering
    \begin{subfigure}[t]{0.47\linewidth}
        \centering
        \begin{tikzpicture}[scale=0.9, baseline=(current bounding box.south)]
        \tikzset{
            qnode/.style={circle, draw, thick, minimum size=14pt, inner sep=2pt, text width=16pt, align=center},
            Aone/.style={->, >=stealth, very thick, red!80, shorten <=11pt, shorten >=11pt},
            Atwo/.style={->, >=stealth, very thick, orange!80!black, shorten <=11pt, shorten >=11pt},
            Athree/.style={->, >=stealth, very thick, blue!70, shorten <=11pt, shorten >=11pt},
            Afour/.style={->, >=stealth, very thick, green!70!black, shorten <=11pt, shorten >=11pt},
        } 
        
        \coordinate (Q11) at (0,4);
        \coordinate (Q12) at (2,4);
        \coordinate (Q13) at (4,4);
        \coordinate (Q14) at (6,4);
        
        \coordinate (Q21) at (0,2);
        \coordinate (Q22) at (2,2);
        \coordinate (Q23) at (4,2);
        \coordinate (Q24) at (6,2);
        
        \coordinate (Q31) at (0,0);
        \coordinate (Q32) at (2,0);
        \coordinate (Q33) at (4,0);
        \coordinate (Q34) at (6,0);
        
        \coordinate (Q41) at (0,-2);
        \coordinate (Q42) at (2,-2);
        \coordinate (Q43) at (4,-2);
        \coordinate (Q44) at (6,-2);
        
        \draw[dashed, thick] (-1,5) -- (7,-3)
            node[below right] {Diagonal of $Q$};
        
        \node[qnode, fill=red!20]   at (Q11) {$Q_{11}$};
        \node[qnode, fill=green!20] at (Q12) {$Q_{12}$};
        \node[qnode, fill=blue!20]  at (Q13) {$Q_{13}$};
        \node[qnode, fill=red!20]   at (Q14) {$Q_{14}$};
        
        \node[qnode, fill=blue!20]   at (Q21) {$Q_{21}$};
        \node[qnode, fill=orange!20] at (Q22) {$Q_{22}$};
        \node[qnode, fill=orange!20] at (Q23) {$Q_{23}$};
        \node[qnode, fill=green!20]  at (Q24) {$Q_{24}$};
        
        \node[qnode, fill=green!20]  at (Q31) {$Q_{31}$};
        \node[qnode, fill=orange!20] at (Q32) {$Q_{32}$};
        \node[qnode, fill=orange!20] at (Q33) {$Q_{33}$};
        \node[qnode, fill=blue!20]   at (Q34) {$Q_{34}$};
        
        \node[qnode, fill=red!20]   at (Q41) {$Q_{41}$};
        \node[qnode, fill=blue!20]  at (Q42) {$Q_{42}$};
        \node[qnode, fill=green!20] at (Q43) {$Q_{43}$};
        \node[qnode, fill=red!20]   at (Q44) {$Q_{44}$};
        
        
        \draw[Aone] (Q11) to[out=30,  in=150] (Q14);
        \draw[Aone] (Q14) to[out=-60, in=60]  (Q44);
        \draw[Aone] (Q44) to[out=210, in=-30] (Q41);
        \draw[Aone] (Q41) to[out=120, in=-120](Q11);
        
        \draw[Athree] (Q13) to[out=-20, in=120]   (Q34);
        \draw[Athree] (Q34) to[out=-120,in=30]    (Q42);
        \draw[Athree] (Q42) to[out=180, in=-60]  (Q21);
        \draw[Athree] (Q21) to[out=120, in=200]  (Q13);
        
        \draw[Atwo] (Q22) to[out=20,  in=160] (Q23);
        \draw[Atwo] (Q23) to[out=-70, in=70]  (Q33);
        \draw[Atwo] (Q33) to[out=200, in=-20] (Q32);
        \draw[Atwo] (Q32) to[out=110, in=-110](Q22);
        
        \draw[Afour] (Q12) to[out=-60, in=150] (Q24);
        \draw[Afour] (Q24) to[out=-120,in=30]  (Q43);
        \draw[Afour] (Q43) to[out=150, in=-60] (Q31);
        \draw[Afour] (Q31) to[out=30,  in=-150](Q12);
        
        \begin{scope}[shift={(8,1)}]
            \draw[Aone]   (-0.8,0)   -- (0.7,0);   \node[right] at (0.3,0)   {$A^1$};
            \draw[Atwo]   (-0.8,-0.8)-- (0.7,-0.8);\node[right] at (0.3,-0.8){$A^2$};
            \draw[Athree] (-0.8,-1.6)-- (0.7,-1.6);\node[right] at (0.3,-1.6){$A^3$};
            \draw[Afour]  (-0.8,-2.4)-- (0.7,-2.4);\node[right] at (0.3,-2.4){$A^4$};
        \end{scope}  
        \path (current bounding box.south west) --
          (current bounding box.north east);
        \end{tikzpicture}
    \caption{}
    \label{fig:diagonal rules}
    \end{subfigure}
\hfill
    \begin{subfigure}[t]{0.47\linewidth} 
        \centering
        \begin{tikzpicture}[scale=0.9, baseline=(current bounding box.south)]
        \definecolor{orbmag}{RGB}{195,110,235}   
        \definecolor{orbgrn}{RGB}{150,200,170}   
        \tikzset{
          qnode/.style={circle, draw, thick, minimum size=14pt, inner sep=2pt, text width=16pt, align=center},
          purpleorb/.style={->, >=stealth, very thick, orbmag,
                            shorten <=11pt, shorten >=11pt},
          greenorb/.style={->, >=stealth, very thick, orbgrn,
                           shorten <=11pt, shorten >=11pt},
        }
        \draw[dashed, thick] (0,6) -- (6,0)
          node[below right] {Diagonal of $Q$};
        \draw[thick] (0,0) rectangle (2.5,2.5);
        \draw[thick] (3.5,3.5) rectangle (6,6);
        \coordinate (Q51) at (0.4,2.1);  
        \coordinate (Q52) at (2.1,2.1);  
        \coordinate (Q61) at (0.4,0.4);  
        \coordinate (Q62) at (2.1,0.4);  
        
        \coordinate (Q15) at (3.9,5.6);  
        \coordinate (Q25) at (3.9,3.9);  
        \coordinate (Q16) at (5.6,5.6);  
        \coordinate (Q26) at (5.6,3.9);  
        
        \node[qnode, fill=orbmag] at (Q15) {$Q_{15}$};
        \node[qnode, fill=orbmag] at (Q52) {$Q_{52}$};
        \node[qnode, fill=orbmag] at (Q61) {$Q_{61}$};
        \node[qnode, fill=orbmag] at (Q26) {$Q_{26}$};
        
        \node[qnode, fill=orbgrn]   at (Q16) {$Q_{16}$};
        \node[qnode, fill=orbgrn]   at (Q62) {$Q_{62}$};
        \node[qnode, fill=orbgrn]   at (Q25) {$Q_{25}$};
        \node[qnode, fill=orbgrn]   at (Q51) {$Q_{51}$};
        
        \draw[purpleorb] (Q15) to[out=-140, in=90]  (Q52);  
        \draw[purpleorb] (Q52) to[out=25,in=-150]   (Q26);  
        \draw[purpleorb] (Q26) to[out=-90, in=-40]  (Q61);  
        \draw[purpleorb] (Q61) to[out=60,  in=-160] (Q15);  
        
        \draw[greenorb] (Q16) to[out=-30, in=-60]   (Q62);  
        \draw[greenorb] (Q62) to[out=50, in=-120] (Q25);  
        \draw[greenorb] (Q25) to[out=-140,  in=60] (Q51);  
        \draw[greenorb] (Q51) to[out=90, in=100](Q16);  
        
        \begin{scope}[shift={(6.1,2)}]
          \draw[purpleorb] (0.0,0) -- (1.5,0);
          \node[right] at (1.05,0) {$A_{[1,5]}=...=A_{[2,6]}$};

          \draw[greenorb] (0.0,-0.7) -- (1.5,-0.7);
          \node[right] at (1.05,-0.7)
            {$A_{[1,6]}=...=A_{[5,1]}$};
        
        \end{scope}
        \path (current bounding box.south west) --
          (current bounding box.north east);
        \end{tikzpicture}
        \caption{}
        \label{fig:off diagonal rules}
    \end{subfigure}
    \caption{Illustration of: (a) the rule defined in (\ref{eq:diagonal_rule}); (b) the rule defined in (\ref{eq:off-diagonal_rule}).}
    \label{fig:connectivity_rules}
\end{figure}

Now we introduce the gates we apply after permutation. We partition the $N^2$ qubits into subsets of four qubits with cyclically alternating indices, and interactions are restricted to qubits within the same subset. We first define a $4\times 4$ diagonal subblock $D_b$ in $Q$ to be a subset containing sixteen qubits,
\begin{align}
    D_b = \{Q_{pq} | p,q\in\{4b-3,4b-2,4b-1,4b\}\}\label{eq:diagonal_subblock}
\end{align}
where $b\in\{1,2,...,\frac{N}{4}\}$. Then we can write down four subsets of $D_b$ that satisfy the cyclic connectivity,
\begin{align}
\begin{split}
    A_b^1 &= \{Q_{4b-3\;4b-3}, Q_{4b-3\;4b}, Q_{4b\;4b}, Q_{4b\;4b-3}\}\\
    A_b^2 &= \{Q_{4b-2\;4b-2}, Q_{4b-2\;4b-1}, Q_{4b-1\;4b-1}, Q_{4b-1\;4b-2}\}\\
    A_b^3 &= \{Q_{4b-3\;4b-1}, Q_{4b-1\;4b}, Q_{4b\;4b-2}, Q_{4b-2\;4b-3}\}\\
    A_b^4 &= \{Q_{4b-3\;4b-2}, Q_{4b-2\;4b}, Q_{4b\;4b-1}, Q_{4b-1\;4b-3}\}.\label{eq:diagonal_rule}
\end{split}
\end{align}
In addition, we define the cyclic connectivity rule for qubits outside such $4 \times 4$ diagonal subblocks to be
\begin{align}
    A_{[i,j]} = \{Q_{ij}, Q_{j \; i+(-1)^{i+1}},  Q_{i+(-1)^{i+1} \; j+(-1)^{j+1}},  Q_{j+(-1)^{j+1} \; i}\}, \label{eq:off-diagonal_rule}
\end{align}
Notice that
\begin{align}
    A_{[i,j]} &= A_{[j,i+(-1)^{i+1}]} = A_{[i+(-1)^{i+1},j+(-1)^{j+1}]} = A_{[j+(-1)^{j+1},i]},\label{eq:qubits_in_same_subset}
\end{align}
and that the union of subsets in the diagonal subblocks in (\ref{eq:diagonal_rule}) and subsets in the off-diagonal subblocks in (\ref{eq:off-diagonal_rule}) form a partition of all $N^2$ qubits in $Q$. 
Figure \ref{fig:connectivity_rules} shows example diagrams for the two cyclic connectivity rules defined in (\ref{eq:diagonal_rule}) and (\ref{eq:off-diagonal_rule}), from which it is clear that the connectivity within each $4\times4$ diagonal subblocks remains confined to that subblock, and the connectivity outside the diagonal subblocks alternates between the off-diagonal $2\times2$ subblocks that are symmetric with respect to the main diagonal. We refer to this partition scheme as Rule 1. In later sections of this paper, we also examine the case where every $2\times2$ subblock of $Q$, including those within the $4\times4$ digaonal subblocks, obeys (\ref{eq:off-diagonal_rule}). We refer to this as Rule 2.  Rule 2 will later be used as a way to simplify an experimental setup.
We will see later that the simplified Rule 2 can potentially lead to some pathologies for certain $N$ (see Appendix \ref{disjoint_sets}).  

Having defined the interacting subsets of qubits, we consider a series of four-qubit gates -- mimicking $\mathrm{tr}(\Phi^4)$ --  on each subset to model the interactions between qubits in the cartoon matrix model.  Overall, the interactions in a given time step correspond to the (schematic) gate
\begin{align}
    U_{\mathrm{int}} &= \prod_{N^2/4 \text{ subsets } a=\lbrace i,j,k,l\rbrace } U_{ijkl} =  \prod_{a=1}^{N^2/4} W_{a},\label{eq:U_int}
\end{align}
where each grouping of $ijkl$ is chosen according to Rule 1 or 2 above, and  $W_{a}$ denotes the four-qubit gate acting on qubits in the interacting subset $A_a$ defined in (\ref{eq:diagonal_rule}) and (\ref{eq:off-diagonal_rule}) with $a$ being the index of the subsets ranging from 1 to $N^2/4$. 

The overall unitary circuit $U$ generating the dynamics at each time step $\Delta t$ is constructed from an index permutation followed by interactions defined through cyclic connectivity and the multi-qubit gates acting on each subset, i.e.
\begin{align}
    U &:= U_{\text{int}} U_{\text{perm}} = \left(\prod_{a=1}^{N^2/4} W_a\right) U_{\mathrm{perm}}.\label{eq:U}
\end{align} Together, these two operations are our cartoon for the dynamics in (\ref{eq:trott_approximation}). 

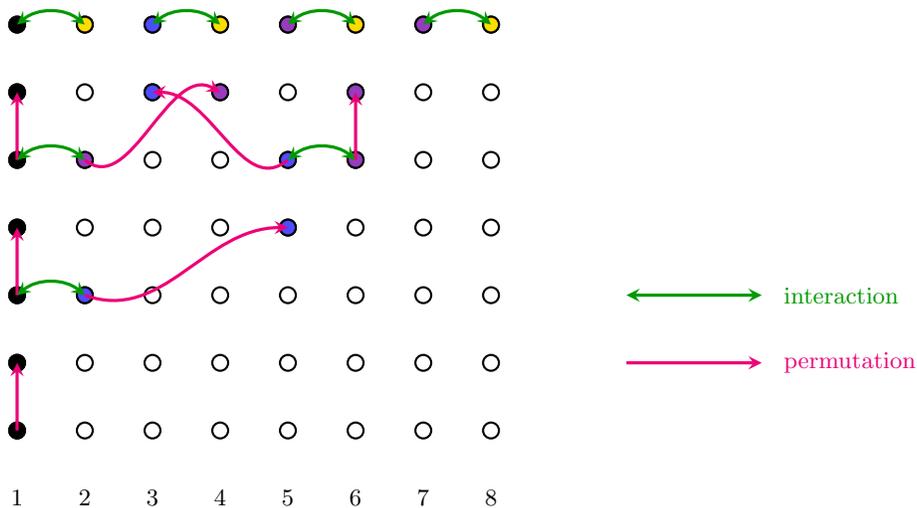
\begin{figure}[t]
    \definecolor{myPurple}{RGB}{150,60,190}
    \centering
    \begin{tikzpicture}[x=0.9cm,y=0.9cm,>=stealth]
    \tikzset{
      site/.style={circle,draw,thick,minimum size=6pt,inner sep=0pt},
      blackdot/.style={site,fill=black},
      bluedot/.style={site,fill=blue!70},
      purpledot/.style={site,fill=myPurple},
      yellowdot/.style={site,fill=yellow!80!orange},
      perm/.style={->,very thick,draw=magenta!80!red},
      inter/.style={<->,very thick,draw=green!60!black}
    }
    
    \begin{scope}
    
    \foreach \x in {1,...,8}
      \node at (\x,0) {\x};
    
    \foreach \y in {1,...,7}{
      \foreach \x in {1,...,8}{
        \node[site] at (\x,\y) {};
      }
    }
    
    
    \node[blackdot] at (1,1) {};
    
    \node[blackdot] at (1,2) {};
    
    \node[blackdot] at (1,3) {};
    \node[bluedot]   at (2,3) {};
    
    \node[blackdot] at (1,4) {};
    \node[bluedot]   at (5,4) {};
    
    \node[blackdot]   at (1,5) {};
    \node[purpledot] at (2,5) {};
    \node[bluedot]    at (5,5) {};
    \node[purpledot] at (6,5) {};
    
    \node[blackdot]   at (1,6) {};
    \node[bluedot]    at (3,6) {};
    \node[purpledot] at (4,6) {};
    \node[purpledot] at (6,6) {};
    
    \node[blackdot]   at (1,7) {};
    \node[yellowdot]  at (2,7) {};
    \node[bluedot]    at (3,7) {};
    \node[yellowdot]  at (4,7) {};
    \node[purpledot] at (5,7) {};
    \node[yellowdot]  at (6,7) {};
    \node[purpledot] at (7,7) {};
    \node[yellowdot]  at (8,7) {};
    
    
    \draw[perm] (1,1) to[out=90,in=-90] (1,2);
    \draw[perm] (1,3) to[out=90,in=-90] (1,4);
    \draw[perm] (1,5) to[out=90,in=-90] (1,6);
    
    \draw[perm] (2,3) to[out=-20,in=180] (5,4);
    
    \draw[perm] (2,5) to[out=-40,in=150] (4,6);
    \draw[perm] (5,5) to[out=-140,in=0]  (3,6);
    \draw[perm] (6,5) to[out=90,in=-90]  (6,6);
    
    
    \draw[inter] (1,3) to[bend left=40] (2,3);
    
    \draw[inter] (1,5) to[bend left=40] (2,5);
    \draw[inter] (5,5) to[bend left=40] (6,5);
    
    \draw[inter] (1,7) to[bend left=40] (2,7);
    \draw[inter] (3,7) to[bend left=40] (4,7);
    \draw[inter] (5,7) to[bend left=40] (6,7);
    \draw[inter] (7,7) to[bend left=40] (8,7);
    
    \end{scope}
    
    
    \draw[perm] (10,2) -- (12,2);
    \node[anchor=west,text=magenta!80!red] at (12.2,2) {permutation};
    
    \draw[inter] (12,3) -- (10,3);
    \node[anchor=west,text=green!60!black] at (12.2,3) {interaction};
    
    \end{tikzpicture}
    \caption{The dynamics of a row/column infection process driven by permutations and cyclic connectivity is demonstrated. At $t=0$, the first qubit is infected, and by $t=3$ all eight qubits are infected, indicating full scrambling of the system.}
    \label{fig:1Dsystem_diagram}
\end{figure}

A diagram for an intuitive picture of how the combined effect of row and column permutations and cyclic connectivity leads to fast scrambling is shown in Figure \ref{fig:1Dsystem_diagram}, in the case of a simple 1D system with $N=8$, i.e. a simple $1\times8$ row of qubits labeled 1, 2 , …, 8 from left to right. Within three time steps of applying index permutation and interactions with cyclic connectivity, the dynamics spreads across the entire system.  It is easy to see (but harder to draw) that the same dynamics causes very fast scrambling in the corresponding $8\times 8$ matrix model.

In the rest of this paper, we present numerical simulations showing that our model, which includes only a subset of the $N^4$ interactions, exhibits fast scrambling and other simple dynamical signatures of chaos and holography.  Different choices of $U_{\text{int}}$ can lead to similar qualitative dynamics, and the rest of the paper explores the further details.

\subsection{Random unitary circuits and infection model}\label{RUC&infection}
Intuitively, we can think of the scrambling of quantum information as an infection process
\cite{Bentsen2019FastScramblingSparseGraphs, NahumRuhmanVijayHaah2017EntanglementGrowth,LashkariStanfordHastingsOsborneHayden2013FastScramblingConjecture,RobertsStanfordStreicher2018OperatorGrowthSYK,QiStreicher2019QuantumEpidemiologySYK}: suppose initially one qubit is infected, representing an initial qubit where we have applied a perturbation.  Intuitively, the gates that we apply to the system cause infected qubits to infect ``healthy qubits", just as a generic gate will spread the information between multiple qubits (as can be seen by considering operator growth in the Heisenberg picture).  Since the chaotic dynamics driven by a sufficiently complicated fixed Hamiltonian can be expected to resemble those of a random unitary circuit, we first consider when the four-qubit gates are Haar random unitaries: at every time step, we randomly choose the four-qubit gate for each interacting subset of qubits in $\mathrm{U}(2^4)$.  It is easiest to analyze the resulting dynamics in the limit where the on-site Hilbert space dimension $q \gg 2$ is very large \cite{NahumRuhmanVijayHaah2017EntanglementGrowth}, as in this regime infected qubits never become uninfected.\footnote{It is simple to undertsand why this is the case.  Consider for simplicity a 2-qudit gate acting on the operator $\mathcal{O}_1I_2$ or $\mathcal{O}_1\mathcal{O}_2$ where $\mathcal{O}_{1,2}$ represent non-identity operators.  Under a random unitary $U$, $U^\dagger \mathcal{O}_1I_2 U$ evolves into a superposition of all $q^4-1$ operators that are not $I_1I_2$; the (Frobenius) weight of each has the same expectation value.  The probability of seeing an operator that has identity one site is $2(q^2-1)/(q^4-1)$ which clearly vanishes as $q\rightarrow\infty$.}  The infection therefore tracks which qubits have a chance to interact with which other qubits, or equivalently the largest possible operator that could appear in Heisenberg time evolution.

\begin{figure}[h]
    \centering
    \begin{subfigure}[h]{0.5\textwidth}
        \centering
        \includegraphics[scale=0.5]{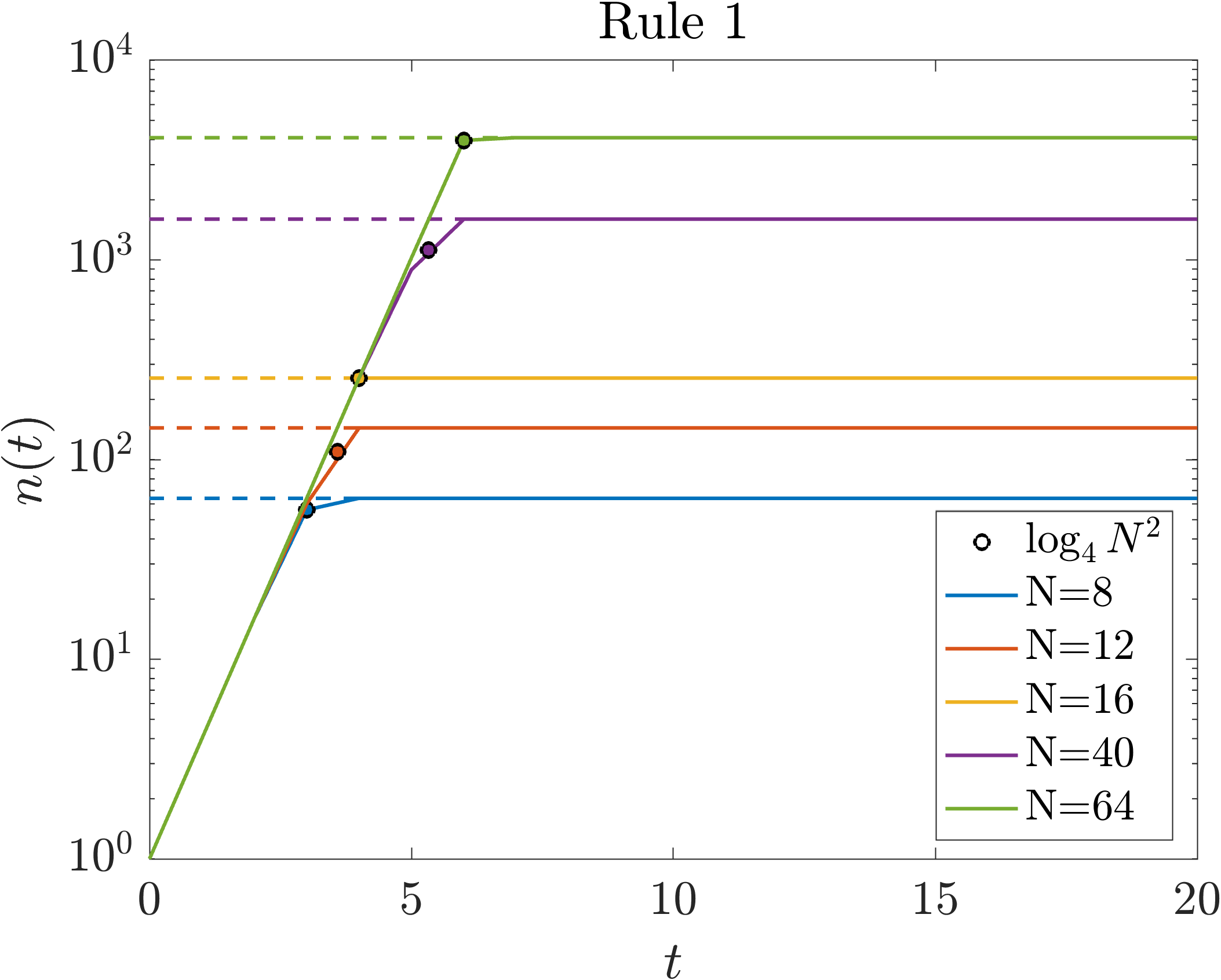}
        \caption{}
        \label{fig:infection_different}
    \end{subfigure}%
    \begin{subfigure}[h]{0.5\textwidth}
        \centering
        \includegraphics[scale=0.5]{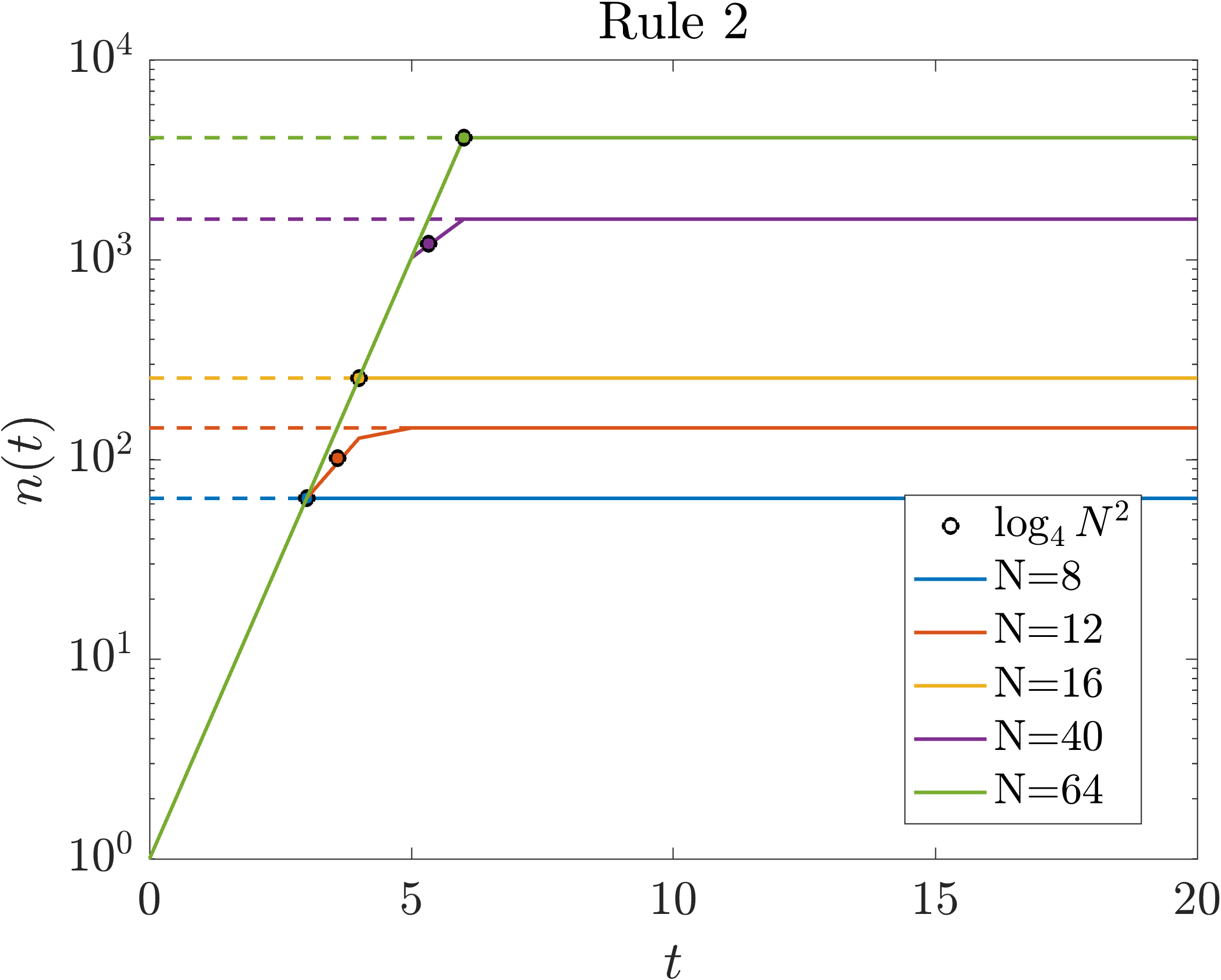}
        \caption{}
        \label{fig:infection_same}
    \end{subfigure}
    \caption{Growth of the number of infected qubits $n(t)$ as a function of time $t$ in random unitary circuits. The onset of full scrambling occurs no earlier than $\log_4 N^2$.}
    \label{fig:RUC_dynamics}
\end{figure}

We can easily confirm numerically that our fixed permutation gate, together with random $U_{\mathrm{int}}$ acting on large-$q$ qudits, exhibits exponentially fast operator growth.   Figure \ref{fig:RUC_dynamics} shows that the number $n(t)$ of infected qubits grows exponentially with time under the proposed permutation $\sigma$ in (\ref{eq:sigma}) for both Rule 1 and Rule 2, and eventually saturates at $N^2$. Since each infected qubit can infect three other qubits (thus operator size can at most increase by 4 in each circuit step), we know the scrambling time $t_{\mathrm{s}}$ required to infect all qubits obeys \begin{equation}
    t_{\mathrm{s}} \ge \log_4N^2.
\end{equation}
In fact, our numerics shows that this bound is essentially as tight as possible for any $N$. In particular, as shown in Figure \ref{fig:infection_same}, the bound is saturated when $N$ is a power of two under Rule 2, the bound is saturated, consistent with the dynamics illustrated in Figure \ref{fig:1Dsystem_diagram}.

There is another way to summarize this result, which will be useful to us later.  For a general gate $W_a$, we expect that the average operator size $n(t)$ grows exponentially with time.  This implies that
\begin{align}
    t_s &= \lambda_{\mathrm{L}}^{-1}\ln{N^2} + \cdots , 
\end{align}
where $\lambda_{\mathrm{L}}$ is a Lyapunov exponent and $\cdots$ is an $N$-independent offset depending on the details of the operator size at which we count $n(t)\sim N^2$ as a ``scrambled" operator. For large-$q$ random unitary circuits,
\begin{align}
    \lambda_{\mathrm{L}} &= \ln{4}\nonumber \\
     &\approx 1.39.
    \label{Lyapunov}
\end{align}

\section{Clifford dynamics}\label{clifford dynamics}
Random unitary circuits are convenient analytical tools, but there are a number of shortcomings if our inspiration is to mock a matrix model.  Most importantly, there is no fixed Hamiltonian and no notion of low temperature states (where one might expect a semiclassical black hole geometry to form).  As a partial step towards simulating an honest Hamiltonian, we might wish to replace the random unitary circuit in time with a fixed unitary $U$ applied repeatedly in time.  This system has discrete time-translation symmetry and is referred to as Floquet dynamics. 

We will therefore look for a simple model of Floquet dynamics, with the same interactions as before, to verify that similar fast scrambling physics arises. To maintain some tractability, however, we must make another assumption, namely that the Floquet time-evolution operator we apply is a Clifford gate. Although we do not claim this is a ``faithful" model of a real matrix model, this simplification has a number of conceptual as well as practical advantages.  Firstly, Clifford dynamics is efficiently simulatable classically, so we can explicitly compute more complex protocols such as a Hayden-Preskill-inspired recovery algorithm for scrambled information \cite{YoshidaKitaev2017} with hundreds of qubits.  Secondly, high-fidelity Clifford gates are an active goal of experimental research in atomic physics \cite{Evered2023HighFidelityNeutralAtoms,Bluvstein2024LogicalProcessor,Senoo2025MultiQubitMapping}
, in particular in studies of neutral atom quantum computing using mobile optical tweezer arrays.   This experimental platform is the most natural (that we are aware of) to realize our desired interactions, as we will discuss in Section \ref{double_layer}.

\subsection{Construction of unitary gates within Clifford framework}\label{clifford setup}

When studying Clifford dynamics, one prefers to study the time evolution of a state $|\psi(t)\rangle$ by keeping track of the Heisenberg time-evolution of its stabilizers $S(t)$: namely the operators such that $S(t)|\psi(t)\rangle = |\psi(t)\rangle$.  If the initial state is $|000\rangle$ --
we will frequently assume there are $Z_1,\ldots, Z_{N^2}$ at $t=0$ -- then because the time-evolution is Clifford, the stabilizers will remain Pauli strings.
Therefore, in our cartoon model we can simply study the time evolution of  the operators $X$ and $Z$ for each qubit (the most general operator is just a product of these). 
As is standard, if we are not interested in the overall phase of these operators, we may further represent an arbitrary such operator acting on $n$ qubits as a binary vector of length $2n$. The first $n$ components correspond to the presence of $X$ operators $X_1, X_2, ..., X_n$, while the second $n$ components correspond to the presence of $Z$ operators $Z_1, Z_2, ..., Z_n$. 
We use lowercase letters to denote the binary representations. For example, in the two-qubit case, we assign the binary representations as follows:
\begin{align}
    \begin{split}
        X_1\rightarrow
            \vec{x_1} = \begin{bmatrix}
                1\\
                0\\
                0\\
                0
            \end{bmatrix},
        Z_1\rightarrow
            \vec{z_1} = \begin{bmatrix}
                0\\
                0\\
                1\\
                0
            \end{bmatrix},
        I_1\rightarrow
            \vec{i_1} = \begin{bmatrix}
                0\\
                0\\
                0\\
                0
            \end{bmatrix},
        Y_1\rightarrow
            \vec{y_1} = \begin{bmatrix}
                1\\
                0\\
                1\\
                0
            \end{bmatrix},\\    
        X_2\rightarrow
            \vec{x_2} = \begin{bmatrix}
                0\\
                0\\
                1\\
                0
            \end{bmatrix},
        Z_2\rightarrow
            \vec{z_2} = \begin{bmatrix}
                0\\
                0\\
                0\\
                1
            \end{bmatrix},
        I_2\rightarrow
            \vec{i_2} = \begin{bmatrix}
                0\\
                0\\
                0\\
                0
            \end{bmatrix},
        Y_2\rightarrow
            \vec{y_2} = \begin{bmatrix}
                0\\
                1\\
                0\\
                1
            \end{bmatrix}.
    \end{split}\label{eq:example_binary_representation}
\end{align}
For any Pauli operators $P_1$ and $P_2$ (not necessarily single-qubit operators as in \eqref{eq:example_binary_representation}), we have
\begin{align}
    P_1 P_2 = (\vec{p_1} + \vec{p_2}) \text{ (mod 2)}.\label{eq:multiplication_binary_rule}
\end{align}
Henceforth we leave the binary addition implicit when we use this ``stabilizer formalism".

We now construct the necessary gates mentioned in Section \ref{cartoon} within the Clifford framework. In our model, each qubits is labeled by an entry in a matrix, and so it is indexed by a pair of indices $ij$.  If under the binary representation a Pauli matrix
\begin{align}
    \mathcal{O}_{ij} &\rightarrow o_{ij},\label{eq:binary representation of single-qubit operator}
\end{align}
then for any Clifford unitary $U_k$ there exists a $2n\times 2n$ binary matrix $\mathcal{U}_k$ such that 
\begin{align} \label{eq:cliffordrule}
    U_k^\dagger \mathcal{O}_{ij} U_k &\rightarrow \mathcal{U}_k o_{ij}.
\end{align}
Let us then describe how to build our gates and express them in the Clifford framework.  We start with the overall qubit permutations which take row/column $i$ to $\sigma(i)$ for permutation $\sigma$.  The unitary which does this is
\begin{align}
\begin{split}
    U_{\mathrm{perm}}
    \ket{q_{11} \dots q_{NN}}
    &=
    \ket{q_{\sigma(1)\sigma(1)} \dots q_{\sigma(N)\sigma(N)}},
    \label{eq:U_perm}
\end{split}
\end{align}
where $\ket{q_{ij}}\in\{\ket{0},\ket{1}\}$ is the state of qubit $Q_{ij}$. 
Any permutation can also be associated with an $N\times N$ permutation matrix $\Sigma$, and it is straightforward to see that $\mathcal{U}_{\mathrm{perm}} = \Sigma\otimes I_2$ in the binary representation.

Next we define $U_{\text{int}}$ in (\ref{eq:U_int}) as to apply the same four-qubit Clifford gate $W$ to all interacting subsets, ensuring that the interaction structure within each subset is identical.
For example, in the numerical simulations in the main text, we set specifically
\begin{align}
    W = \text{S}_{q_4} \text{CNOT}_{q_4\rightarrow q_1} \text{CNOT}_{q_1\rightarrow q_2} \text{CNOT}_{q_2\rightarrow q_3} \text{CNOT}_{q_3\rightarrow q_4} \text{H}_{q_1}\label{eq:4-qubit_gate}
\end{align}
where $q_1,q_2,q_3,q_4$ are the indices of qubits in a interacting subset.  Our general conclusions are not sensitive to this choice: see Appendix \ref{app:altgate}. We remind the reader that in the stabilizer formalism these gates correspond to: 
\begin{equation}
    \begin{split}
        \mathrm{H} &= \left[\begin{array}{cc} 0 &\ 1 \\ 1 &\ 0 \end{array}\right]\\
        \mathrm{S} &= \left[\begin{array}{cc} 1 &\ 0 \\ 1 &\ 1 \end{array}\right]\\
        \mathrm{CNOT}_{1\rightarrow 2} &= \left[\begin{array}{cccc} 1 &\ 1 &\ 0 &\ 0 \\ 0 &\ 1 &\ 0 &\ 0 \\ 0 &\ 0 &\ 1 &\ 0 \\ 0 &\ 0 &\ 1 &\ 1 \end{array}\right].
    \end{split}
\end{equation}

The key conceptual difference between this set-up and the random unitary circuit studied before is that time-evolution is always by the same gate $U$, rather than a random gate.   We aim to understand how much, if any, this reduces scrambling.    In the following subsections, we explore three possible measures that characterize scrambling in the system.

\subsection{Operator size growth}\label{operator size growth}
From the infection model analysis, we have seen that in the random unitary circuit case the number of infected qubits grows exponentially, resembling fast scrambling behavior. We expect that the Clifford circuit case shows similar dynamics, since the basic structure of the evolution is similar in both cases.

The main difference between the random unitary circuit and the Clifford Floquet circuit is that the multi-qubit Clifford circuit in  (\ref{eq:4-qubit_gate}) will grow $XIII \rightarrow ZIIZ$ -- in other words sometimes operators do not grow as large as they ``could have" based on connectivity.
In addition, some qubits that become ``infected" may later be ``disinfected" through cancellations between Pauli operators on the same qubit during the time evolution, for example a four-qubit operator $ZZZZ\rightarrow YYII$ shrinks to a two-qubit operator under $W_a$. 

At the early stage of the Clifford dynamics,  we expect that the number of infected qubits is small compared to the total system size, and moreover the permutation gate is exchanging qubits in such a way that every gate acts on mostly identity Pauli strings; hence it should grow small operators into larger ones in each subset $A_a$, rather than shrinking large operators, in analogy to the dynamics in Figure \ref{fig:1Dsystem_diagram}. As time progresses, however, the Pauli string will act on $\mathrm{O}(N^2)$ qubits which begins to lead to more frequent cancellations between Pauli operators on each qubit. We therefore expect that the operator size grows exponentially at early times, and at late times it saturates at a finite value consistent with a random Clifford circuit.  For an $N^2$-qubit system there are $4^{N^2}$ possible Pauli strings, since each qubit carries exactly one operator from the set $\{I,X,Y,Z\}$. If the initial Pauli string is the identity $I_{11}...I_{NN}$, then it remains invariant under the application of Clifford gates and therefore its operator size stays zero. This is the unique Pauli string with zero operator size. Every other Pauli string $A$ necessarily contains at least one non-identity single-qubit operator, and the probability that under a random (large) Clifford unitary $U$ we map $A$ to any other Pauli string $A^\prime$ is
\begin{align}
    P(UAU^\dagger = A^\prime) &= \frac{1}{4^{N^2} - 1} = \frac{1}{4^{N^2}} + O(4^{-{N^2}})\label{eq:probability of nonzero size operator}
\end{align}
for every $A^\prime$.  
Given that each qubit allows four choices, we can approximate the average probability of obtaining each single-qubit operator for qubit $j$ to be
\begin{align}
    P(I_j\otimes \cdots ) = P(X_j\otimes \cdots ) = P(Y_j\otimes \cdots ) = P(Z_j\otimes \cdots ) = \frac{1}{4}.
    \label{eq:probability of single-qubit operator}
\end{align}
So for large $N$, we expect that the typical operator size is
\begin{align}
    \langle n\rangle &= \frac{3}{4}N^2.\label{eq:operator size upper bound}
\end{align}
where $\langle n\rangle$ denotes the number of $X/Y/Z$ Paulis in the operator $A(t)$.

\begin{figure}[t]
    \centering
    \begin{subfigure}[h]{0.5\textwidth}
        \centering
        \includegraphics[scale=0.5]{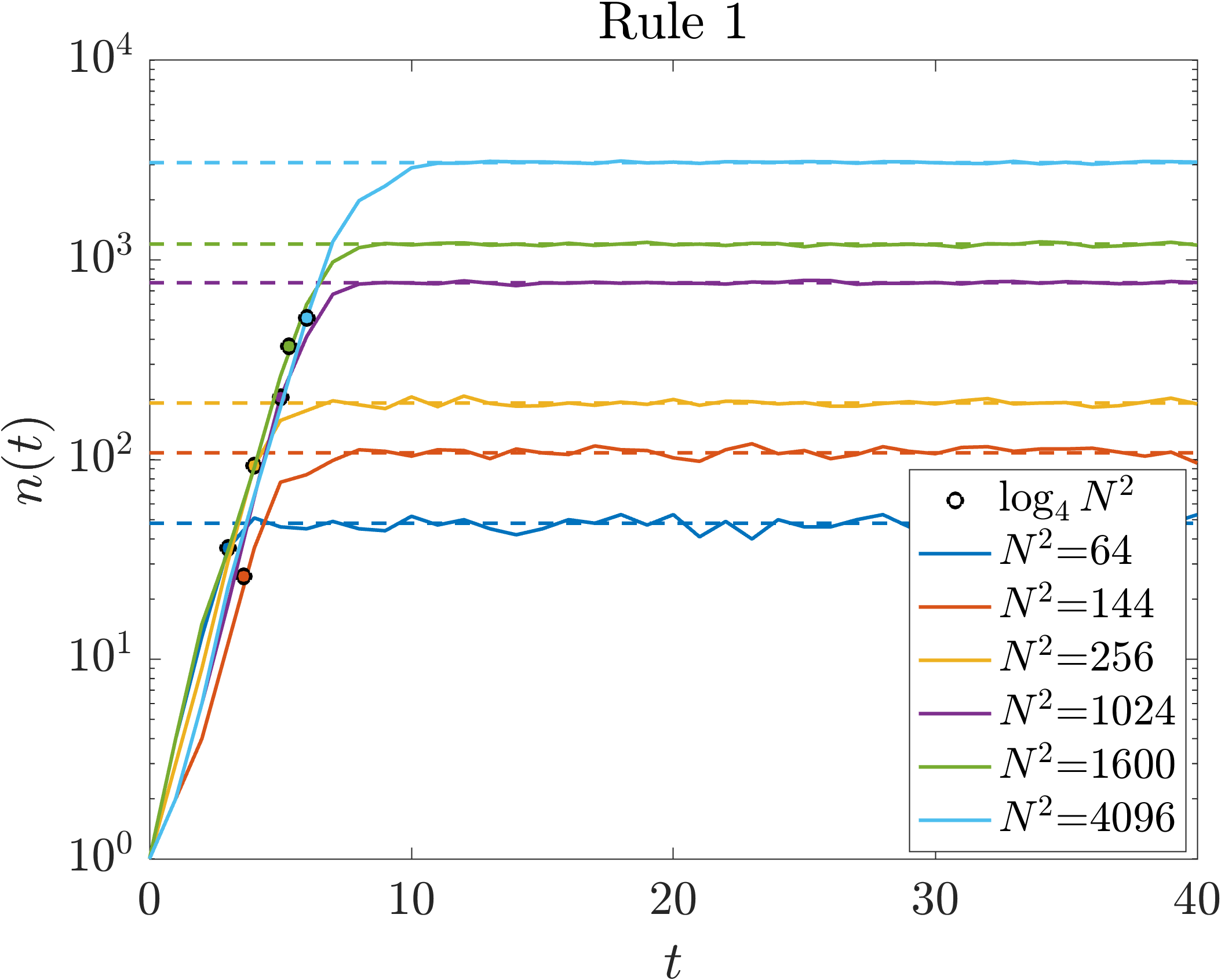}
        \caption{}
        \label{fig:OSG_different}
    \end{subfigure}%
    \begin{subfigure}[h]{0.5\textwidth}
        \centering
        \includegraphics[scale=0.5]{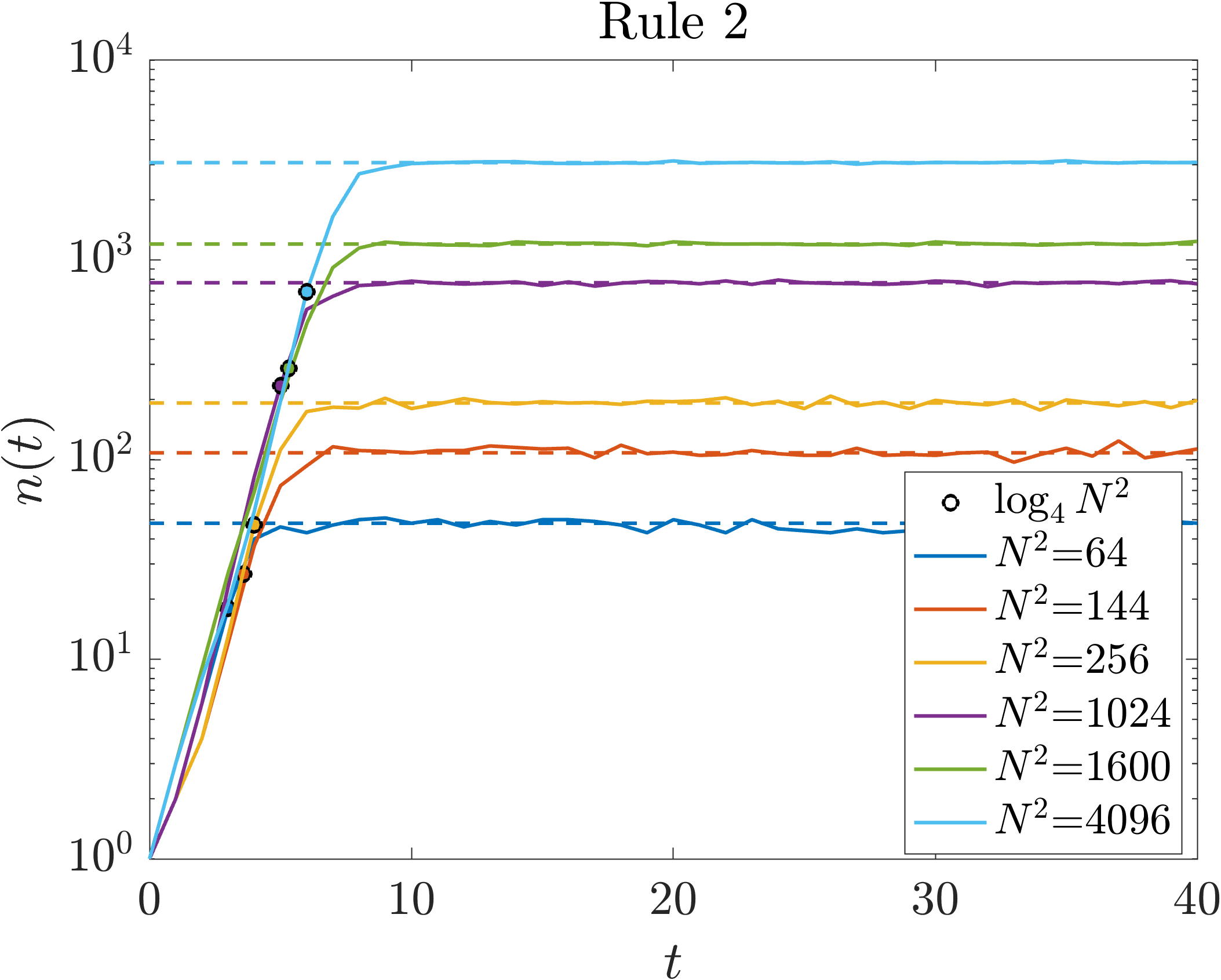}
        \caption{}
        \label{fig:OSG_same}
    \end{subfigure}
    \caption{
    Growth of the operator size $n(t)$ as a function of time $t$ in fixed Clifford dynamics. The onset of full scrambling occurs no earlier than $\log_4 N^2$; the time at which this bound is reached is depicted as a solid circle in each plot.
    }
    \label{fig:OSG}
\end{figure}

\begin{figure}[t]
    \centering
    \begin{subfigure}[h]{0.5\textwidth}
        \centering
        \includegraphics[scale=0.5]{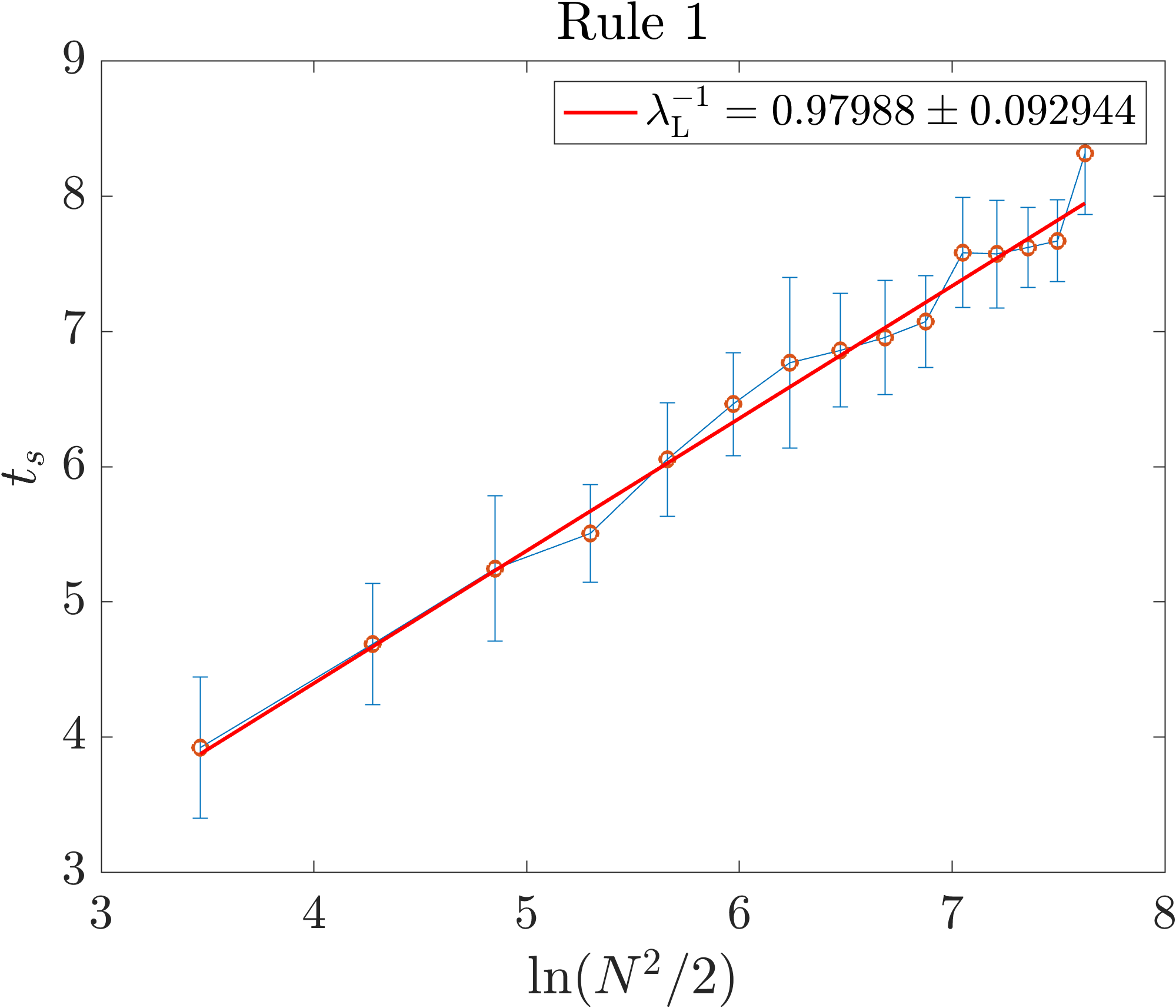}
        \caption{}
        \label{fig:lyapunov_rule1}
    \end{subfigure}%
    \begin{subfigure}[h]{0.5\textwidth}
        \centering
        \includegraphics[scale=0.5]{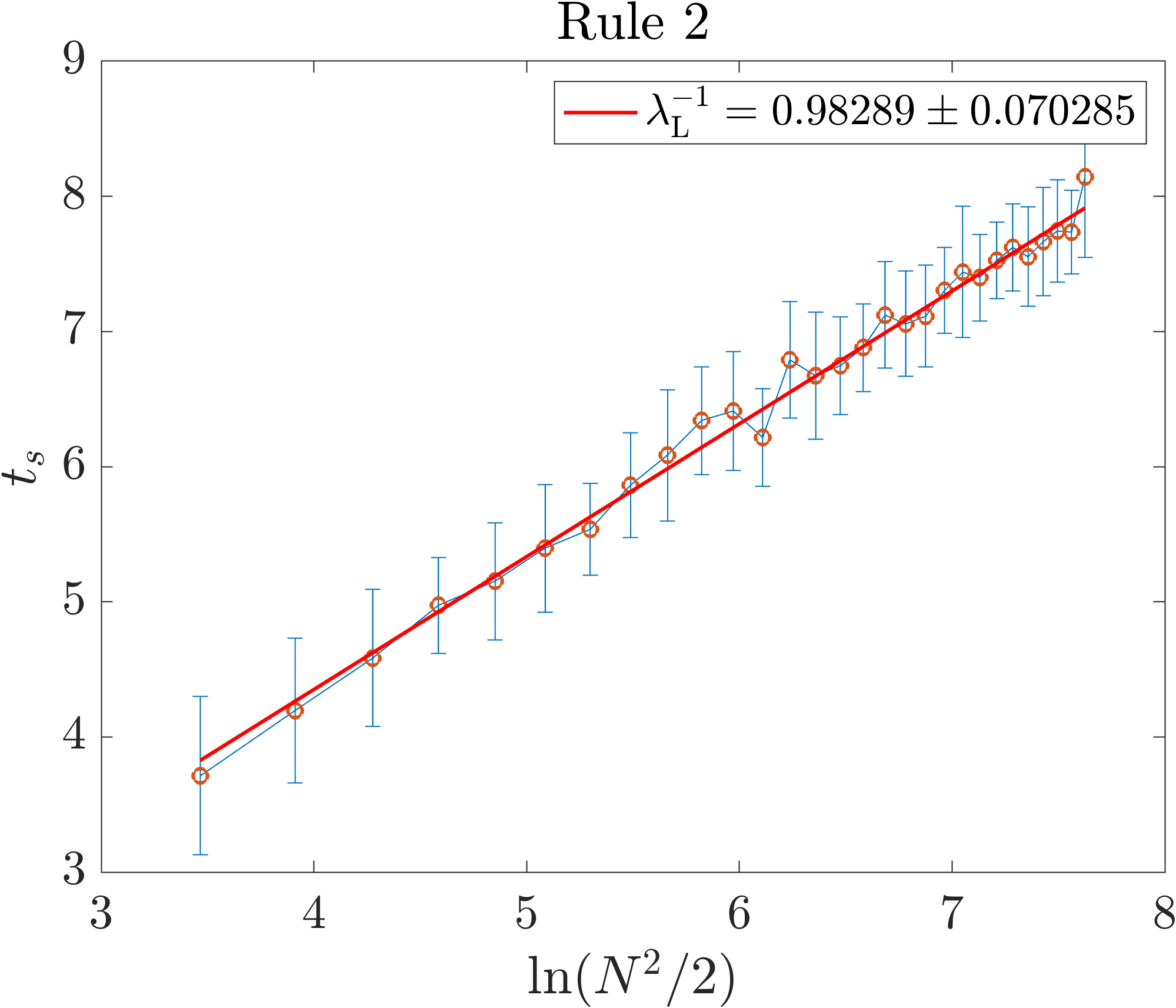}
        \caption{}
        \label{fig:lyapunov_rule2}
    \end{subfigure}
    \caption{Growth of the scrambling time $t_s$ as a function of $\ln(N^2/2)$. For Rule 2, we use $2|N$, since it is defined generically for any even $N$. 
    }
    \label{fig:lyapunov_clifford}
\end{figure}

For the simulation, we choose an initial single-qubit operator $\mathcal{O}$ taken to be either an $X$ or $Z$ operator on a randomly selected qubit.  We then observe the time dependence of its operator size. As shown in Figure \ref{fig:OSG}, the operator size $n(t)$ exhibits exponential growth at early times and saturates at $\langle n\rangle$, which is consistent with the full scrambling observed in random Clifford circuit dynamics.  It is also clear, however, that the Floquet dynamics does not reach ``maximal" operator size quite as fast due to the ``imperfections" in operator growth described above. When estimating the Lyapunov exponent in the simulations, we determine it from the time at which the operator size reaches $\log_4 (N^2/2)$. This choice avoids inaccuracies caused by fluctuations that appear as the operator size approaches full scrambling at $\log_4 (3N^2/4)$. This does not affect the estimate of $\lambda_{\mathrm{L}}$, as they only change the vertical intercept in Figure \ref{fig:lyapunov_clifford} while leaving the slope unchanged. So we read off $\lambda_{\mathrm{L}}$, defined in (\ref{Lyapunov}). From the data in Figure \ref{fig:lyapunov_clifford}, we find that
\begin{align}
    \lambda_{\mathrm{L}} \approx 1.02\pm0.10
    \label{eq:lyapunov_for_clifford_rule1}
\end{align}
for Rule 1 and
\begin{align}
    \lambda_{\mathrm{L}} \approx 1.02\pm0.07
    \label{eq:lyapunov_for_clifford_rule2}
\end{align}
for Rule 2. Thus, the chosen Clifford dynamics yield a smaller Lyapunov exponent than the random unitary case in which $\lambda_{\mathrm{L}}\approx1.39$, consistent with the presence of ``imperfections". We note that the numerical value $\lambda_{\mathrm{L}}\approx \frac{3}{4}\ln{4}\approx1.04$ is close to the average probability of obtaining a $X/Y/Z$ Pauli on a qubit in (\ref{eq:probability of single-qubit operator}). This can be interpreted as each qubit infecting three out of the four qubits in an interacting set per unit time in the infection model, rather than all four as in random unitary circuit dynamics.

\subsection{Entanglement entropy growth}\label{entropy growth}
In addition to the growth of the operator size, we also evaluate scrambling in the system using entanglement entropy, which has been argued to be a better metric for scrambling \cite{Bentsen2019FastScramblingSparseGraphs,Lucas2019QuantumManyBodyDynamicsStarGraph,Harrow2021SeparationOTOCEntanglement}.
Specifically, we partition the system into a subsystem $A$ and its complement $A^c$, and consider an initial state $\ket{\psi}$ taken to be
\begin{align}
    \ket{\psi} = \ket{0...0}_{A}\otimes \ket{0...0}_{A^c}\label{eq:initial_state}.
\end{align}
We can study the dynamics of this state in the Clifford formalism by keeping track of $N^2$ commuting and independent stabilizers of this state:  $Z_1,Z_2,...,Z_{N^2}$, as they evolve in time.  Let 
\begin{align}
    \mathcal{S} &= \begin{bmatrix}
          \vec{z_1} & \vec{z_2}  & ... &  \vec{z_{N^2}}
        \end{bmatrix}\label{eq:stabilizer_matrix}
\end{align}
denote the binary matrix of independent stabilizers, whose time evolution is
\begin{align}
    \mathcal{S}(t) &= \begin{bmatrix}
          \vec{z_1}(t) & \vec{z_2}(t) & ... & \vec{z_{N^2}}(t)
    \end{bmatrix}\nonumber\\
    &\equiv \left(\begin{array}{c} 
    \mathcal{A} \\ \mathcal{D} \end{array}\right),\label{eq:stabilizer_matrix(t)}
\end{align}
where the block matrices
$\mathcal{A}$ and $\mathcal{D}$ are defined as the rows of $\mathcal{S}(t)$ corresponding to the qubits in subsystem $A$ and the qubits in subsystem $A^c$ respectively. The entanglement entropy $S_A$ of subsystem $A$ in the stabilizer formalism is
\begin{align}
    S_A(t) =\text{rank}(\mathcal{A}) - \abs{A},\label{eq:S_A(t)}
\end{align}
where $\abs{A}$ is the number of qubits in subsystem $A$.\footnote{(\ref{eq:S_A(t)}) is interpreted as the number of Bell pairs shared between $A$ and $A^c$, up to unitary transformations within both subsystems.} 

 If we randomly choose qubits in $A$, then it is possible to already have maximal entanglement after the first time step. Therefore, to ask whether our Floquet circuit quickly generates entanglement, we should choose the ``worst" possible set $A$ where we know that maximal entanglement cannot be generated right away.  We choose a set $A$ for which entanglement cannot saturate right away by first calculating the set of qubits $A(t_f)$ that have interacted with some reference qubit qubit $Q_{ij}$ at various times between the initial time $t=0$ and the final time $t=t_f$. 
 In other words,
 \begin{align}
     A(t_f) := \bigcup_{t=0}^{t_f}\text{partners}\bigl(\mathcal{O}_{ij}(t)\bigr)\label{eq:choice of A}
 \end{align}
where $\mathrm{partners}(\mathcal{O}_{ij}(t))$ denotes the sites where the operator $\mathcal{O}_{ij}(t)$ could have spread under arbitrary (e.g. random) unitary dynamics with our ``matrix model" connectivity.\footnote{This support is the set of infected sites after time $t$ if the infection starts at $ij$, using the analogy from Section \ref{RUC&infection}.}\footnote{If $|A(t_{f-1})| < |A| < |A(t_f)|$, the set $A$ is constructed as the union of $A(t_f-1)$ and a subset of the qubits added in the final time growth step. In the implementation, this subset is chosen according to index order in the code rather than random sampling.}
Notice that for $t\le t_f$, $Q_{ij}$ evolves entirely within $A(t_f)$, so the stabilizer $Z_{ij}(t)$ has support entirely within $A$.  Therefore, $S_A(t)\le |A|-1$ for $0\le t \le t_f$.   However, other qubits in $A$ may still interact with those in $A^c$ during the evolution, and thus we still expect the entanglement entropy $S_A$ to grow with time.  Indeed, if there is scrambling, at sufficiently late times $t_f$, for any choice of set $A$ with $\abs{A}\leq N^2/2$ we expect
\begin{align}
    S_A(t_f) &= \abs{A}.\label{eq:final_entropy}
\end{align} 
 
\begin{figure}[t]
    \centering
    \begin{subfigure}[h]{0.5\textwidth}
        \centering
        \includegraphics[scale=0.5]{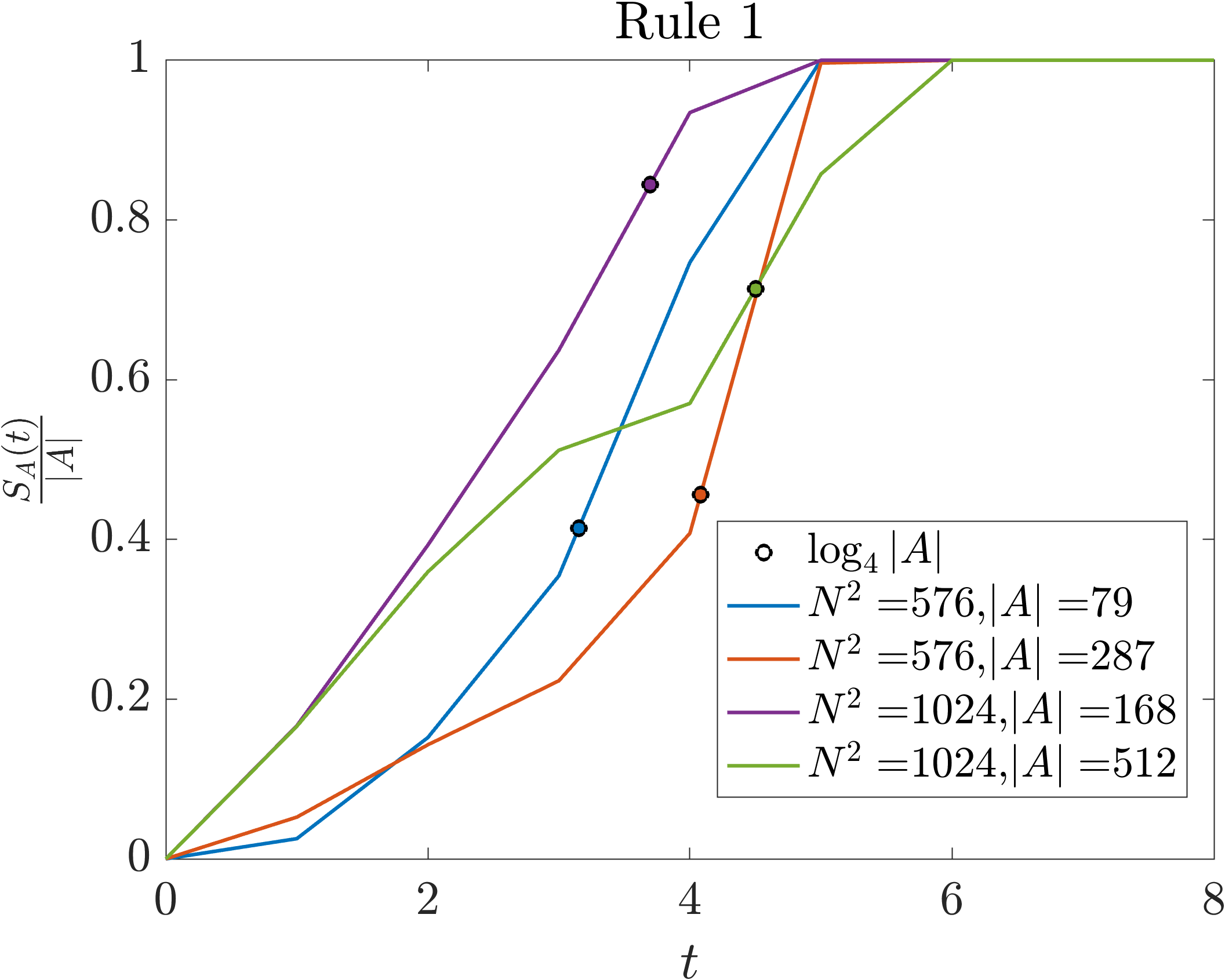}
        \caption{}
        \label{fig:EE_different}
    \end{subfigure}%
    \begin{subfigure}[h]{0.5\textwidth}
        \centering
        \includegraphics[scale=0.5]{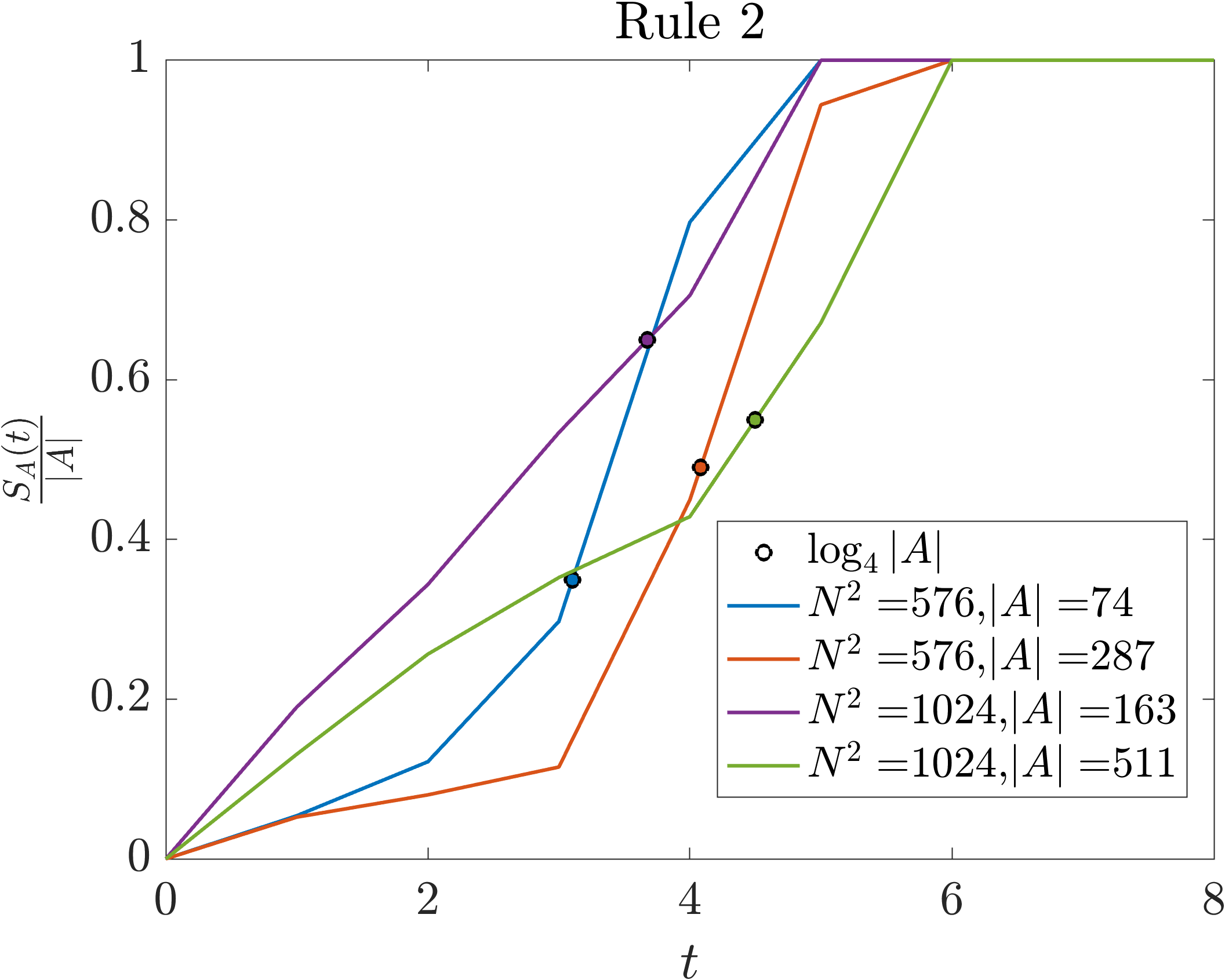}
        \caption{}
        \label{fig:EE_same}
    \end{subfigure}
    \caption{Growth of the entanglement entropy $S_A$ of subsystem $A$ as a function of time $t$, for $|A|\leq N^2/2$. The onset of maximal entanglement occurs no earlier than $\log_4 |A|$.}\label{fig:EE}
\end{figure}

 In the simulations shown in Figure \ref{fig:EE}, we choose $A$ as specified above. 
At the initial time $t=0$, our initial conditions imposed
\begin{align}
    S_A(0) &= 0. \label{eq:initial_entropy}
\end{align} 
From the construction of set $A$ above, we see that at time $t\le t_f$, for any $kl \in A(t_f-t)$, $Z_{kl}(t)$ will lie entirely within $A(t_f)$.  Therefore we expect that \begin{equation}
    S_A(t) \lesssim  (1-4^{-t})|A|,
\end{equation} 
using the estimate $|A(t)|\lesssim 4^t$.  
The saturation time $t_{\text{saturation}}$ of $S_A$ satisfies
\begin{align}
    t_{\text{saturation}}\geq \log_4{\abs{A}}. \label{eq:entropy_time_dependence}
\end{align}

From Figure \ref{fig:EE} we can see that for different choices of $N$ and $\abs{A}$ values, (\ref{eq:final_entropy}) and (\ref{eq:initial_entropy}) are satisfied. The saturation time exceeds $\log_4\abs{A}$, which aligns with our expectation in (\ref{eq:entropy_time_dependence}).  However, the saturation time is not qualitatively different from the lower bound in \eqref{eq:entropy_time_dependence}, suggesting that our Floquet circuit is still a fast scrambler. 

Perhaps as a side remark, because our construction of $A$ always includes the reference qubit, that qubit remains in $A$ at all subsequent times (e.g., t=1). As a result, small but nonzero entanglement may appear at early times due to the presence of this additional qubit. The small early-time entanglement observed in some cases is therefore likely due to particular combinations of reference qubit choices and subsystem sizes $|A|$.

\subsection{Recovery of quantum information in the Hayden-Preskill protocol}\label{HP recovery}
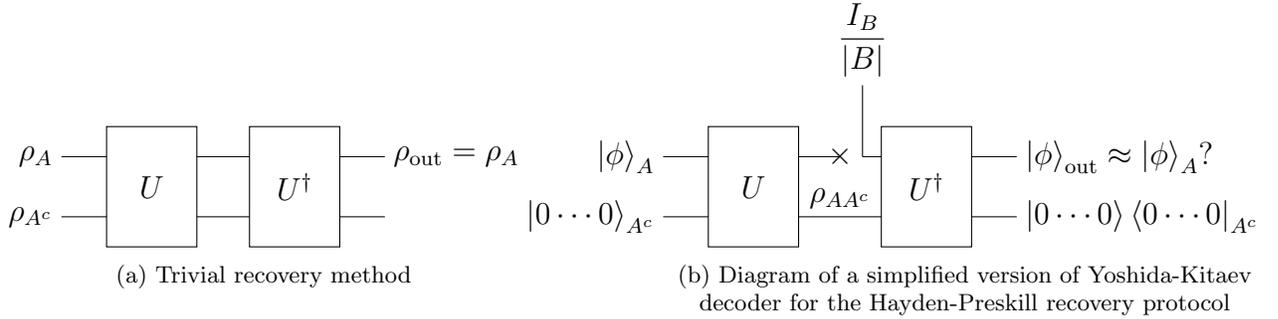
\begin{figure}[t]
    \centering
    \begin{subfigure}[t]{0.48\textwidth}
    \centering
    \begin{tikzpicture}[scale=1.0, baseline=(basept),
      every node/.style={font=\large}
    ]

    \draw (-0.1,0.8) -- (0.5,0.8);
    \draw (-0.1,0.0) -- (0.5,0.0);

    \node[left] at (-0.1,0.8) {$\rho_A$};
    \node[left] at (-0.1,0.0) {$\rho_{A^c}$};

    \draw (0.5,-0.4) rectangle (1.7,1.2);
    \coordinate (basept) at (0,1.2); 
    \node at (1.1,0.4) {$U$};

    \draw (1.7,0.8) -- (2.4,0.8);
    \draw (1.7,0.0) -- (2.4,0.0);

    \draw (2.4,-0.4) rectangle (3.6,1.2);
    \node at (3.0,0.4) {$U^\dagger$};

    \draw (3.6,0.8) -- (4.2,0.8);
    \draw (3.6,0.0) -- (4.2,0.0);

    \node[right] at (4.2,0.8) {$\rho_{\text{out}} = \rho_A$};

    \end{tikzpicture}
    \caption{Trivial recovery method}
    \label{fig:unitary_transf}
\end{subfigure}%
\hfill
\begin{subfigure}[t]{0.48\textwidth}
    \centering
    \begin{tikzpicture}[scale=1.0, baseline=(basept),
      every node/.style={font=\large},trim left=-0.1cm
    ]

    \def\yinTop{0.8}
    \def\yinBot{0.0}

    \def\xInL{-0.1}

    \def\xUoneL{0.5}
    \def\xUoneR{1.7}   
    \def\xUtwoL{2.8}
    \def\xUtwoR{4.0}   
    \def\yGateBot{-0.4}
    \def\yGateTop{1.2}

    \def\xX{2.25}
    \def\xInsert{2.55}   

    \def\xOutR{4.6}

    \coordinate (basept) at (0,\yGateTop);

    \draw (\xInL,\yinTop) -- (\xUoneL,\yinTop);
    \draw (\xInL,\yinBot) -- (\xUoneL,\yinBot);

    \node[left] at (\xInL,\yinTop) {$\ket{\phi}_A$};
    \node[left] at (\xInL,\yinBot) {$\ket{0\cdots 0}_{A^c}$};

    \draw (\xUoneL,\yGateBot) rectangle (\xUoneR,\yGateTop);
    \node at ({(\xUoneL+\xUoneR)/2},{(\yGateBot+\yGateTop)/2}) {$U$};

    \draw (\xUoneR,\yinBot) -- (\xUtwoL,\yinBot);
    \node[above] at ({(\xUoneR+\xUtwoL)/2},\yinBot) {$\rho_{AA^c}$};

    \draw (\xUoneR,\yinTop) -- (\xX,\yinTop);

    \node at (\xX,\yinTop) {$\times$};

    \draw (\xInsert,\yinTop) -- (\xUtwoL,\yinTop);

    \draw (\xInsert,\yinTop) -- (\xInsert,1.75);
    \node[above] at (\xInsert,1.75) {$\dfrac{I_B}{|B|}$};

    \draw (\xUtwoL,\yGateBot) rectangle (\xUtwoR,\yGateTop);
    \node at ({(\xUtwoL+\xUtwoR)/2},{(\yGateBot+\yGateTop)/2}) {$U^\dagger$};

    \draw (\xUtwoR,\yinTop) -- (\xOutR,\yinTop);
    \draw (\xUtwoR,\yinBot) -- (\xOutR,\yinBot);

    \node[right] at (\xOutR,\yinTop) {$\ket{\phi}_{\text{out}} \approx \ket{\phi}_A$?};
    \node[right] at (\xOutR,\yinBot) {$\ket{0\cdots 0}\bra{0\cdots 0}_{A^c}$};

    \end{tikzpicture}
    \caption{Diagram of a simplified version of Yoshida-Kitaev decoder for the Hayden-Preskill recovery protocol}
    \label{fig:decoder_diagram}
\end{subfigure}
    \caption{Quantum information recovery in a trivial (a) or simplified Hayden-Preskill (b) recovery protocol. Here $B$ is the set of the erased $r$ qubits, which may not necessarily equal $A$.
    }
    \label{fig:diagram for HP}
\end{figure}

Here, we aim to test whether the system acts as a good scrambler, drawing on the Yoshida–Kitaev decoding scheme for the Hayden–Preskill protocol \cite{HaydenPreskill2007,YoshidaKitaev2017}.  A minimalistic protocol which tests this was recently presented in \cite{Vikram2026BidirectionalTeleportation}: we partition the system into a subsystem $A$ containing $q$ qubits and its complement $A^c$ with $N^2 - q$ qubits.  One can trivially recover the quantum information initially encoded in $A$ by simply reversing the unitary, namely applying $U^\dagger = U^{-1}$, as illustrated in Figure \ref{fig:unitary_transf}. What happens if access to the quantum information on $r$ of the qubits is lost?   A key idea is that a holographic theory should be such an efficient scrambler that -- very quickly -- $A$ and $A^c$ (and indeed all subparts of the system) become highly entangled, such that it is still possible to recover the information originally in $A$, even after removing  $r$ of the qubits.   

  Following \cite{Vikram2026BidirectionalTeleportation}, we suppose the initial state of the system is
\begin{align}
    \ket{\psi} = \ket{\phi}_{A}\otimes \ket{0...0}_{A^c},\label{eq:initial_state}
\end{align}
where $\ket{\phi}_A$ is an arbitrary logical state. If we know how to recover all the single-Pauli operators acting on qubits in $A$, we can recover all quantum information stored in this state.  

More concretely, we know that $|\psi\rangle$ is always stabilized by $Z_{kl}$ for each $kl\notin A$; call this set of operators $\mathcal{S}$.   Logical operators acting in $A$ correspond to $\lbrace X_{kl}, Z_{kl}\rbrace$ for $kl\in A$, and we call this set $\mathcal{L}$.   Given any Pauli string $S\in\mathcal{S}$ and any logical $L\in \mathcal{L}$, we know that \begin{equation}
    L\ket{\psi} = SL\ket{\psi} = LS\ket{\psi}.
\end{equation}  
So any $LS$ is a valid logical operator, as in the theory of stabilizer quantum codes \cite{gottesman1997,calderbank1997quantum}.

Ideally what happens is that each logical $L\in\mathcal{L}$ is evolved by the unitary $U$ to a complicated operator:  $ULU^\dagger = L(t)$, which has support outside the $r$ qubits which will be erased.  We aim to find a stabilizer $S(t)$ such that $L(t)S(t)$ has no support on the $r$ qubits which are erased.   This means that $L(t)S(t)$ acts identically on the quantum (mixed) state after and before erasing the qubits in set $B$, namely the logical information is preserved and after applying $U^\dagger$, we can act with the logical operation by applying $LS$, as opposed to simply $L$.

To carry this out in practice it is useful to evolve the stabilizer set $\mathcal{S}$ and the logical set $\mathcal{L}$ in time, which we can do in the binary representation.   With a slight abuse of notation, let us use $\mathcal{S}$ and $\mathcal{L}$ to denote the $2N^2 \times (N^2-q)$ and $2N^2\times 2q$ matrices that encode all such operators in the binary represenation.  After time $t$, let us denote \begin{subequations}
    \begin{align}
        \mathcal{S}(t) &= \left(\begin{array}{c} \mathcal{S}_r \\ \mathcal{S}_{\text{safe}} \end{array}\right), \\
        \mathcal{L}(t) &= \left(\begin{array}{c} \mathcal{L}_r \\ \mathcal{L}_{\text{safe}} \end{array}\right),
    \end{align}
\end{subequations}
where e.g. $\mathcal{S}_r$ is a $2r\times (N^2-q)$ matrix which encodes the rows of $\mathcal{S}(t)$ that have support on $X$ and $Z$ Paulis on the $r$ qubits that are deleted after time evolution, while $\mathcal{S}_{\mathrm{safe}}$ contains the remaining data.  We can always carry out the recovery protocol above if the column space of $\mathcal{S}_r$ contains the column space of $\mathcal{L}_r$, or in other words if \begin{equation}
    \mathrm{rank}\left( \left[ \begin{array}{ll} \mathcal{S}_r &\ \mathcal{L}_r\end{array}\right]\right) = \mathrm{rank}(\mathcal{S}_r). \label{eq:rankcondition}
\end{equation} 

After a long time where operators are very scrambled, we should expect that $\mathrm{rank}(\mathcal{S}_r)=2r$ so long as $2r \le N^2-q$.  As this is maximal rank, \eqref{eq:rankcondition} would be trivially satisfied. In our simplified Hayden-Preskill protocol setting, then, we could recover at most $q \le N^2 -2r$ qubits, 
which is consistent with the well-known result in quantum information theory that for the quantum erasure channel with erasure probability $p<1/2$, the quantum capacity is given by $1-2p$ \cite{BennettShorSmolinThapliyal1999EntanglementAssisted,BennettDiVincenzoSmolin1997CapacitiesQuantumErasure,HaydenPreskill2007}
, where $p=r/N^2$ is the fraction of erased qbits.  

\begin{figure}[t]
    \centering
    \begin{subfigure}[h]{0.5\textwidth}
        \centering
        \includegraphics[scale=0.5]{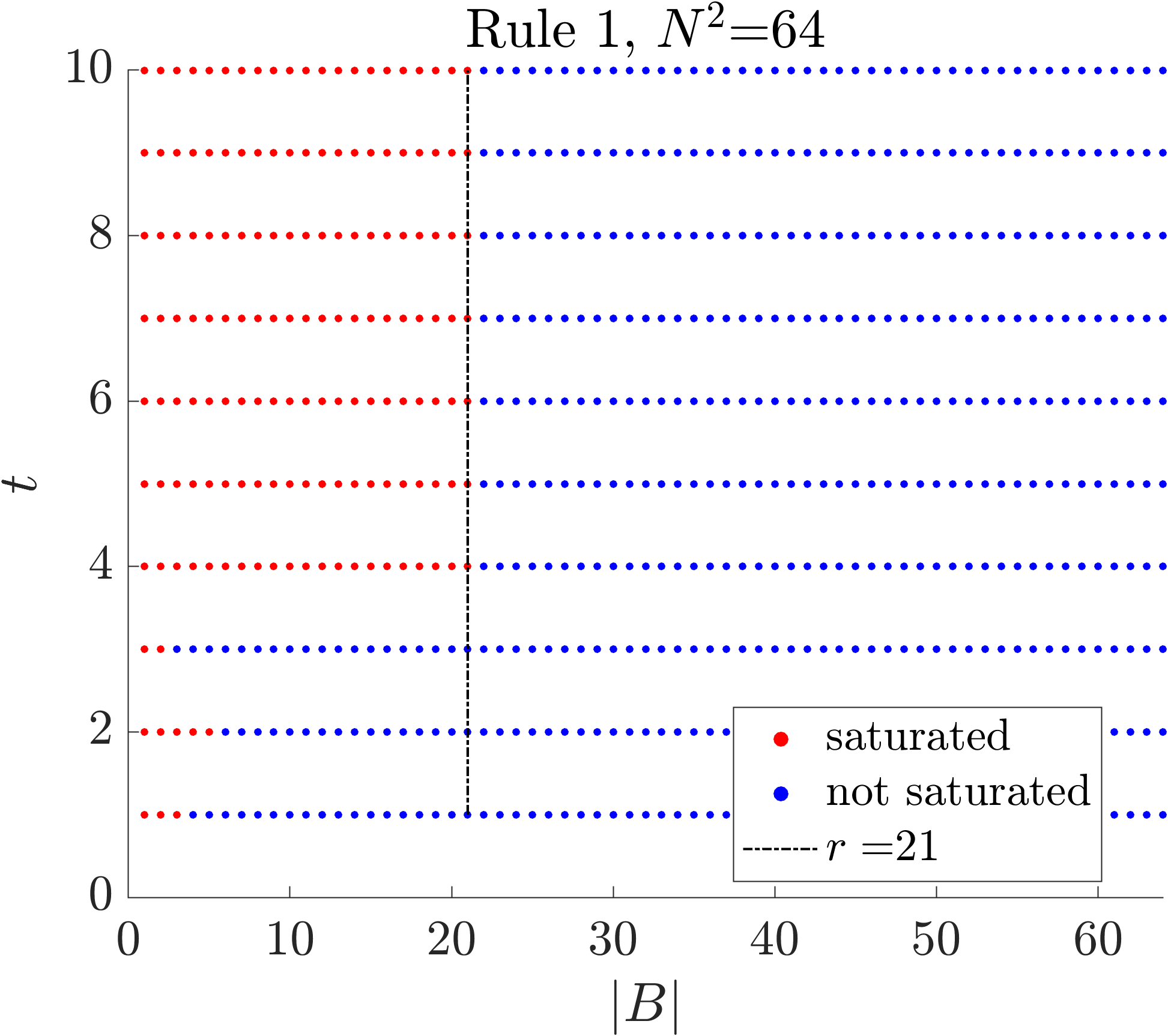}
        \caption{}
        \label{fig:HP_different}
    \end{subfigure}%
        \begin{subfigure}[h]{0.5\textwidth}
        \centering
        \includegraphics[scale=0.5]{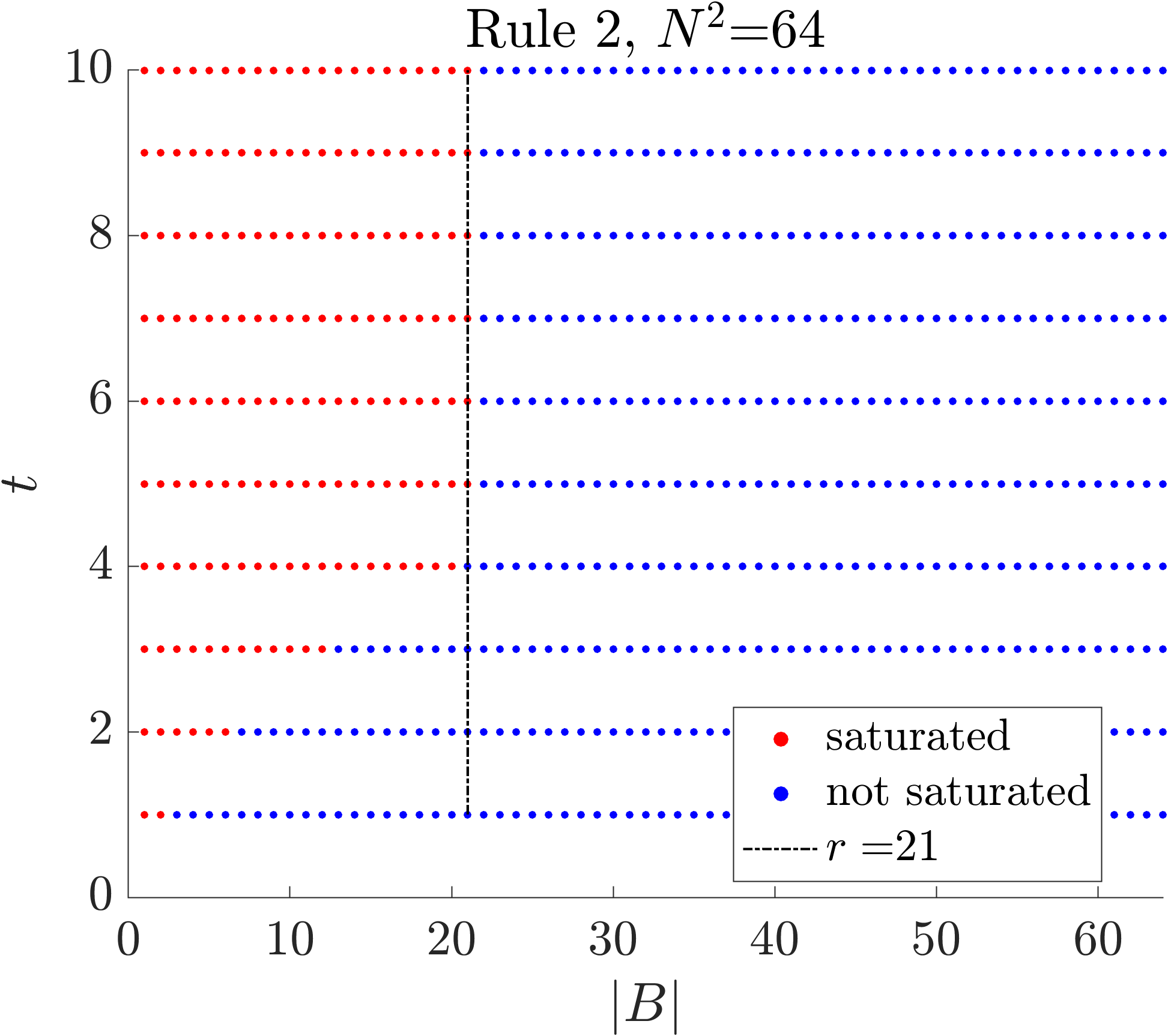}
        \caption{}
        \label{fig:HP_same}
    \end{subfigure}
    \caption{Satisfaction of the sufficient condition in (\ref{eq:rankcondition}) for quantum information recovery for different final times $t$ and erased set sizes $r$, where the erased set $B$ is generated by a randomly selected reference qubit.}\label{fig:HP}
\end{figure}

For simplicity in what follows, we erase the same $r=q$ qubits that we started by encoding logical information into.  Then we conclude that we must take \begin{equation}
    q \le \frac{N^2}{3}\label{eq:boundary_of_recovery}
\end{equation}
in order to have a chance at recovering logical information. To test how this protocol works in practice, we construct sets $A$ with $|A|=q$ analogously to in the previous subsection, and run the protocol described above.  Figure \ref{fig:HP} shows what happens with $r=q$, and indeed at late times we see that recovery is always achieved so long as \eqref{eq:boundary_of_recovery} is obeyed. 

Note that in \cite{YoshidaKitaev2017} and even in \cite{Vikram2026BidirectionalTeleportation}, the setup appears at face value to be more complex than the one in Figure \ref{fig:diagram for HP}.  To compare our simplified procedure to \cite{Vikram2026BidirectionalTeleportation}, one can simply copy $A$ (so the Hilbert space is $\mathcal{H}_A \otimes \mathcal{H}_{A^c}\otimes \mathcal{H}_{A^\prime}$, if set $A^\prime$ is the copy) and apply $U$ to $AA^c$ while applying $U^\dagger$ to $A^\prime A^c$ -- this has the effect of teleporting $|\phi\rangle$ to $A^\prime$ if the recovery protocol is successful.  In turn, this extended protocol is analogous to the probabilistic decoder in \cite{YoshidaKitaev2017} if one rearranges their circuit so that $U$ and $U^\dagger$ act sequentially in time, rather than at the same time on two entangled subsystems.  See \cite{Vikram2026BidirectionalTeleportation} for more discussion on this point.  In this paper, we worked with the absolute simplest protocol to illustrate the essential point.  From an error-correcting perspective, the crucial point is that the Floquet circuit $U$ creates an error-correcting code that can correct arbitrary errors on the $r$ erased qubits.

Indeed, the simplest Hayden-Preskill recovery protocol requires measurement of $A^c$ and postselection, as was depicted in Figure \ref{fig:decoder_diagram}.  In experiment this is prohibitively difficult as the system size grows.  In Clifford dynamics in particular, the probability of measuring a specific output state will decay exponentially with $r$ and $q$.   An advantage of our simpler Clifford dynamics is that we can actually deduce the quantum error correction needed to avoid postselection altogether, as we explained above:  one simply needs to access the logical information via $SL$, rather than just $L$, after the protocol.

\begin{figure}[t]
    \centering
    \begin{subfigure}[h]{0.5\textwidth}
        \centering
        \includegraphics[scale=0.5]{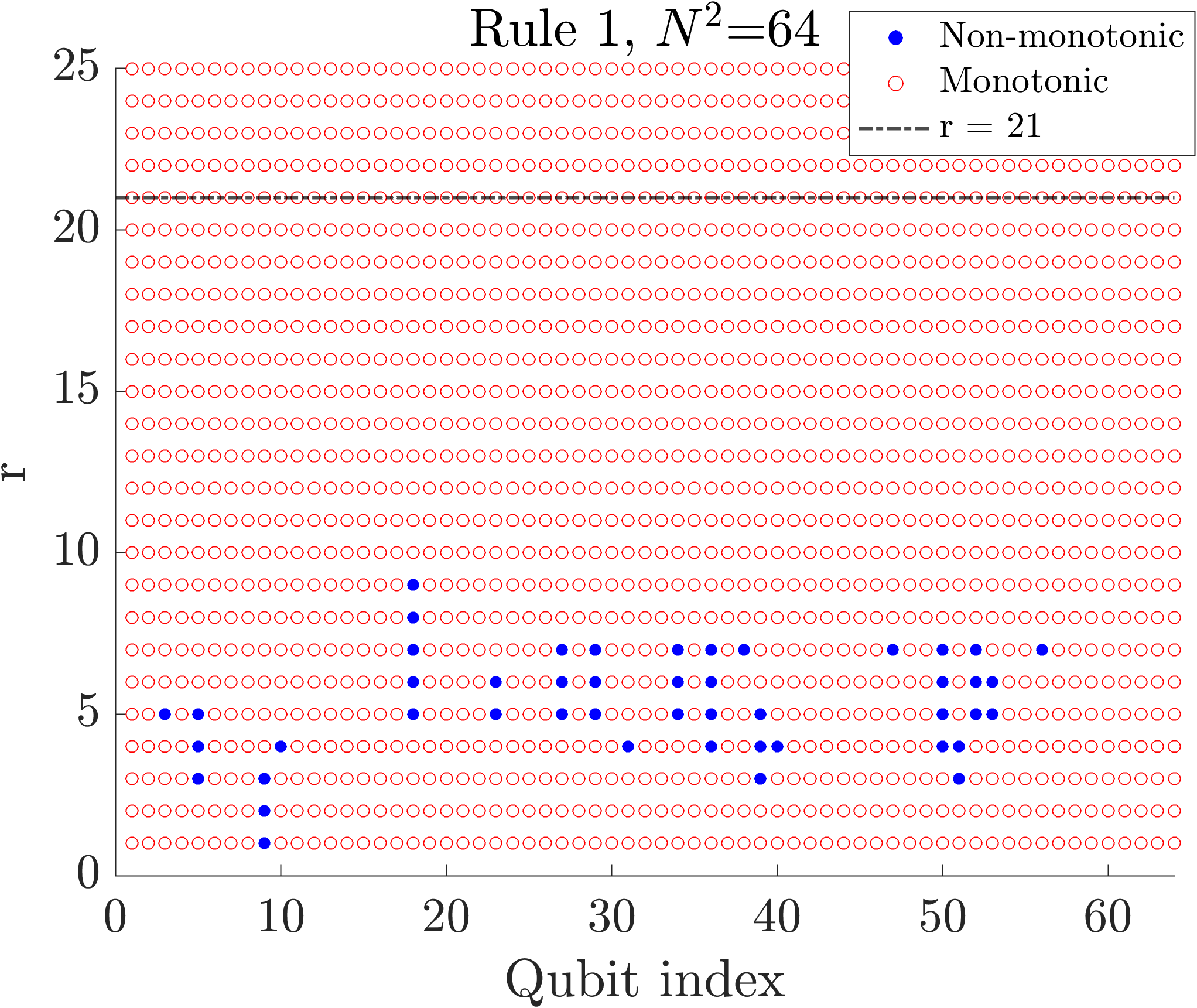}
        \caption{}
        \label{fig:diff_monotonicity_r}
    \end{subfigure}%
        \begin{subfigure}[h]{0.5\textwidth}
        \centering
        \includegraphics[scale=0.5]{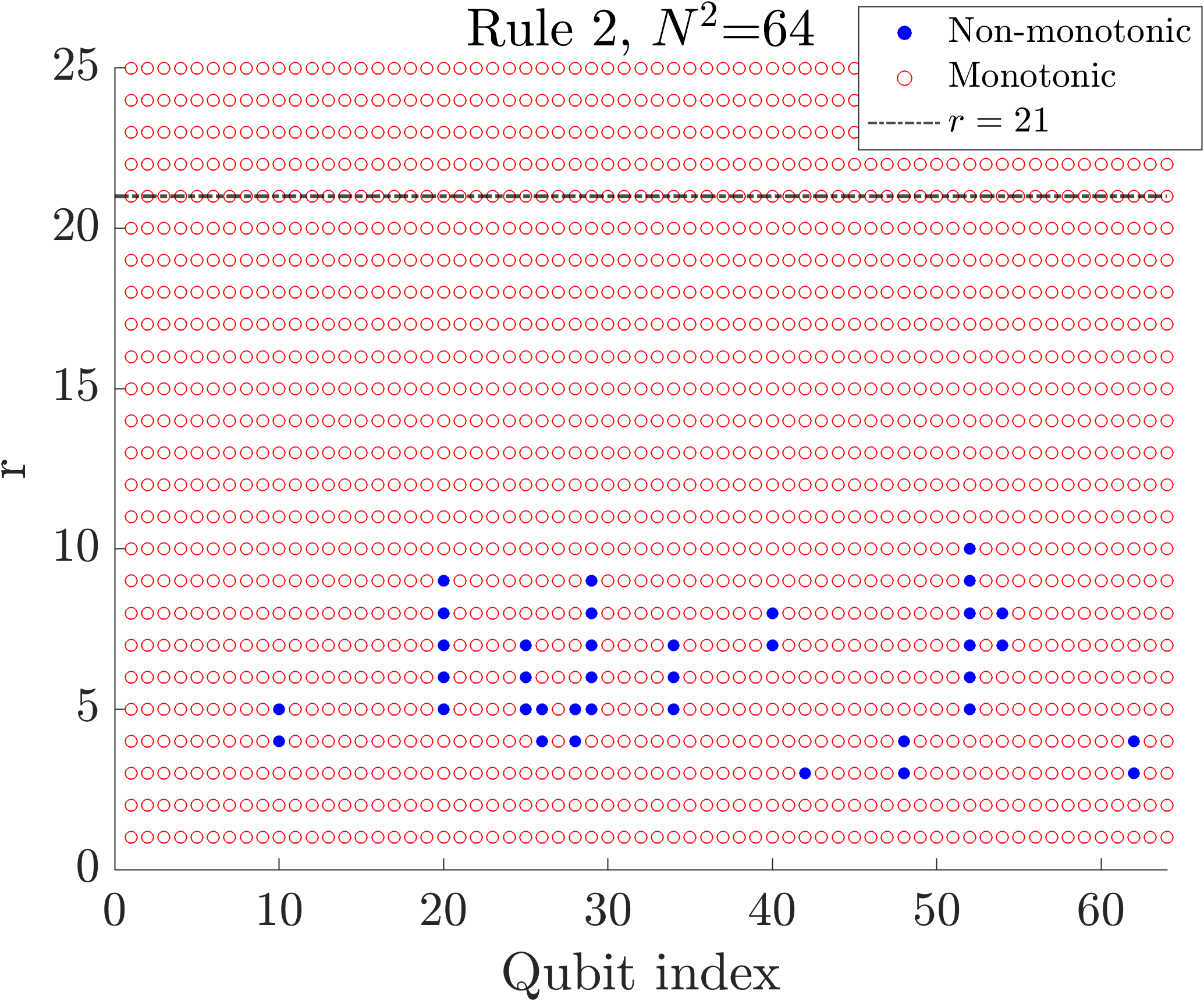}
        \caption{}
        \label{fig:same_monotonicity_r}
    \end{subfigure}
    \begin{subfigure}[h]{0.5\textwidth}
        \centering
        \includegraphics[scale=0.5]{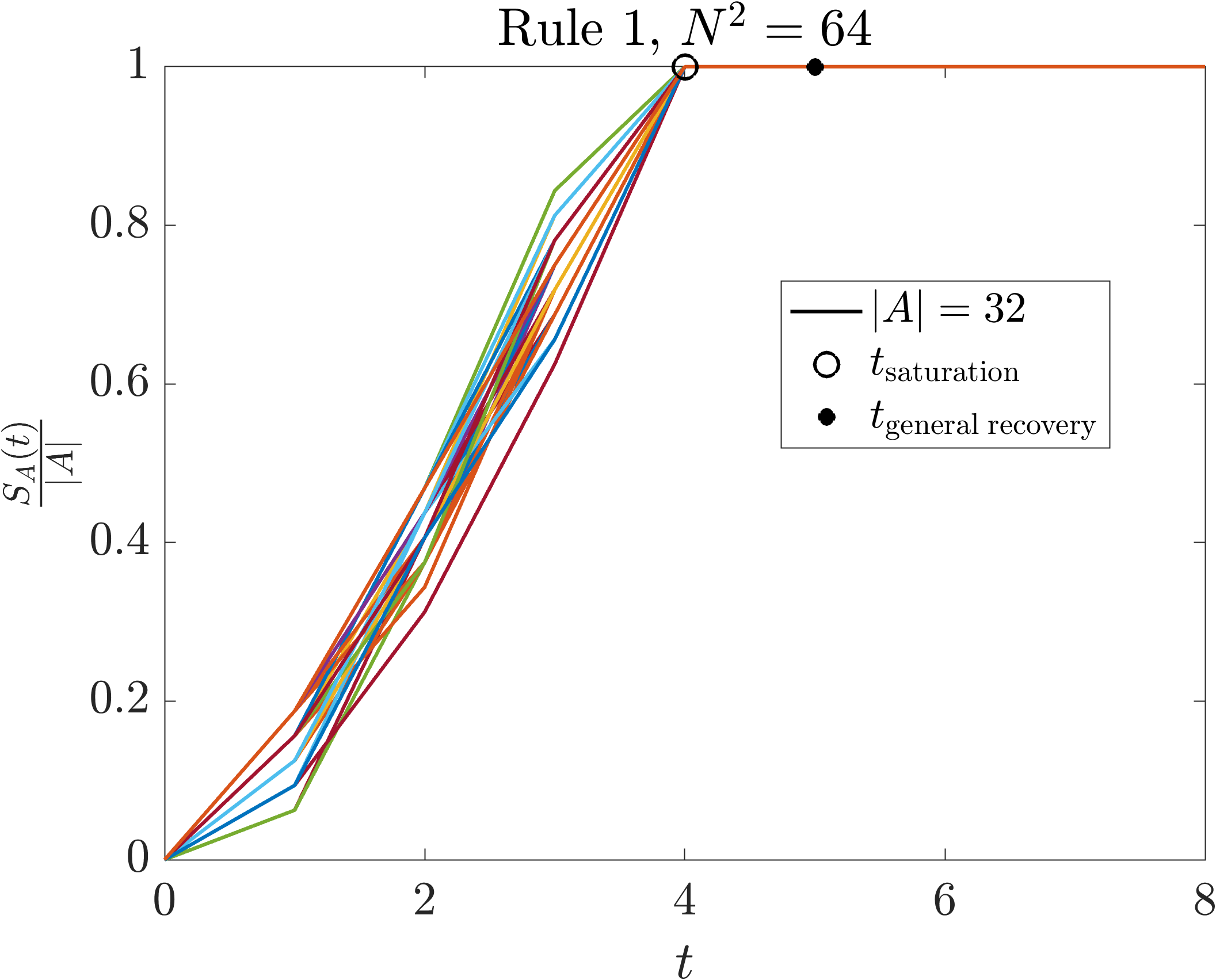}
        \caption{}
        \label{fig:rule_1_comparison}
    \end{subfigure}%
        \begin{subfigure}[h]{0.5\textwidth}
        \centering
        \includegraphics[scale=0.5]{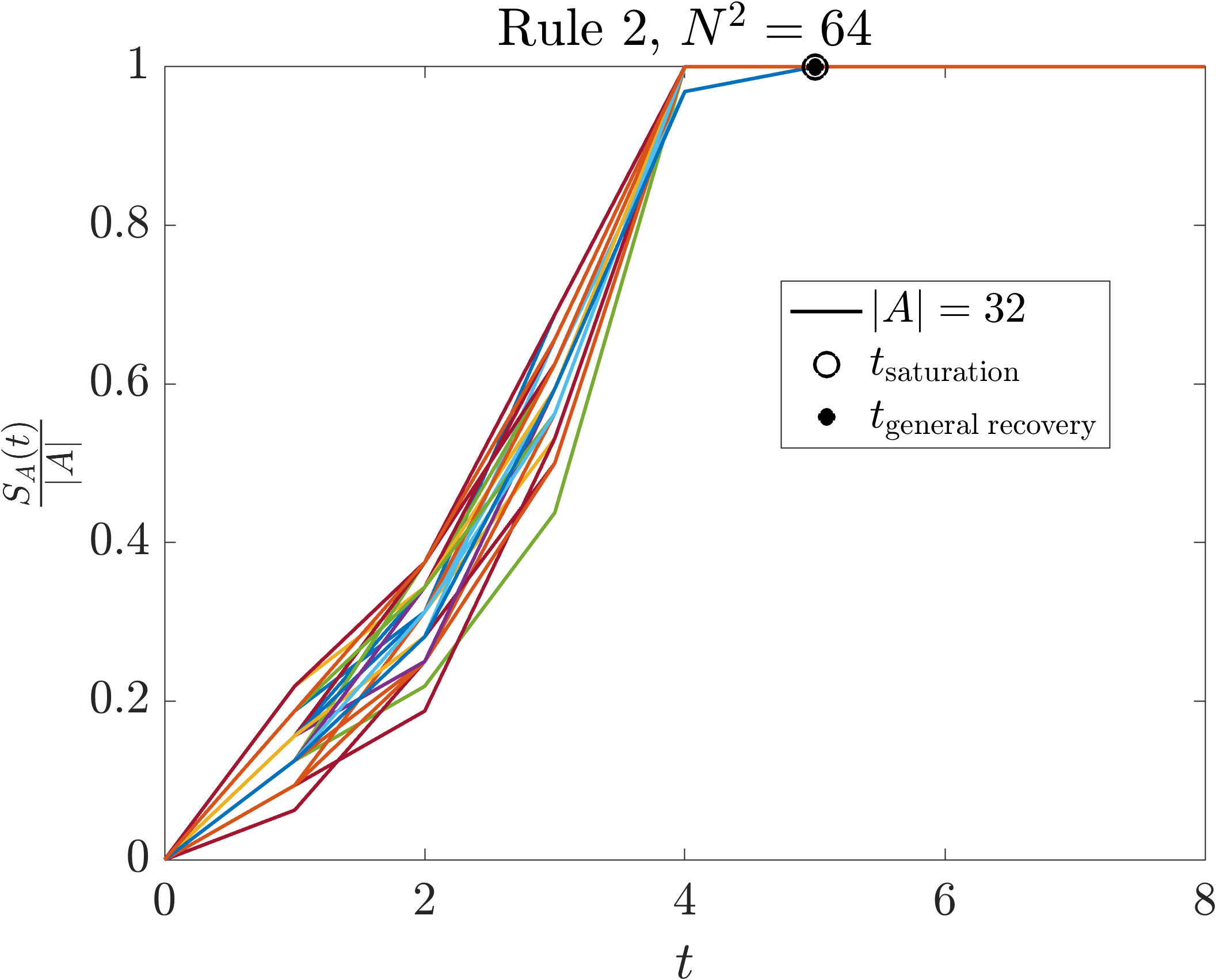}
        \caption{}
        \label{fig:rule_2_comparison}
    \end{subfigure}
    \caption{Monotonicity/non-monotonicity behaviors of the scrambling system. (a) and (b): Behavior for different choices of reference qubit, erased size $r$, and final times $t\in[1,10]$. (c) and (d): The comparison of general recovery time and saturation time for the entanglement entropy. Note that general recovery occurs no earlier than the onset of maximal entanglement.}
    \label{fig:non-monotonicity}
\end{figure}

A minor note is that we sometimes observe that the rank condition in (\ref{eq:rankcondition}) saturates at early times, but becomes unsaturated later during the time evolution, and then saturates again, exhibiting a non-monotonic behavior. Figure \ref{fig:non-monotonicity} provides a more complete picture of which choices of qubits used to define subsystem $A$ exhibit this type of non-monotonic behavior for some values of $r$ under the same conditions as in Figure \ref{fig:HP}, and which do not. We believe that this non-monotonicity arises from the shuffling dynamics induced by $\sigma$, which can move qubits into and out of the erased region before the system reaches maximal entanglement. As shown in Figure \ref{fig:rule_1_comparison} and Figure \ref{fig:rule_2_comparison}, the general recovery time--defined as the earliest time after which with any reference qubit the system remains fully recoverable at all reasonable time scales used in our simulation within the recurrence time (i.e. no non-monotonicity occurs afterwards)--is no earlier than the scrambling time of the entanglement entropy, consistent with maximal entanglement being sufficient for recovery.


\section{Double-layer construction for neutral atom arrays}\label{double_layer}
In this section, we will return to the question of a potential experimental implementation of our cartoon matrix model circuits.

\subsection{Motivating the double layer model}\label{experimental_implementation}
\begin{figure}[h]
    \centering
    \begin{subfigure}{0.8\linewidth}
        \centering
        \begin{tikzpicture}[scale=1.2, >=stealth]
            \tikzset{
              hbar/.style={draw, thick},
              vbar/.style={draw, thick},
              dotA/.style={circle, fill=red!80,        minimum size=7pt, inner sep=0pt},
              dotB/.style={circle, fill=green!70!black,minimum size=7pt, inner sep=0pt},
              dotC/.style={circle, fill=cyan!70!blue,  minimum size=7pt, inner sep=0pt},
              dotD/.style={circle, fill=violet!70,     minimum size=7pt, inner sep=0pt},
            }
            
            \draw[hbar] (0, 1.2) rectangle (6, 1.8);   
            \draw[hbar] (0,-0.8) rectangle (6, -0.2);   
            
            \draw[vbar] (2.2, -2.5) rectangle (2.8, 3); 
            \draw[vbar] (4.2, -2.5) rectangle (4.8, 3); 
            
            \node[left]  at (0, 1.5) {$i$};
            \node[left]  at (0,-0.5) {$k$};
            
            \node[above] at (2.5, 3) {$j$};
            \node[above] at (4.5, 3) {$\ell$};
            
            \node[dotA] at (2.5, 1.5) {};
            \node[dotC] at (4.5, 1.5) {};
            \node[dotB] at (2.5,-0.5) {};
            \node[dotD] at (4.5,-0.5) {};
            
            \begin{scope}[shift={(6.5,0)}]
              \node[anchor=west] at (0,1.4)
                {$\color{red!80}{A}_{ij}\,
                  \color{green!70!black}{B}_{jk}\,
                  \color{violet!70}{C}_{k\ell}\,
                  \color{cyan!70!blue}{D}_{\ell i}$};
            
              \draw[->, thick] (0.7,1.1) -- (0.7,0.3);
              \node[anchor=west] at (1.0,0.7)
                {transpose $\;\color{green!70!black}{B},\,\color{cyan!70!blue}{D}$};
            
              \node[anchor=west] at (0,-0.3)
                {$\color{red!80}{A}_{ij}\,
                  \color{green!70!black}{(B^{\mathsf T})}_{kj}\,
                  \color{violet!70}{C}_{k\ell}\,
                  \color{cyan!70!blue}{(D^{\mathsf T})}_{i\ell}$};
            \end{scope}
            \end{tikzpicture}   
        \caption{Illustration of the trace term $\tr(ABCD)$ with matrices $A$ and $C$ and transposed matrices $B^{\text{T}}$ and $D^{\text{T}}$. This representation simplifies the experimental implementation.}
        \label{fig:double layer trace}
     \end{subfigure}
    \vspace{1.0em}
    \begin{subfigure}{0.8\linewidth}
        \centering
    \begin{tikzpicture}[scale=0.88, >=stealth
    ,trim left=1.5cm
    ]
    \tikzset{
      bar/.style={draw, thick, minimum width=0.5cm, minimum height=3.5cm},
      ghostbar/.style={draw, thick, dashed, minimum width=0.5cm, minimum height=3.5cm},
    }
    
    \def\ybot{0}
    \def\ytop{3.5}
    \def\dx{1.2}
    
    \newcommand{\drawcircles}[2]{
        \foreach \k in {0.7,1.4,2.1,2.8}{
            \draw (#1+0.25,#2+\k) circle (0.09);
        }
    }
    
    \def\circlemid{1.75}
    
    \foreach \lab/\i in {1/0, 2/1, 3/2, 4/3} {
      \pgfmathsetmacro{\x}{\i * \dx}
      \draw[bar] (\x,\ybot) rectangle ++(0.5,\ytop);
      \drawcircles{\x}{\ybot}
      \node[above] at (\x+0.25,\ytop) {\lab};
    }
    
    \draw[->, thick] (4.5,2) -- (5.0,2);
    
    \def\xshift{5.8}
    \pgfmathsetmacro{\ydisp}{0.35}
    
    \foreach \lab/\slot in {1/0, 2/2, 3/4, 4/6} {
      \pgfmathsetmacro{\x}{\xshift + \slot * \dx}
      \draw[bar] (\x,\ybot) rectangle ++(0.5,\ytop);
      \node[above] at (\x+0.25,\ytop) {\lab};
    }
    
    \foreach \slot/\cy in {
      0/0,
      2/\ydisp,
      4/0,
      6/\ydisp
    } {
      \pgfmathsetmacro{\x}{\xshift + \slot * \dx}
      \drawcircles{\x}{\cy}
    }
    
    \foreach \slot in {1,3} {
      \pgfmathsetmacro{\x}{\xshift + \slot * \dx}
      \draw[ghostbar] (\x,\ybot) rectangle ++(0.5,\ytop);
    }
    
    
    \pgfmathsetmacro{\xGhostA}{\xshift + 1*\dx}
    \pgfmathsetmacro{\xGhostB}{\xshift + 3*\dx}
    
    \pgfmathsetmacro{\yTopArc}{\ytop + 0.25}
    
    \coordinate (StartA) at ({\xshift + 4*\dx + 0.25},{\ybot + \circlemid});
    \coordinate (StartB) at ({\xshift + 6*\dx + 0.25},{\ydisp + \circlemid});
    
    \draw[->, thick]
      (StartA)
      to[out=150,in=30,looseness=1.4]
      (\xGhostA+0.25,\yTopArc);
    
    \draw[->, thick]
      (StartB)
      to[out=150,in=30,looseness=1.4]
      (\xGhostB+0.25,\yTopArc);
    
    \draw[->, thick] (\xshift+8.3,2) -- (\xshift+8.8,2);
    
    \begin{scope}[shift={(2,0)}]
    
    \def\xshiftC{13.4}
    
    \foreach \lab/\i in {1/0, 3/1, 2/2, 4/3} {
      \pgfmathsetmacro{\x}{\xshiftC + \i * \dx}
      \draw[bar] (\x,\ybot) rectangle ++(0.5,\ytop);
      \drawcircles{\x}{\ybot}
      \node[above] at (\x+0.25,\ytop) {\lab};
    }
    
    \end{scope}
    
    \end{tikzpicture}
        \caption{Schematic illustration of the column permutation under $\sigma$ in AMO experiments.
        }
        \label{fig:AMO permutation}
    \end{subfigure}
    \caption{Experimental implementation of the trace term and permutations.
    }
    \label{fig:experimental implementation}
\end{figure}
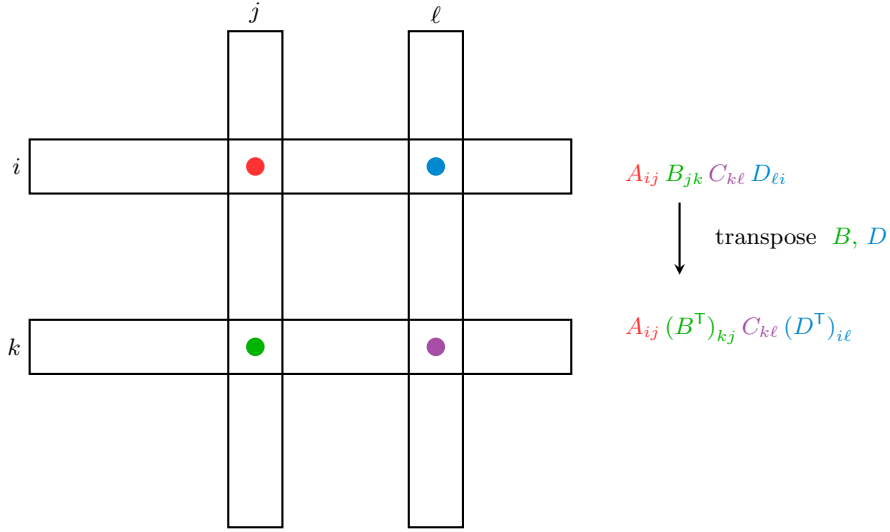
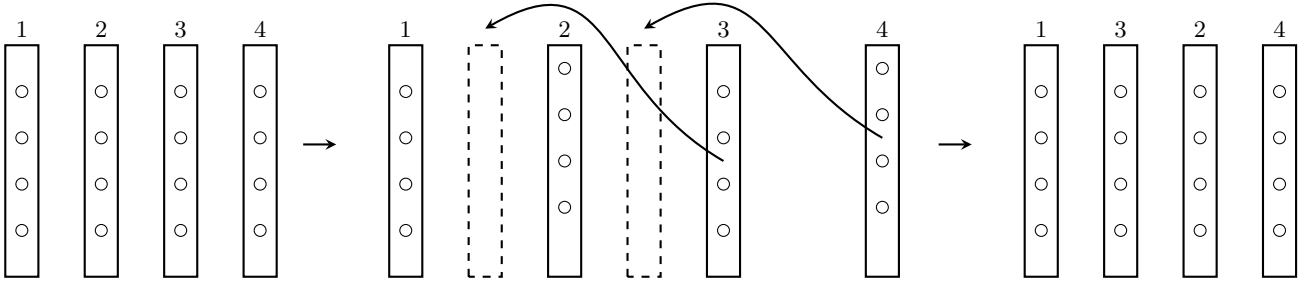

Recall from Section \ref{cartoon} that our goal is to simulate a term like $\tr(\Phi^4)$.  More generally, suppose we wished to simulate $\tr(ABCD)$ where $A,B,C,D$ represent four (possibly distinct) matrices. From (\ref{eq:trotterization}), the corresponding interaction takes the form  $A_{ij}B_{jk}C_{kl}D_{li}$. As summarized in the introduction, if we transpose matrices $B$ and $D$, the interaction becomes $A_{ij}(B^\mathsf{T})_{kj}C_{kl}(D^\mathsf{T})_{il}$.   As shown in Figure \ref{fig:double layer trace}, if $A/C$ and $B^{\mathsf{T}}/D^{\mathsf{T}}$ were aligned as arrays, then this interaction would take place between qubits in very specific locations in the array.   If the qubits that \emph{initially} started in these locations were later permuted to be next to each other (which could be achieved via permutations of entire rows and columns at once), then we could actually use a local interaction in space to achieve the interaction in the matrix model, combined with a simple pattern of row and column permutations to implement our desired gates.

So we see that the cyclic connectivity rules given in (\ref{eq:diagonal_rule}) and (\ref{eq:off-diagonal_rule}) are not \emph{quite} enough for the cartoon matrix model to be efficiently implemented in experiment.   It turns out that the rules we stated above can still require nonlocal interactions between qubits beceause we have not cleverly alternated a matrix with its transpose in the interaction Hamiltonian.  Since this can be rather unnatural from a microscopic perspective --  $\tr(\Phi \Phi^{\mathsf{T}} \Phi \Phi^{\mathsf{T}})$ is not invariant under the same symmetries of $\tr(\Phi^4)$ -- we therefore propose an alternative microscopic model that is more amenable to our experiment.  Our alternative cartoon consists of two interacting matrices $B$ and $T$, with $2N^2$ total qubits, and a Hamiltonian with the interaction pattern \begin{equation}
    H \sim \tr(BTBT).
\end{equation}
Notice that if we now store the matrix $B$ (bottom) and $T$ (top) next to each other as 2d arrays, but with $T$ transposed relative to $B$, we can realize the four-qubit interactions needed in $H$ by the simple motions depicted in Figure \ref{fig:experimental implementation}.

We also adjust the connectivity pattern from (\ref{eq:diagonal_rule}) and (\ref{eq:off-diagonal_rule}) to alternate between the two layers in the sequence: $BTBT$ and $TBTB$. In this configuration, the cyclically interacting qubits in each subset can be placed spatially close to one another, as shown in Figure \ref{fig:experimental implementation}. 
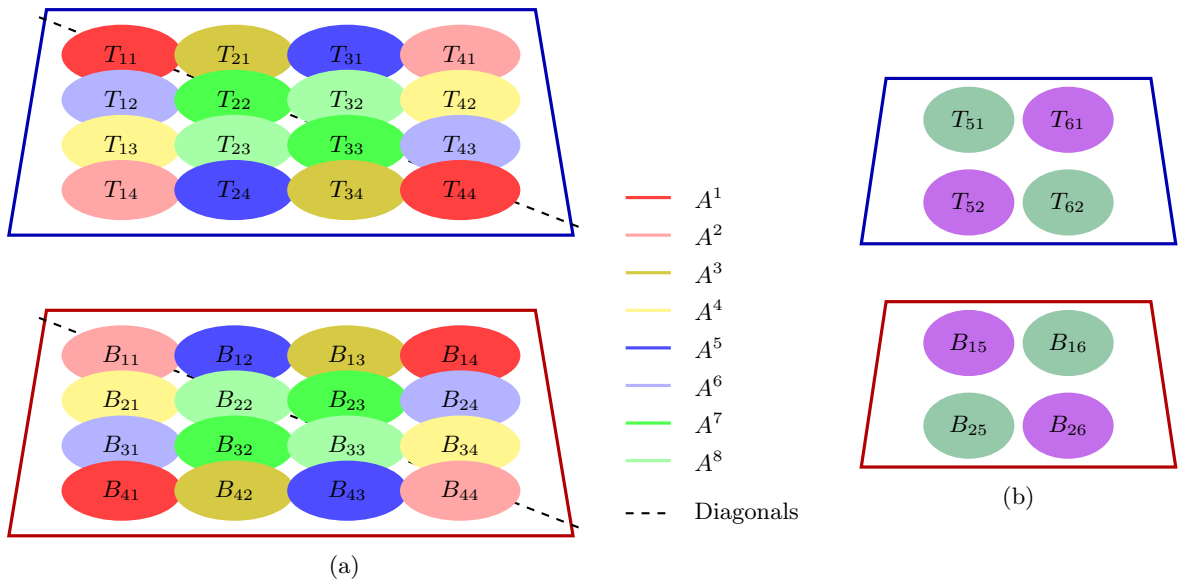
\begin{figure}[h]
    \centering
    \begin{subfigure}[h]{0.5\textwidth}
        \centering
        \begin{tikzpicture}[scale=1.0]
        \colorlet{Aone}   {red!75!white}          
        \colorlet{Atwo}   {red!35!white}          
        \colorlet{Athree} {yellow!80!black}       
        \colorlet{Afour}  {yellow!55!white}       
        \colorlet{Afive}  {blue!70!white}         
        \colorlet{Asix}   {blue!30!white}         
        \colorlet{Aseven} {green!70!white}        
        \colorlet{Aeight} {green!35!white}        
        
        \tikzset{
          trapezoidT/.style={draw=blue!70!black, very thick},
          trapezoidB/.style={draw=red!70!black,  very thick},
          diagline/.style={dashed, thick},
          Tellipse/.style={draw=none},
          Bellipse/.style={draw=none},
          Tlabel/.style={font=\small},
          Blabel/.style={font=\small},
        }
        
        \def\colxA{1.5}
        \def\colxB{3.0}
        \def\colxC{4.5}
        \def\colxD{6.0}
        
        \def\rowyOneTop{2.4}
        \def\rowyTwoTop{1.8}
        \def\rowyThreeTop{1.2}
        \def\rowyFourTop{0.6}
        
        \def\rowyOneBot{-1.6}
        \def\rowyTwoBot{-2.2}
        \def\rowyThreeBot{-2.8}
        \def\rowyFourBot{-3.4}
        
        \def\rx{0.8}
        \def\ry{0.4}
        
        \draw[trapezoidT]
          (0,0) -- (7.5,0) -- (7.0,3) -- (0.5,3) -- cycle;
        \draw[diagline] (0.4,2.9) -- (7.6,0.1);
        
        
        \draw[Tellipse, fill=Aone]   (\colxA,\rowyOneTop) ellipse [x radius=\rx, y radius=\ry];
        \node[Tlabel] at (\colxA,\rowyOneTop) {$T_{11}$};  
        
        \draw[Tellipse, fill=Athree] (\colxB,\rowyOneTop) ellipse [x radius=\rx, y radius=\ry];
        \node[Tlabel] at (\colxB,\rowyOneTop) {$T_{21}$};  
        
        \draw[Tellipse, fill=Afive]  (\colxC,\rowyOneTop) ellipse [x radius=\rx, y radius=\ry];
        \node[Tlabel] at (\colxC,\rowyOneTop) {$T_{31}$};  
        
        \draw[Tellipse, fill=Atwo]   (\colxD,\rowyOneTop) ellipse [x radius=\rx, y radius=\ry];
        \node[Tlabel] at (\colxD,\rowyOneTop) {$T_{41}$};  
        
        \draw[Tellipse, fill=Asix]   (\colxA,\rowyTwoTop) ellipse [x radius=\rx, y radius=\ry];
        \node[Tlabel] at (\colxA,\rowyTwoTop) {$T_{12}$};  
        
        \draw[Tellipse, fill=Aseven] (\colxB,\rowyTwoTop) ellipse [x radius=\rx, y radius=\ry];
        \node[Tlabel] at (\colxB,\rowyTwoTop) {$T_{22}$};  
        
        \draw[Tellipse, fill=Aeight] (\colxC,\rowyTwoTop) ellipse [x radius=\rx, y radius=\ry];
        \node[Tlabel] at (\colxC,\rowyTwoTop) {$T_{32}$};  
        
        \draw[Tellipse, fill=Afour]  (\colxD,\rowyTwoTop) ellipse [x radius=\rx, y radius=\ry];
        \node[Tlabel] at (\colxD,\rowyTwoTop) {$T_{42}$};  
        
        \draw[Tellipse, fill=Afour]  (\colxA,\rowyThreeTop) ellipse [x radius=\rx, y radius=\ry];
        \node[Tlabel] at (\colxA,\rowyThreeTop) {$T_{13}$};  
        
        \draw[Tellipse, fill=Aeight] (\colxB,\rowyThreeTop) ellipse [x radius=\rx, y radius=\ry];
        \node[Tlabel] at (\colxB,\rowyThreeTop) {$T_{23}$};  
        
        \draw[Tellipse, fill=Aseven] (\colxC,\rowyThreeTop) ellipse [x radius=\rx, y radius=\ry];
        \node[Tlabel] at (\colxC,\rowyThreeTop) {$T_{33}$};  
        
        \draw[Tellipse, fill=Asix]   (\colxD,\rowyThreeTop) ellipse [x radius=\rx, y radius=\ry];
        \node[Tlabel] at (\colxD,\rowyThreeTop) {$T_{43}$};  
        
        \draw[Tellipse, fill=Atwo]   (\colxA,\rowyFourTop) ellipse [x radius=\rx, y radius=\ry];
        \node[Tlabel] at (\colxA,\rowyFourTop) {$T_{14}$};  
        
        \draw[Tellipse, fill=Afive]  (\colxB,\rowyFourTop) ellipse [x radius=\rx, y radius=\ry];
        \node[Tlabel] at (\colxB,\rowyFourTop) {$T_{24}$};  
        
        \draw[Tellipse, fill=Athree] (\colxC,\rowyFourTop) ellipse [x radius=\rx, y radius=\ry];
        \node[Tlabel] at (\colxC,\rowyFourTop) {$T_{34}$};  
        
        \draw[Tellipse, fill=Aone]   (\colxD,\rowyFourTop) ellipse [x radius=\rx, y radius=\ry];
        \node[Tlabel] at (\colxD,\rowyFourTop) {$T_{44}$};  
        
        \draw[trapezoidB]
          (0,-4) -- (7.5,-4) -- (7.0,-1) -- (0.5,-1) -- cycle;
        \draw[diagline] (0.4,-1.1) -- (7.6,-3.9);
        
        
        \draw[Bellipse, fill=Atwo]   (\colxA,\rowyOneBot) ellipse [x radius=\rx, y radius=\ry];
        \node[Blabel] at (\colxA,\rowyOneBot) {$B_{11}$};  
        
        \draw[Bellipse, fill=Afive]  (\colxB,\rowyOneBot) ellipse [x radius=\rx, y radius=\ry];
        \node[Blabel] at (\colxB,\rowyOneBot) {$B_{12}$};  
        
        \draw[Bellipse, fill=Athree] (\colxC,\rowyOneBot) ellipse [x radius=\rx, y radius=\ry];
        \node[Blabel] at (\colxC,\rowyOneBot) {$B_{13}$};  
        
        \draw[Bellipse, fill=Aone]   (\colxD,\rowyOneBot) ellipse [x radius=\rx, y radius=\ry];
        \node[Blabel] at (\colxD,\rowyOneBot) {$B_{14}$};  
        
        \draw[Bellipse, fill=Afour]  (\colxA,\rowyTwoBot) ellipse [x radius=\rx, y radius=\ry];
        \node[Blabel] at (\colxA,\rowyTwoBot) {$B_{21}$};  
        
        \draw[Bellipse, fill=Aeight] (\colxB,\rowyTwoBot) ellipse [x radius=\rx, y radius=\ry];
        \node[Blabel] at (\colxB,\rowyTwoBot) {$B_{22}$};  
        
        \draw[Bellipse, fill=Aseven] (\colxC,\rowyTwoBot) ellipse [x radius=\rx, y radius=\ry];
        \node[Blabel] at (\colxC,\rowyTwoBot) {$B_{23}$};  
        
        \draw[Bellipse, fill=Asix]   (\colxD,\rowyTwoBot) ellipse [x radius=\rx, y radius=\ry];
        \node[Blabel] at (\colxD,\rowyTwoBot) {$B_{24}$};  
        
        \draw[Bellipse, fill=Asix]   (\colxA,\rowyThreeBot) ellipse [x radius=\rx, y radius=\ry];
        \node[Blabel] at (\colxA,\rowyThreeBot) {$B_{31}$};  
        
        \draw[Bellipse, fill=Aseven] (\colxB,\rowyThreeBot) ellipse [x radius=\rx, y radius=\ry];
        \node[Blabel] at (\colxB,\rowyThreeBot) {$B_{32}$};  
        
        \draw[Bellipse, fill=Aeight] (\colxC,\rowyThreeBot) ellipse [x radius=\rx, y radius=\ry];
        \node[Blabel] at (\colxC,\rowyThreeBot) {$B_{33}$};  
        
        \draw[Bellipse, fill=Afour]  (\colxD,\rowyThreeBot) ellipse [x radius=\rx, y radius=\ry];
        \node[Blabel] at (\colxD,\rowyThreeBot) {$B_{34}$};  
        
        \draw[Bellipse, fill=Aone]   (\colxA,\rowyFourBot) ellipse [x radius=\rx, y radius=\ry];
        \node[Blabel] at (\colxA,\rowyFourBot) {$B_{41}$};  
        
        \draw[Bellipse, fill=Athree] (\colxB,\rowyFourBot) ellipse [x radius=\rx, y radius=\ry];
        \node[Blabel] at (\colxB,\rowyFourBot) {$B_{42}$};  
        
        \draw[Bellipse, fill=Afive]  (\colxC,\rowyFourBot) ellipse [x radius=\rx, y radius=\ry];
        \node[Blabel] at (\colxC,\rowyFourBot) {$B_{43}$};  
        
        \draw[Bellipse, fill=Atwo]   (\colxD,\rowyFourBot) ellipse [x radius=\rx, y radius=\ry];
        \node[Blabel] at (\colxD,\rowyFourBot) {$B_{44}$};  
        
        \begin{scope}[shift={(8.2,-3)}]
          \draw[very thick, Aone]   (0,3.5) -- (0.6,3.5); \node[right] at (0.8,3.5) {$A^1$};
          \draw[very thick, Atwo]   (0,3.0) -- (0.6,3.0); \node[right] at (0.8,3.0) {$A^2$};
          \draw[very thick, Athree] (0,2.5) -- (0.6,2.5); \node[right] at (0.8,2.5) {$A^3$};
          \draw[very thick, Afour]  (0,2.0) -- (0.6,2.0); \node[right] at (0.8,2.0) {$A^4$};
          \draw[very thick, Afive]  (0,1.5) -- (0.6,1.5); \node[right] at (0.8,1.5) {$A^5$};
          \draw[very thick, Asix]   (0,1.0) -- (0.6,1.0); \node[right] at (0.8,1.0) {$A^6$};
          \draw[very thick, Aseven] (0,0.5) -- (0.6,0.5); \node[right] at (0.8,0.5) {$A^7$};
          \draw[very thick, Aeight] (0,0.0) -- (0.6,0.0); \node[right] at (0.8,0.0) {$A^8$};
        
          \draw[diagline] (0,-0.7) -- (0.6,-0.7);
          \node[right] at (0.8,-0.7) {Diagonals};
        \end{scope}
        \end{tikzpicture}  
        \caption{}
        \label{fig:double diagonal rules}
    \end{subfigure}%
        \begin{subfigure}[h]{0.5\textwidth}
        \centering
        \begin{tikzpicture}[scale=1.1]
        \definecolor{orbmag}{RGB}{195,110,235}   
        \definecolor{orbgrn}{RGB}{150,200,170}   
        
        \tikzset{
          trapezoidT/.style={draw=blue!70!black, very thick},
          trapezoidB/.style={draw=red!70!black,  very thick},
          Tellipse/.style={draw=none},
          Bellipse/.style={draw=none},
          Ttext/.style={font=\small},
          Btext/.style={font=\small},
        }
        
        \def\xL{2.0}
        \def\xR{3.2}
        \def\yTtop{1.7}
        \def\yTbot{0.7}
        \def\yBtop{-1.0}
        \def\yBbot{-2.0}
        \def\rx{0.55}
        \def\ry{0.4}
        
        \draw[trapezoidT]
          (0.7,0.2) -- (4.5,0.2) -- (4.2,2.2) -- (1.0,2.2) -- cycle;
        
        \draw[Tellipse, fill=orbgrn] (\xL,\yTtop) ellipse [x radius=\rx, y radius=\ry];
        \node[Ttext] at (\xL,\yTtop) {$T_{51}$};
        
        \draw[Tellipse, fill=orbmag] (\xR,\yTtop) ellipse [x radius=\rx, y radius=\ry];
        \node[Ttext] at (\xR,\yTtop) {$T_{61}$};
        
        \draw[Tellipse, fill=orbmag] (\xL,\yTbot) ellipse [x radius=\rx, y radius=\ry];
        \node[Ttext] at (\xL,\yTbot) {$T_{52}$};
        
        \draw[Tellipse, fill=orbgrn] (\xR,\yTbot) ellipse [x radius=\rx, y radius=\ry];
        \node[Ttext] at (\xR,\yTbot) {$T_{62}$};
        
        \draw[trapezoidB]
          (0.7,-2.5) -- (4.5,-2.5) -- (4.2,-0.5) -- (1.0,-0.5) -- cycle;
        
        \draw[Bellipse, fill=orbmag] (\xL,\yBtop) ellipse [x radius=\rx, y radius=\ry];
        \node[Btext] at (\xL,\yBtop) {$B_{15}$};
        
        \draw[Bellipse, fill=orbgrn] (\xR,\yBtop) ellipse [x radius=\rx, y radius=\ry];
        \node[Btext] at (\xR,\yBtop) {$B_{16}$};
        
        \draw[Bellipse, fill=orbgrn] (\xL,\yBbot) ellipse [x radius=\rx, y radius=\ry];
        \node[Btext] at (\xL,\yBbot) {$B_{25}$};
        
        \draw[Bellipse, fill=orbmag] (\xR,\yBbot) ellipse [x radius=\rx, y radius=\ry];
        \node[Btext] at (\xR,\yBbot) {$B_{26}$};
        
        
        
        \end{tikzpicture}
        \caption{}
        \label{fig:double off-diagonal rule}
    \end{subfigure}
    \caption{Visualization of adjusted connectivity rules for double-layer system.}
    \label{fig:double_connectivity_rules}
\end{figure}

In addition to the cyclic connectivity rules, we apply identical row and column permutations to both the bottom layer and the transposed top layer, which can be implemented easily through the procedures illustrated in Figure \ref{fig:AMO permutation} in experiments, as we discuss more in the next paragraph.  
For example, for $N = 8$, (\ref{eq:sigma}) defines the index permutation $2 \rightarrow 5$. Accordingly, the second row and column in each layer are permuted: $B_{12}$ at position $(1,2)$ in the bottom layer is mapped to position $(1,5)$, while $T_{12}$ at position $(2,1)$ in the top layer is mapped to position $(5,1)$. Hence, in both layers the same permutation $\sigma$ in (\ref{eq:sigma}) is applied consistently to all labeled qubits. 

Recent work proposes a similar move set of permutations in Figure \ref{fig:AMO permutation} in neutral-atom arrays, motivated by quantum error correction \cite{Xu2024ConstantOverheadAtomArrays,Guo2025SelfCorrectingNeutralAtoms}.  The proposal is to hold neutral atoms in spatial light modulator (SLM) traps, while acousto-optic deflector (AOD) optical tweezers are used to transfer atoms between trap sites \cite{Bluvstein2024LogicalQuantumProcessor}. The movable optical tweezers can then implement our ``shuffling" permutation on both rows and columns as follows. First, all even-indexed atoms are released from their SLM traps. After being transported in parallel by AOD tweezers to a set of unoccupied SLM traps (“vacant sites”) located at the rightmost end of the array, the spacing between these atoms is rescaled. 
During the transfer, the ordering of atoms is preserved to avoid trajectory intersections that could disturb them.  Next, all odd-indexed atoms except for the first atom are released from their SLM traps. After the spacing between these atoms is rescaled, they are shifted left using AOD tweezers and placed into their target SLM traps, again preserving their ordering. See Figure \ref{fig:AMO permutation} for a detailed illustration, where atoms in the even-indexed columns are vertically displaced relative to those in the odd-indexed columns to avoid collisions \cite{Guo2025SelfCorrectingNeutralAtoms}. 
Finally, the even-indexed atoms are transferred back from the ``vacant sites" and placed adjacent to the odd-indexed atoms, with the spacing rescaled once again and all atoms restored to their original positions (up to the permutation, of course). Thus, the desired shuffling dynamics is realized. The row/column permutation of atoms thus follows naturally.  

The key observation is that the atomic motions described here correspond to global translations and/or rescaling of just a few distinct sublattices of the atoms.  One does not need single-atom control in order to implement this circuit, which will make it more appealing for experimental implementation vs. a generic few-body circuit with gates between randomly chosen qubits; see \cite{Xu2024ConstantOverheadAtomArrays,Constantinides:2024rwb} for work on optimizing routing in this setting.

\subsection{Scrambling in the double layer model}
In the double layer model, the $2N^2$ qubits are partitioned into $N^2/2$ subsets of four qubits each. The rest follows the same construction as in \eqref{eq:U_int}, \eqref{eq:U}, and (\ref{eq:U_perm}) for the single-layer model. Since the total number of qubits doubles, the operator size is expected to grow to 
\begin{align}
    \langle n\rangle_{\text{double}} &= \frac{3}{4}\cdot 2N^2\nonumber\\
    &= \frac{3}{2}N^2\label{eq:double operator size upper bound}
\end{align}
from (\ref{eq:operator size upper bound}). Similarly, (\ref{eq:final_entropy}) and (\ref{eq:initial_entropy}) should be satisfied for set $A$ with
\begin{align}
    \abs{A}&\leq \frac{2N^2}{2}\nonumber\\
    &= N^2, \label{eq: double_EE}
\end{align}
and the boundary of recovery is at
\begin{align}
    q_{\text{double}}=r_{\text{double}}=\frac{2N^2}{3}\label{eq:double_boundary_of_recovery}
\end{align}
from (\ref{eq:boundary_of_recovery}).

\begin{figure}[t]
    \centering
    \begin{subfigure}[h]{0.5\textwidth}
        \centering
        \includegraphics[scale=0.5]{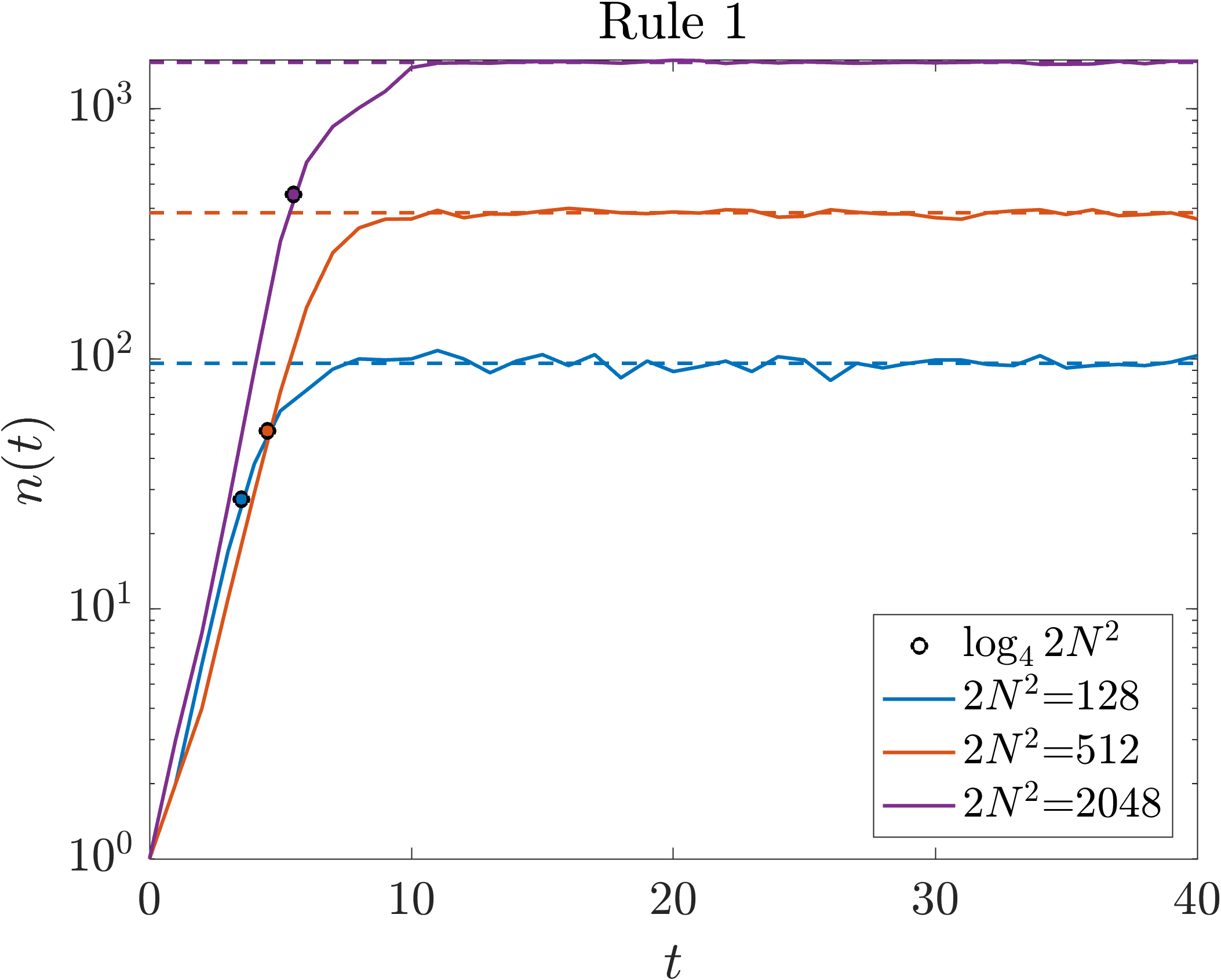}
        \caption{}
        \label{fig:OSG_different_even}
    \end{subfigure}%
    \begin{subfigure}[h]{0.5\textwidth}
        \centering
        \includegraphics[scale=0.5]{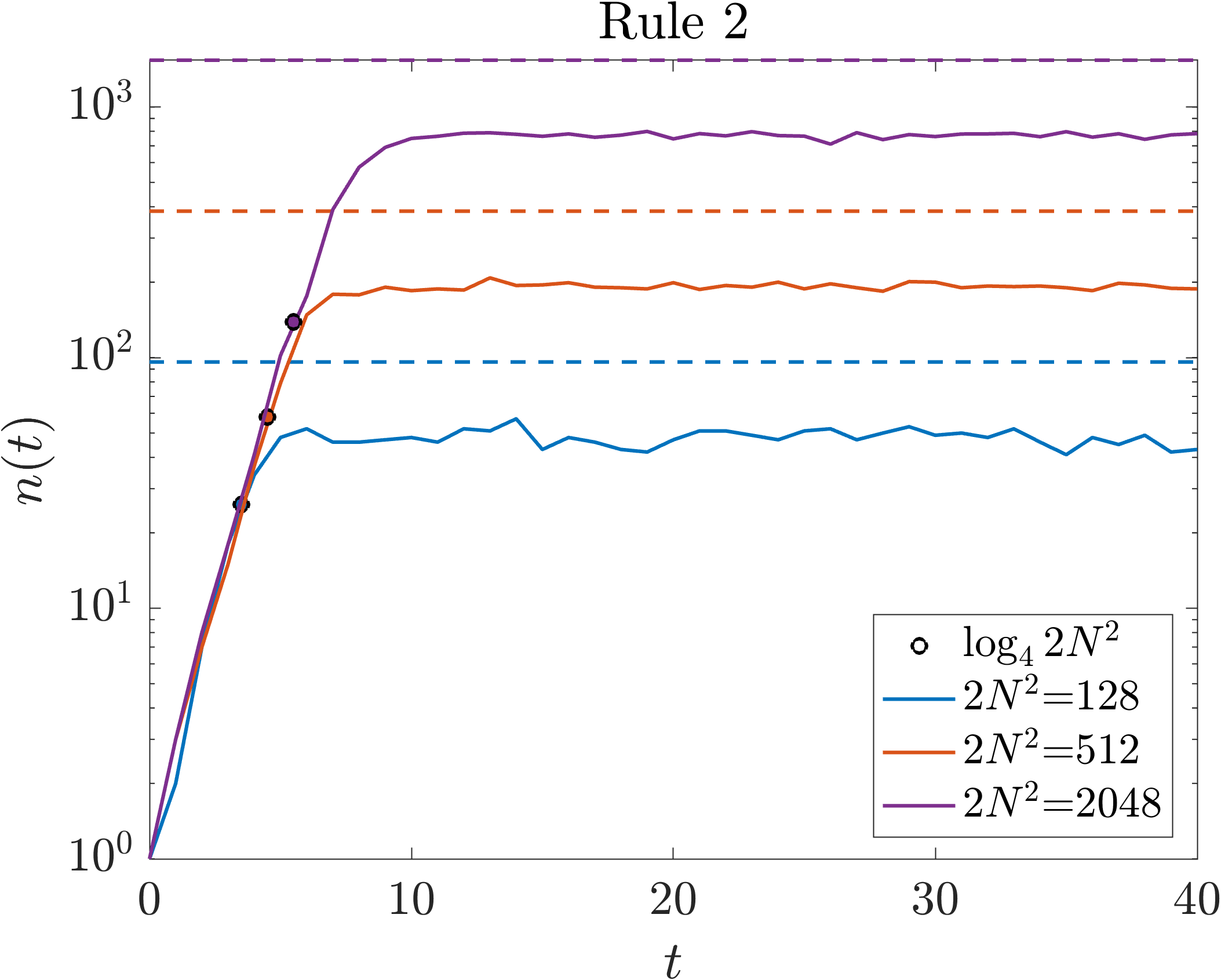}
        \caption{}
        \label{fig:OSG_same_even}
    \end{subfigure}
    \begin{subfigure}[h]{0.5\textwidth}
        \centering
        \includegraphics[scale=0.5]{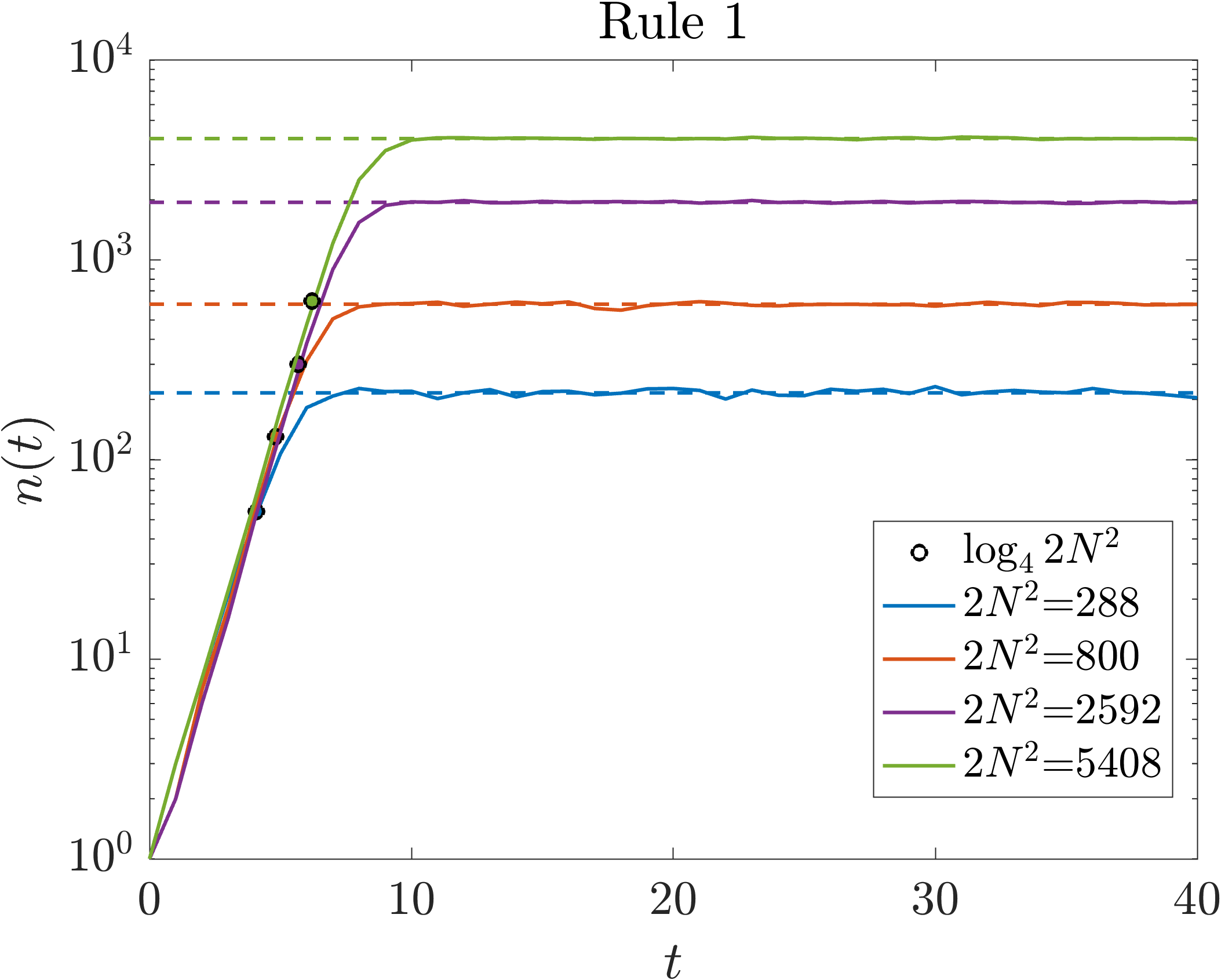}
        \caption{}
        \label{fig:OSG_different_odd}
    \end{subfigure}%
    \begin{subfigure}[h]{0.5\textwidth}
        \centering
        \includegraphics[scale=0.5]{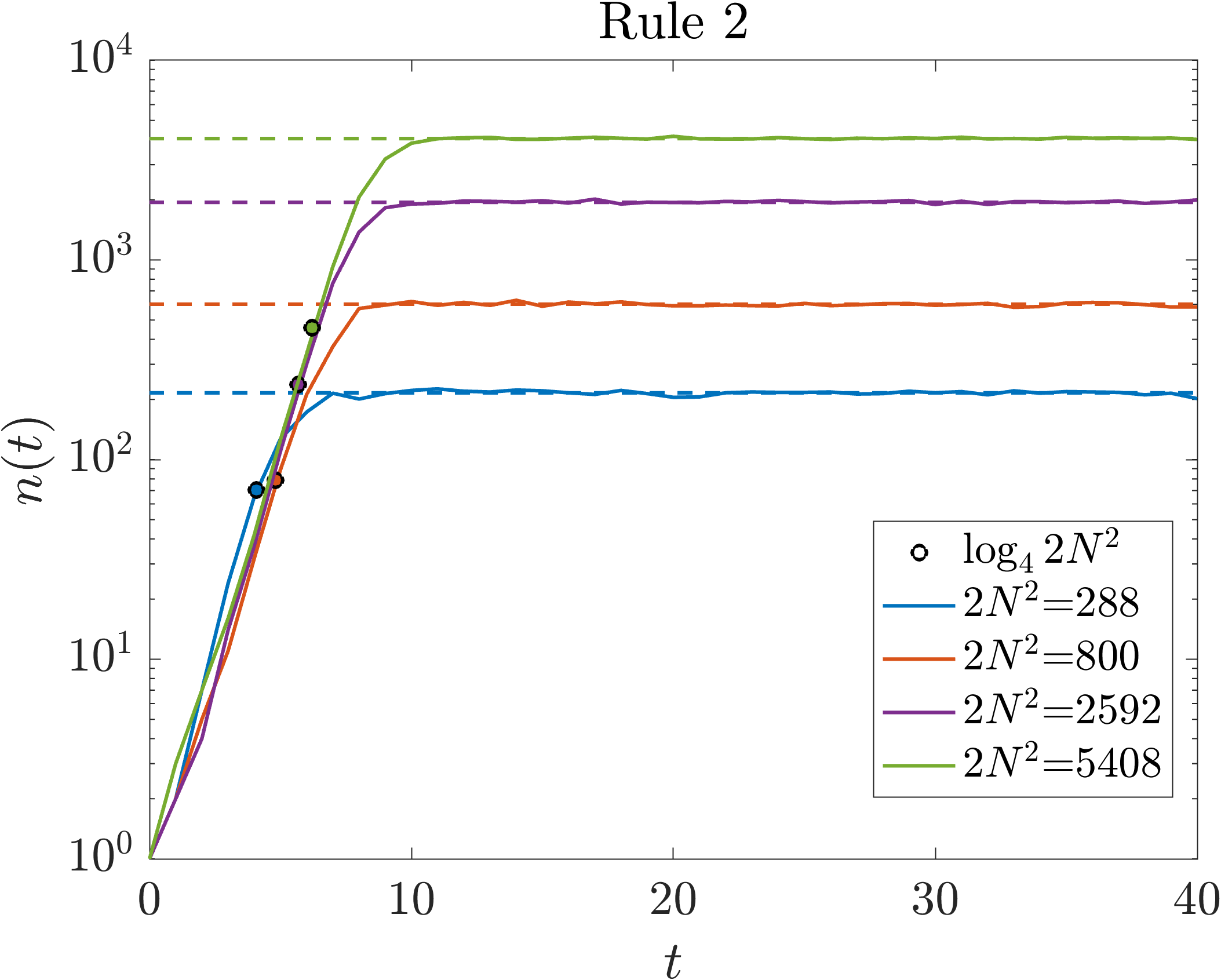}
        \caption{}
        \label{fig:OSG_same_odd}
    \end{subfigure}
    \caption{Growth of the operator size $n(t)$ as a function of time $t$ in the double layer model. The dynamics resemble those of the single layer model, except that in (b) the interacting qubits split into two disjoint subsets, so that the effective number of qubits involved in the interaction is half of the total number of qubits in the two layers.}\label{fig:double_OSG}
\end{figure}

\begin{figure}[h]
    \centering
    \begin{subfigure}[h]{0.5\textwidth}
        \centering
        \includegraphics[scale=0.5]{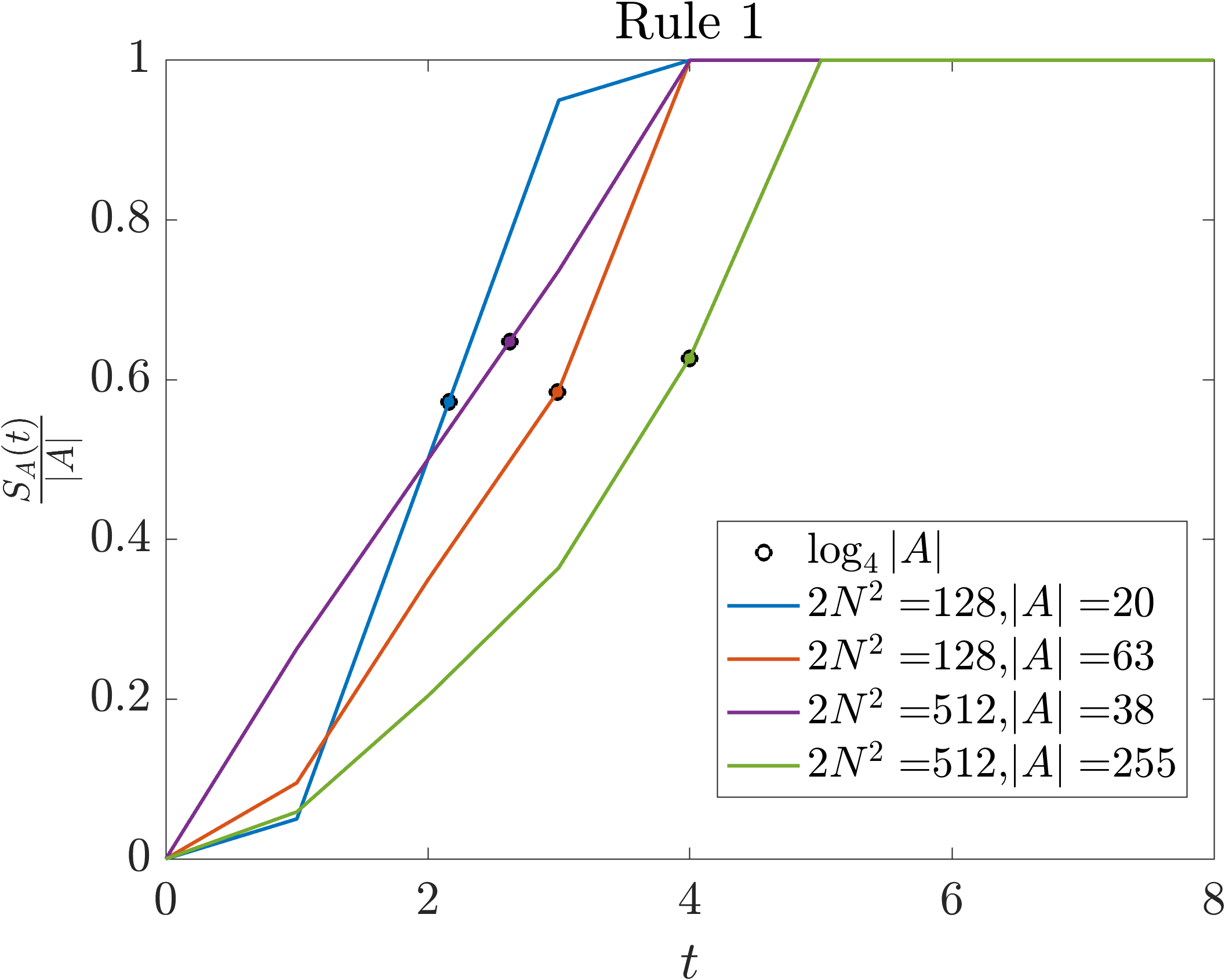}
        \caption{}
        \label{fig:EE_diff_even}
    \end{subfigure}%
    \begin{subfigure}[h]{0.5\textwidth}
        \centering
        \includegraphics[scale=0.5]{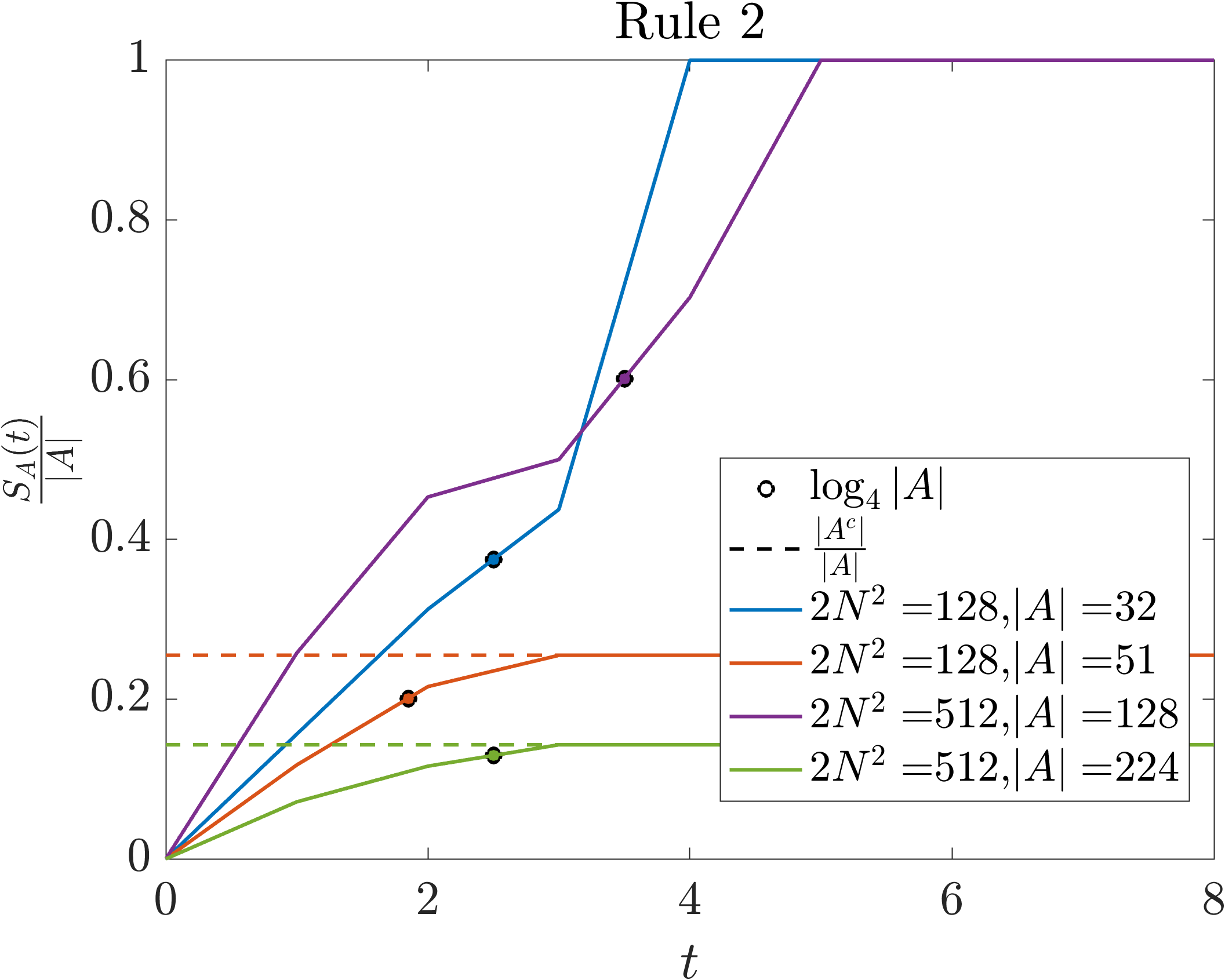}
        \caption{}
        \label{fig:EE_same_even}
    \end{subfigure}
    \begin{subfigure}[h]{0.5\textwidth}
        \centering
        \includegraphics[scale=0.5]{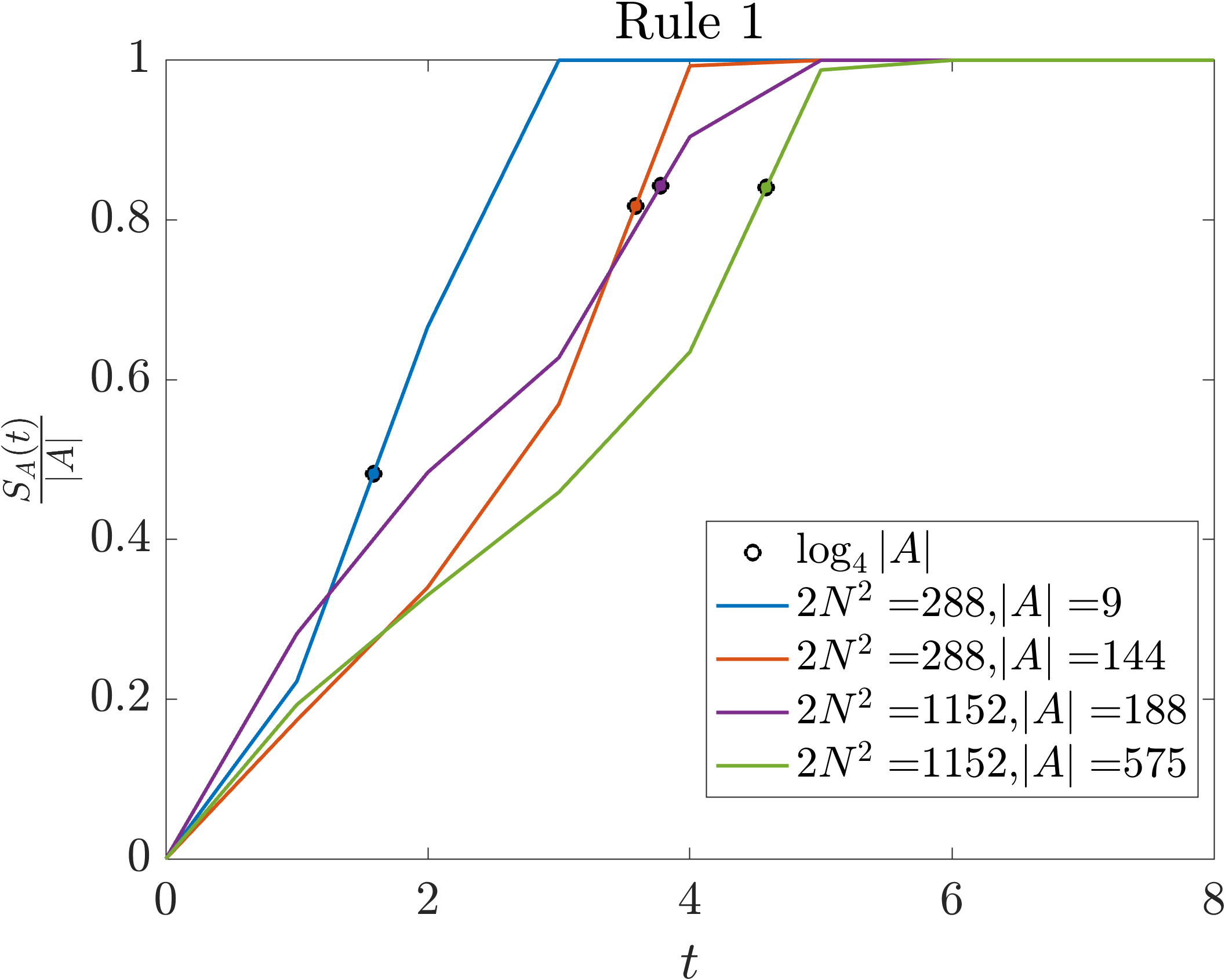}
        \caption{}
        \label{fig:EE_diff_odd}
    \end{subfigure}%
    \begin{subfigure}[h]{0.5\textwidth}
        \centering
        \includegraphics[scale=0.5]{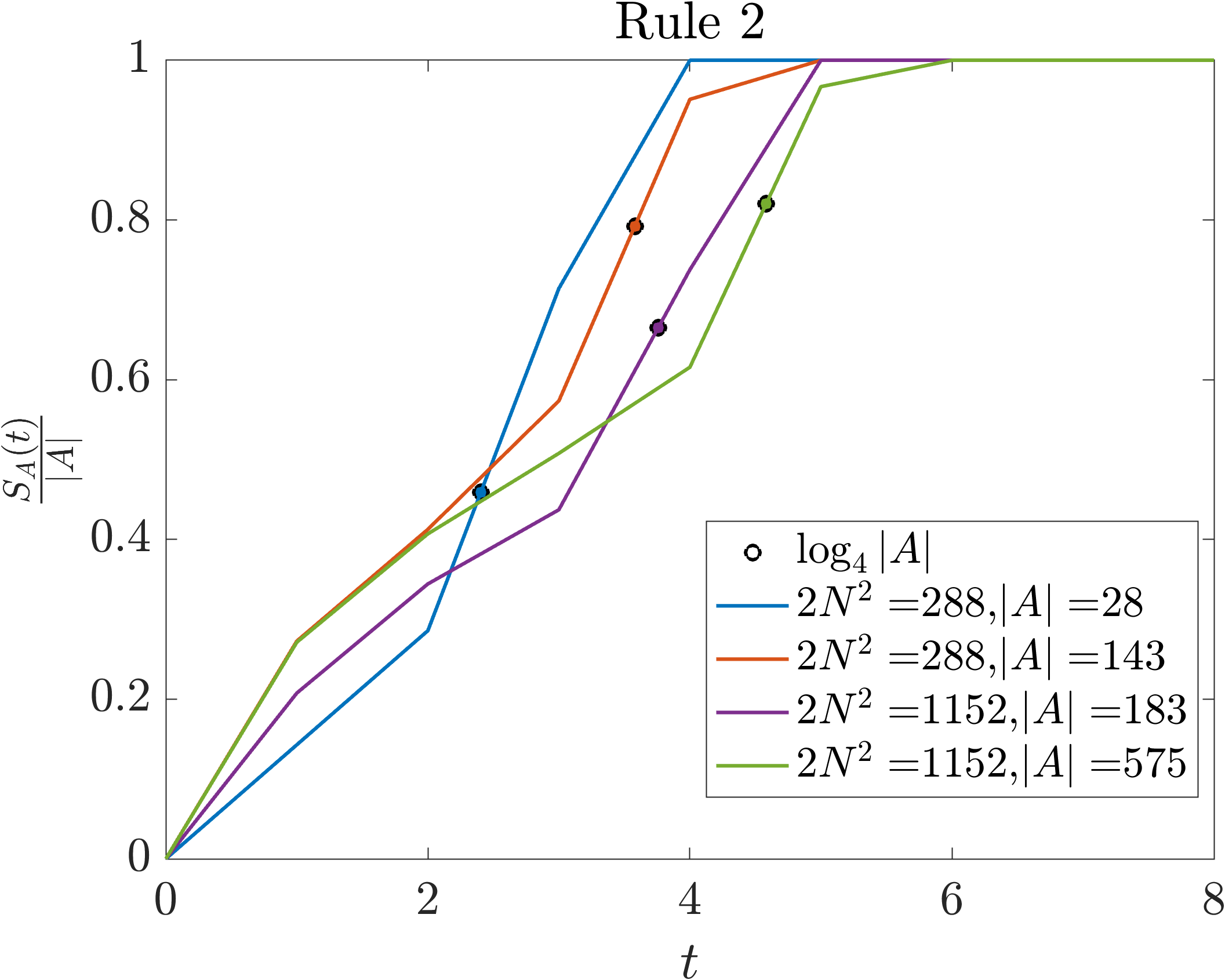}
        \caption{}
        \label{fig:EE_same_odd}
    \end{subfigure}
    \caption{Growth of the entanglement entropy $S_A$ of subsystem $A$ as a function of time $t$ in the double layer model for $|A|\leq N^2$. The dynamics resemble those of the single layer model, except that in (b) the interacting qubits split into two disjoint subsets, and thus $|A|+|A^c| = N^2$. As a consequence, for $|A|>N^2/2$ the late-time entanglement satisfies $S_A/|A^c|=1$, leading to $S_A/|A|=|A^c|/|A|$.}\label{fig:double_EE}
\end{figure}

\begin{figure}[h]
    \centering
    \begin{subfigure}[h]{0.5\textwidth}
        \centering
        \includegraphics[scale=0.5]{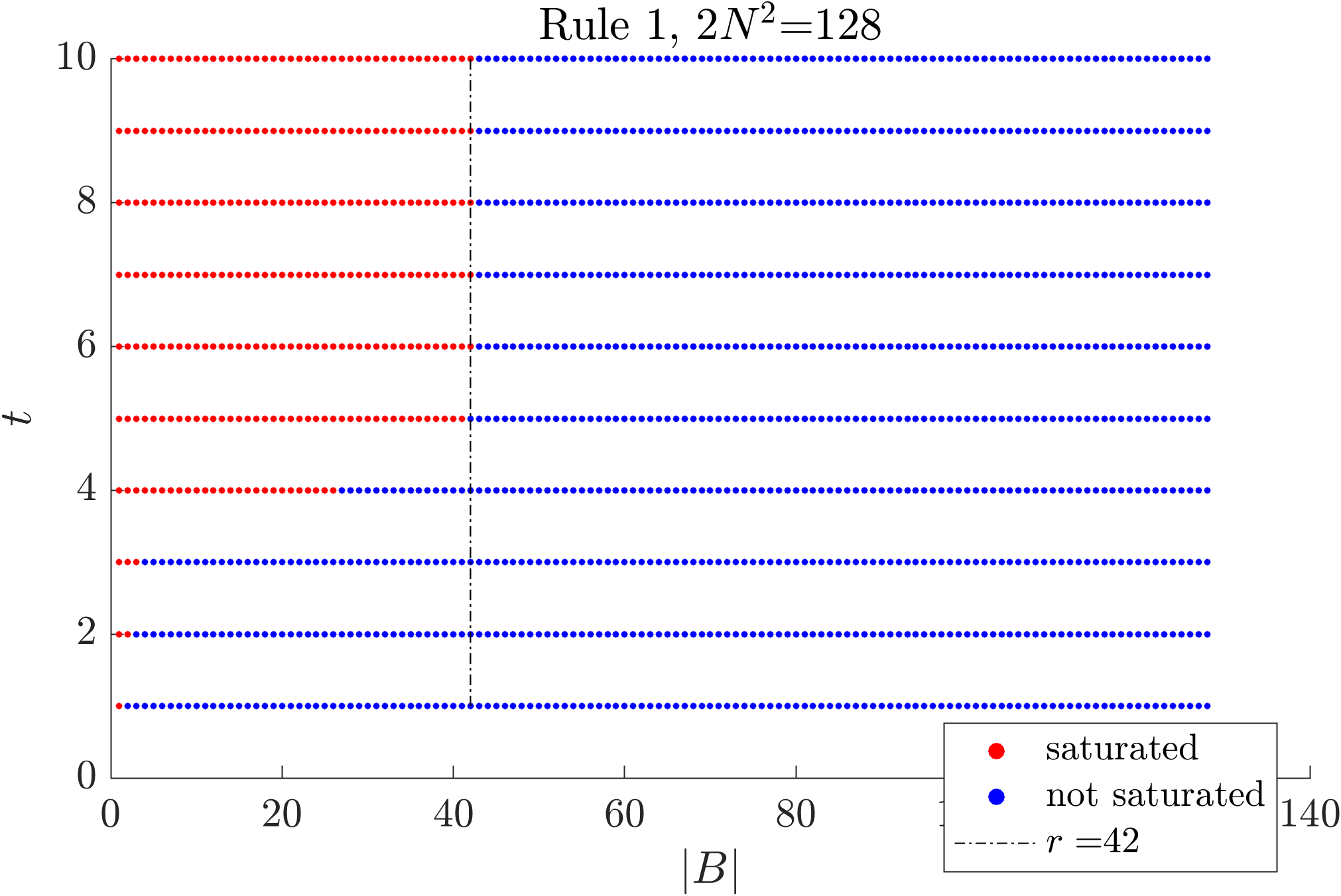}
        \caption{}
        \label{fig:HP_different_even}
    \end{subfigure}%
    \begin{subfigure}[h]{0.5\textwidth}
        \centering
        \includegraphics[scale=0.5]{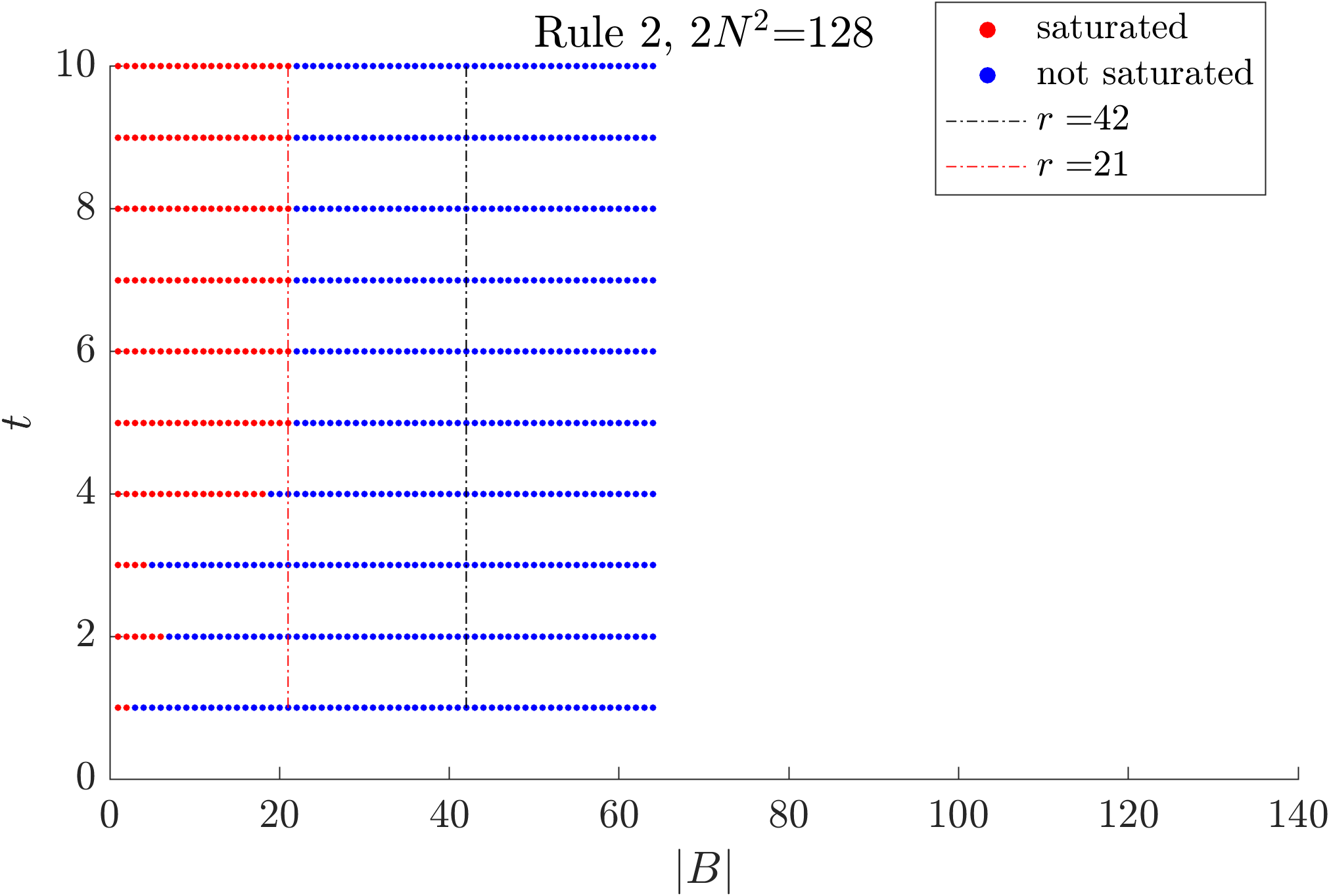}
        \caption{}
        \label{fig:HP_same_even}
    \end{subfigure}
    \begin{subfigure}[h]{0.5\textwidth}
        \centering
        \includegraphics[scale=0.5]{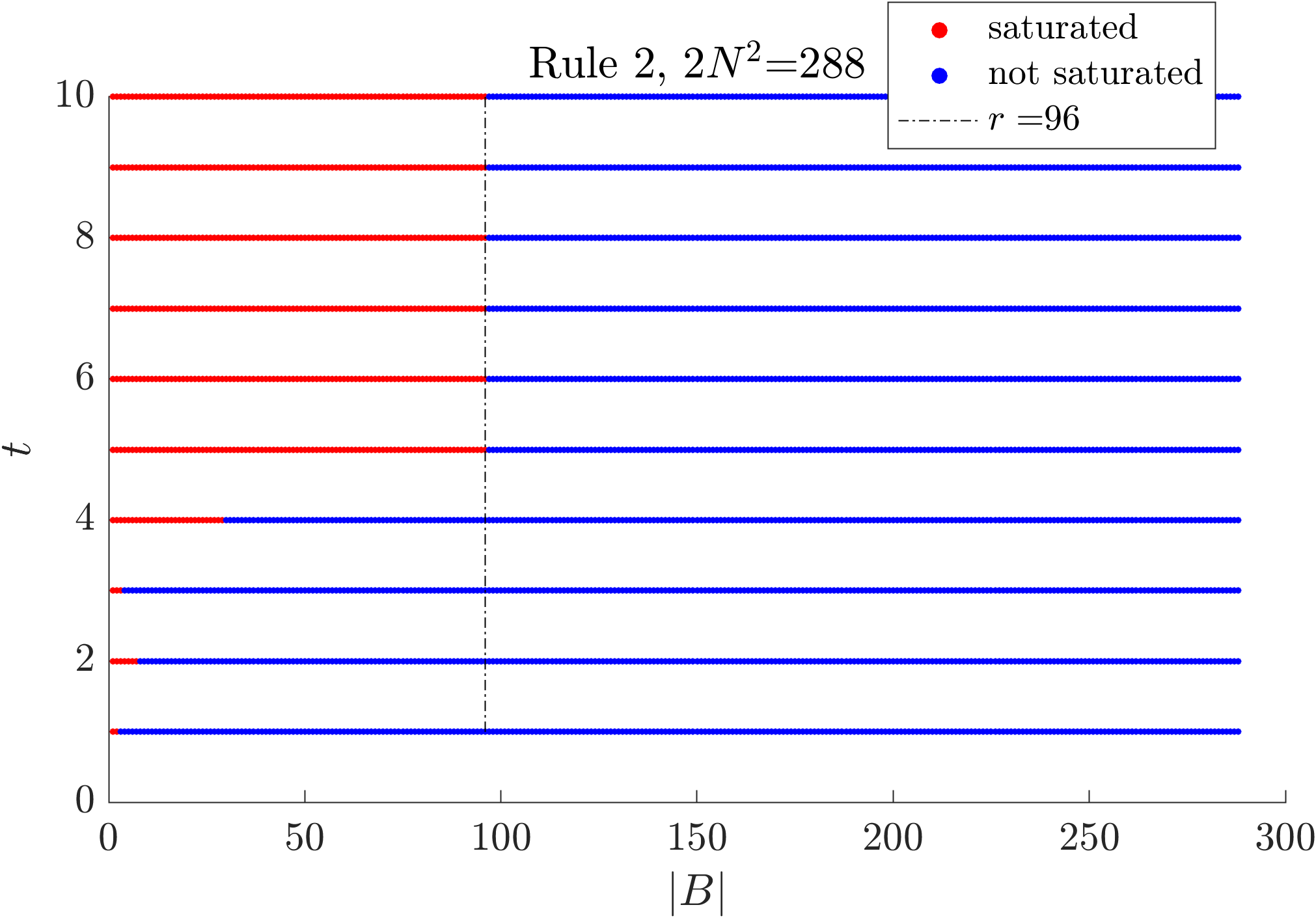}
        \caption{}
        \label{fig:HP_different_odd}
    \end{subfigure}%
    \begin{subfigure}[h]{0.5\textwidth}
        \centering
        \includegraphics[scale=0.5]{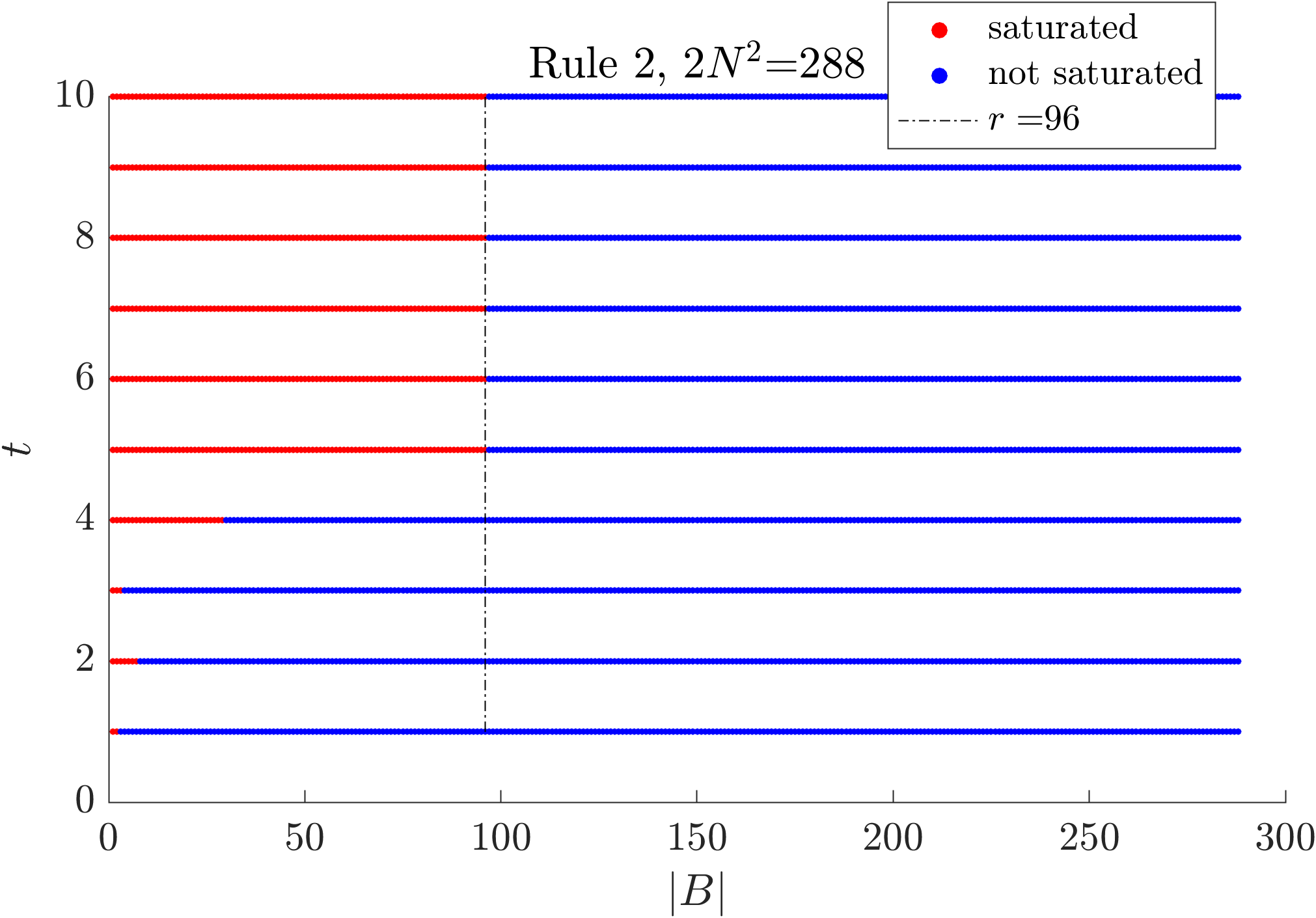}
        \caption{}
        \label{fig:HP_same_odd}
    \end{subfigure}
    \caption{Satisfaction of the sufficient condition in (\ref{eq:rankcondition}) for quantum information recovery in the double layer model for different final times $t$ and erased set sizes $r$, where the erased set $B$ is generated by a randomly selected reference qubit.
    }\label{fig:double_HP}
\end{figure}

We simulate the three measures for double-layer construction, with the results shown in Figure \ref{fig:double_OSG}, Figure \ref{fig:double_EE}, and Figure \ref{fig:double_HP} respectively. When $N$ has an odd factor, the behavior of all three measures closely resembles that of the single-layer model discussed in Section.(\ref{clifford dynamics}), no matter if Rule 1 or Rule 2 is applied to all $2N^2$ qubits. The same features appear when $N$ is a power of two under Rule 1, but when Rule 2 is applied, all measures show the exact same behavior as single-layer system with $N^2$ qubits, rather than doubling the corresponding bounds as one would expect for double-layer system with $2N^2$ qubits. We explain the origin of this peculiarity in Appendix \ref{disjoint_sets}: it is a consequence of an accidental decoupling of the degrees of freedom into 2 decoupled subsets (each of which still scramble as much as possible).

\section{Conclusion}
In this paper, we have described a cartoon Floquet Clifford circuit that is a fast scrambler (in that it rapidly grows operators and generates entanglement).  The interaction graph of the circuit mocks a holographic matrix model, and is designed to be accesible in near-term experiments using neutral atom arrays.  We believe that this work represents a first step along a path towards potential simulations of holographic matrix models in experiment.
Of course, we should strongly emphasize that the Floquet circuits we have described are clearly not meant to actually simulate a model of quantum gravity -- much further work is necessary to deduce the resources needed to perform a serious simulation of a matrix model in a neutral atom quantum processor (or any other hardware).    The original holographic matrix models, for example, exhibit supersymmetry \cite{BanksFischlerShenkerSusskind1997MatrixModel} and multiple interacting matrix degrees of freedom, all of which are too complicated for a first experimental demonstration of our idea.  Still, a number of interesting recent papers \cite{BerensteinHanadaHartnoll2009MultiMatrixEmergentGeometry,Anninos2015LargeNMatrices,Hartnoll2018TopologicalOrderMatrixIsing,Hartnoll2017MatrixQMFromQubits,Frenkel:2023aft} point out that even simplified quantum matrix models without supersymmetry can exhibit many non-trivial signatures of physics in an emergent spacetime.   For this reason, we believe that quantum simulations of a matrix model along the lines we sketched out are worthwhile.   

Although we have focused on the potential implementation of this circuit in a neutral atom array, we should also stress that the permutation gates we use are quite similar to ``power-of-2" interaction graphs which have been realized in cavity QED experiments  \cite{Bentsen:2019rlr, Periwal_2021}.  It would be interesting if this represents an alternative strategy for simulating matrix models.  Another potential modification of our approach would use Gaussian bosonic states, as in the recent work \cite{JeffreyGandariasSchleierSmithSwingle2026BuildingHolographicEntanglement}, rather than stabilizer states of qubits.

In the future, it will be important to understand the generalization of our simple model to non-Clifford ``magic" state dynamics.   This is important to realize any kind of genuine holographic pattern of entanglement.  However, implementing certain holographic protocols such as Hayden-Preskill may become prohibitively difficult in the absence of stabilizer error correction.  Understanding the balance between these competing desires is a task we leave to future work.  

\section*{Acknowledgements}
We thank Monika Schleier-Smith and Amit Vikram for useful discussions.  This work was supported by the Heising-Simons Foundation under Grant 2024-4848.

\begin{appendix}
\section{Decoupling dynamics under Rule 2 when $N=2^k$}\label{disjoint_sets}
When $N=2^k$, every integer $x\in\{1,2,..,N\}$ can be mapped uniquely to some $k$-bit binary string
\begin{align}
    y &= \alpha_{k-1}...\alpha_1 \alpha_0
\end{align}
with digits $\{\alpha_0,...,\alpha_{k-1}\}\in\lbrace 0,1\rbrace$, by choosing
\begin{align}
    x &= \left(\sum_{i=0}^{k-1}\alpha_i 2^i\right) + 1.
\end{align} 
Define $y'$ the cyclic rotation of $y$ that moves the rightmost bit to the leftmost position
\begin{align}
    y' &= \alpha_0\alpha_{k-1}...\alpha_1.
\end{align}
Let $x'$ be the integer corresponding to $y'$, then
\begin{align}
    x' &= \left(\sum_{i=1}^{k-1}\alpha_i 2^{i-1}\right) + \alpha_0 2^{k-1}  + 1.
\end{align}
If $x$ is odd, then $\alpha_0=0$ and
\begin{align}
    \begin{split}
        x' &= \left(\sum_{i=1}^{k-1}\alpha_i 2^{i-1}\right) + 0\cdot 2^{k-1}  + 1 = \left(\sum_{i=1}^{k-1}\alpha_i 2^{i-1}\right) + 1 = \frac{\left(\sum_{i=1}^{k-1}\alpha_i 2^{i}\right)+1}{2} + \frac{1}{2} = \frac{x+1}{2}.
    \end{split}
\end{align}
If $x$ is even, then $\alpha_0=1$, and we instead find 
\begin{align}
    \begin{split}
        x' &= \left(\sum_{i=1}^{k-1}\alpha_i 2^{i-1}\right) + 1\cdot 2^{k-1}  + 1 = (\alpha_1+1)+\left(\sum_{i=2}^{k-1}\alpha_i 2^{i-1}\right) + 2^{k-1} = \frac{2(\alpha_1+1)+\left(\sum_{i=2}^{k-1}\alpha_i 2^i\right) + 2^k}{2}\\
        &=\frac{x+N}{2}.
    \end{split}
\end{align}
Thus, the permutation $\sigma$ defined in (\ref{eq:sigma}) is equivalent as cyclically moving the rightmost bit of a binary string to the leftmost position. Consequently, the cycles generated by $\sigma$ correspond to cyclic rotations of these bit strings. For each $n\in\{0,...,k\}$, let 
\begin{align}
    \begin{split}
        S_n &:= \{\text{binary strings of length $k$ with exactly $n$ ones}\} \label{eq:Sn}.
    \end{split}
\end{align}
Each $S_n$ decomposes into disjoint cycles. 

Now pick any binary string in some cycle and assign it color 1. Since $\sigma$ permutes the strings within that cycle, every string in the same cycle is also assigned color 1. 
Under Rule 2, any qubit $Q_{ij}$ interacts with $Q_{(i\pm 1)(j\pm 1)}$ depending on whether $i$ and $j$ are odd or even, so without loss of generality we can first focus on just one index. Given that assumption, each string of the form
\begin{align}
    \alpha_{k-1}\alpha_{k-2}...\alpha_10
\end{align}
infects its companion
\begin{align}
    \alpha_{k-1}\alpha_{k-2}...\alpha_11
\end{align}
and conversely. For each such pair, if one of them is assigned color 1, then we assign the companion string color 2. Thus, by combining the action of $\sigma$ and the connectivity rule, the two colors propagate throughout all binary strings.

Also note that, because $\sigma$ cyclically rotates the bits, every cycle in $S_{n\geq1}$ contains at least one string whose rightmost bit is 1.

We now restrict to $0<n<N-1$ since the strings in $S_0$ and $S_k$ do not permute. Among the strings in such a $S_n$, exactly $\binom{k-1}{n}$ have rightmost bit 0, and among the strings in $S_{n+1}$, exactly $\binom{k-1}{n}$ have the rightmost bit 1. Rule 2 gives a bijection between these two sets: each rightmost-0 string in $S_n$ infects a unique rightmost-1 string in $S_{n+1}$. By assigning opposite colors to each such pair, and using the fact that every cycle in $S_{n+1}$ contains a rightmost-1 string due to the cyclic rotation of bits under $\sigma$, we see that all strings in receive color 2 if the strings in $S_n$ carry color 1, and vice versa.

By inductively applying this argument over $n$, we find that all $S_n$ with odd $n$ share one color, and all $S_n$ with even $n$ share the other color. Using the identity
\begin{align}
    \sum_{0\leq n \leq k} ^{ n\text{ odd}} \binom{k}{n} &= \sum_{0\leq n \leq k} ^{n\text{ even}} \binom{k}{n},
\end{align}
we conclude that the $2^k$ binary strings split evenly, with half carrying color 1 and half carrying color 2.

Because both the row and column indices follow this pattern, once a pair of indices $(i,j)$ is assigned a color, the colors of all qubits are fixed. In particular, the qubits in each row/column split evenly between the two colors, and therefore all qubits are partitioned into two disjoint color classes of equal size. 
As a direct consequence, Figure \ref{fig:double off-diagonal rule} shows that for odd $i$ and even $j=i+1$, qubits in the top layer $T_{ii}$ and $T_{jj}$ carry the same color, while qubits in the bottom layer $B_{ii}$ and $B_{jj}$ likewise share a common color. Since the corresponding subsets are disconnected, color mixing is forbidden.

\section{Another four-qubit gate choice}\label{app:altgate}
\begin{figure}[t]
    \centering
    \begin{subfigure}[h]{0.5\textwidth}
        \centering
        \includegraphics[scale=0.5]{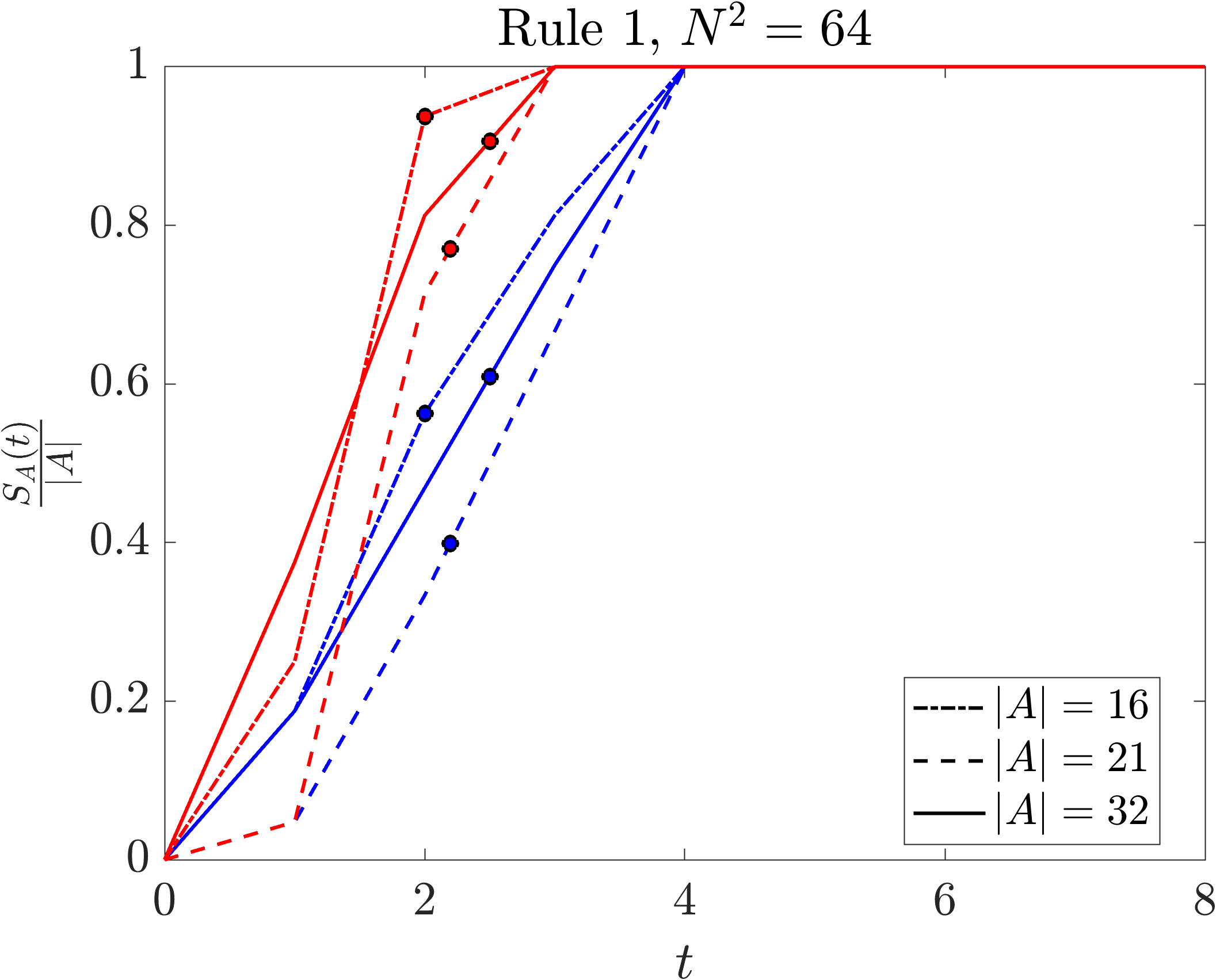}
        \caption{}
        \label{fig:rule1_two_gates}
    \end{subfigure}%
        \begin{subfigure}[h]{0.5\textwidth}
        \centering
        \includegraphics[scale=0.5]{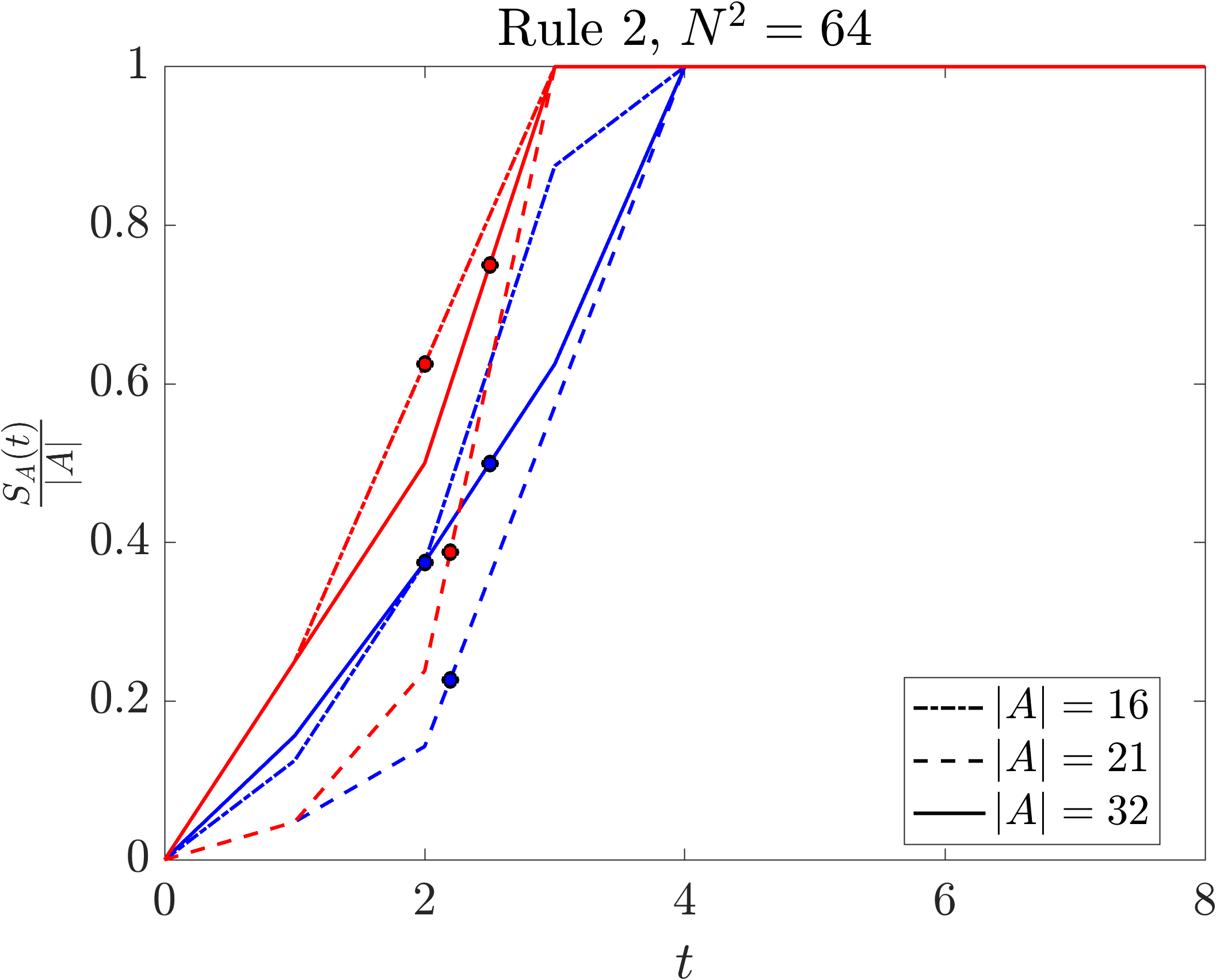}
        \caption{}
        \label{fig:rule2_two_gates}
    \end{subfigure}
    \caption{Entanglement entropy growth for two distinct four-qubit gates in the single layer model. The blue curve corresponds to the gate defined in (\ref{eq:4-qubit_gate}), while the red curve corresponds to $W_{\text{new}}$ in (\ref{new_gate}). }\label{fig:two_gates}
\end{figure}

In Figure \ref{fig:two_gates} we simulate the entanglement entropy growth using an alternative four-qubit Clifford gate,
\begin{align}
    W_{\text{new}}=\text{CNOT}_{q_3\rightarrow q_2} \text{S}_{q_3}\text{CNOT}_{q_2\rightarrow q_4}\text{CNOT}_{q_1\rightarrow q_3} \text{H}_{q_3}\text{CNOT}_{q_4\rightarrow q_1}\text{CNOT}_{q_1\rightarrow q_2}\text{CNOT}_{q_2\rightarrow q_3} \text{CNOT}_{q_3\rightarrow q_4}\text{H}_{q_1}\label{new_gate}
\end{align}
and compare the results results with those obtained using our chosen gate in (\ref{eq:4-qubit_gate}). We find that the two gates exhibit similar behavior, indicating that our conclusions regarding the dynamics of the proposed model extend to other four-qubit gates.  To avoid clutter we have only shown a subset of the multiple diagnostics in Figure \ref{fig:two_gates}, compared to what was discussed in the main text.

\end{appendix}

\bibliography{thebib}
\end{document}